\documentclass[12pt,preprint]{aastex}

\usepackage{amsmath}

\def\ltsima{$\; \buildrel < \over \sim \;$}
\def\simlt{\lower.5ex\hbox{\ltsima}}            
\def\gtsima{$\; \buildrel > \over \sim \;$}
\def\simgt{\lower.5ex\hbox{\gtsima}}            

\newcommand{\asca}{{\it ASCA}}
\newcommand{\rosat}{{\it ROSAT}}
\newcommand{\einstein}{{\it Einstein}}

\newcommand{\heao}{{\it HEAO1}}
\newcommand{\chandra}{{\it Chandra}}
\newcommand{\xmm}{{\it XMM-Newton}}
\newcommand{\exosat}{{\it EXOSAT}}

\newcommand{\logn}{log $N$ - log $S$ relation}

\newcommand{\etal}{et al.}

\newcommand{\erg}{erg s$^{-1}$}
\newcommand{\ergs}{erg cm$^{-2}$ s$^{-1}$}

\newcommand{\de}{deg$^2$}
\newcommand{\nh}{$N_{\rm H}$}
\newcommand{\lx}{$L_{\rm X}$}

\newcommand{\ec}{$E_{\rm c}$}
\newcommand{\lxz}{$L_{\rm X}$-$z$}
\newcommand{\fx}{$F_{\rm X}$}
\newcommand{\cosmo}{($H_0$, $\Omega_{\rm m}$, $\Omega_{\lambda}$)}
\newcommand{\mb}{$M_B$}

\newcommand{\xtI}{{\it X-ray type-I}}
\newcommand{\xtII}{{\it X-ray type-II}}
\newcommand{\otI}{{\it optical type-I}}
\newcommand{\otII}{{\it optical type-II}}

\newcommand{\ciii}{{C~III]$\lambda$1909}}

\newcommand{\mgii}{{Mg~II$\lambda$2800}}

\slugcomment{Accepted for Publication in ApJ}

\lefthead{Ueda et al.}
\righthead{Hard X-ray AGN Luminosity Function}

\begin{document}

\title{Cosmological Evolution of the Hard X-ray AGN Luminosity
Function and the Origin of the Hard X-ray Background}

\author{Yoshihiro Ueda}
\affil{Institute of Space and Astronautical Science,
Kanagawa 229-8510, Japan,
ueda@astro.isas.ac.jp}

\author{Masayuki Akiyama}
\affil{Subaru Telescope, 
National Astronomical Observatory of Japan,
Hilo, HI, 96720}
\email{akiyama@subaru.naoj.org}

\author{Kouji Ohta}
\affil{Department of Astronomy,
Kyoto University, Kyoto, 606-8502, Japan}
\email{ohta@kusastro.kyoto-u.ac.jp}

\and

\author{Takamitsu Miyaji}
\affil{
Department of Physics, 
Carnegie Mellon University
5000 Forbes Ave., Pittsburgh, PA15213, USA}
\email{miyaji@cmu.edu}

\begin{abstract}

We investigate the cosmological evolution of the hard X-ray luminosity
function (HXLF) of Active Galactic Nuclei (AGNs) in the 2--10 keV
luminosity range of $10^{41.5} - 10^{46.5}$ \erg\ as a function of
redshift up to 3. From a combination of surveys conducted at photon
energies above 2 keV with \heao, \asca, and \chandra, we construct a
highly complete ($>$96\%) sample consisting of 247 AGNs over the wide
flux range of $10^{-10} - 3.8\times10^{-15}$ \ergs\ (2--10 keV). For
our purpose, we develop an extensive method of calculating the {\em
intrinsic} (before-absorption) HXLF and the absorption ($N_{\rm H}$)
function. This utilizes the maximum likelihood method fully correcting
for observational biases with consideration of the X-ray spectrum of
each source. We find that (i) the fraction of X-ray absorbed AGNs
decreases with the intrinsic luminosity and (ii) the evolution of the
HXLF of all AGNs (including both type-I and type-II AGNs) is best
described with a luminosity dependent density evolution (LDDE) where
the cutoff redshift increases with the luminosity. Our results
directly constrain the evolution of AGNs that produce a major part of
the hard X-ray background, thus solving its origin quantitatively. A
combination of the HXLF and the \nh\ function enables us to construct
a purely ``observation based'' population synthesis model. We present
basic consequences of this model, and discuss the contribution of
Compton-thick AGNs to the rest of the hard X-ray background.

\end{abstract}


\section{Introduction}

Revealing the cosmological evolution of the AGN luminosity function,
which is directly linked to the accretion history of the universe and
hence the formation history of supermassive black holes in galaxies,
has been a main goal of X-ray surveys. Previously, most of these
studies \citep*[e.g.,][]{mac91,boy93,jon97,pag97,miy00} were made in
the soft X-ray band (\simlt 3 keV) which could be subject to biases
against absorbed AGNs. Hard X-ray surveys (above 2 keV) are a key
ingredient to trace the luminosity function of the whole AGN
population including obscured (type-II) AGNs, main contributors to the
cosmic X-ray background (CXB or XRB; for reviews see e.g.,
\citealt{bol87} and \citealt{fab92}).

Earlier work on the hard X-ray luminosity function (HXLF) by
\citet{boy98} and \citet{laf02} indicates strong evolution of the
HXLF, like that seen in soft X-rays. \citet{cow03} has constrained the
2--8 keV HXLF at two redshift bins ($z$=0.1--1 and $z$=2--4) mainly
from \chandra\ surveys. They argued that the AGN number density for
luminosities lower than $\approx 10^{44}$ \erg\ seems to peak at a
closer redshift than those of higher luminosity, unless they assumed a
very extreme redshift distribution for unidentified sources. This is
consistent with the soft X-ray luminosity function (SXLF) result from
a combined analysis of \rosat\ and \chandra\ data
\citep{has03}. However, the X-ray samples used in the above studies of
the HXLF are still limited in completeness and size. A further study
using a much larger, highly complete sample over a wide
luminosity-redshift range is necessary to unambiguously constrain the
evolution of the luminosity functions of both type-I and type-II AGNs.

Recently Akiyama \etal\ (2003; hereafter A03) have completed an
optical identification program of a 2--10 keV selected sample in the
\asca\ Medium Sensitivity Survey (AMSS; \citealt{ued01}) from an area
of 69 degree$^2$ in the northern sky (Dec.\ $>-10^{\circ}$) with a
flux limit of $3\times10^{-13}$ \ergs\ (AMSSn sample). The AMSSn
sample contains 85 identified AGNs, which is currently the largest
hard-band selected AGN sample that covers the flux range above $\sim
10^{-13}$ \ergs\ (2--10 keV). In this paper we also utilize another
sample from an extension of the identification program of the AMSS in
the southern sky. In total we now have 141 hard X-ray selected AGNs
solely from \asca\ surveys with completeness of 97\% including the
AMSS, the \asca\ Large Sky Survey (=ALSS; \citealt{ued99a};
\citealt{aki00}), the deep \asca\ surveys in the Lockman Hole
\citep{ish01}, and in the Lynx field \citep{oht03}. These highly
complete \asca\ samples provide an ideal opportunity to constrain the
nature of the AGNs, in particular, at the intermediate redshift
universe below $z\approx1$. Furthermore, combining brighter AGN
samples from \heao\ surveys and fainter ones from \chandra\ surveys
enables us to trace the cosmological evolution of AGNs in the 2--10
keV luminosity range of $10^{41.5}-10^{46.5}$ \erg\ at redshifts up to
3.

In this paper we directly constrain the evolution of AGNs that
constitute a major part of the 2--10 keV X-ray background. We
construct a hard X-ray (2--10, 2--8, or 2--7 keV) selected sample with
an extremely high degree of completeness from \heao , \asca , and
\chandra\ surveys. The sample consists of 247 sources in total. In the
analysis we fully consider the detector response and the X-ray
spectrum of each source making use of the best information
available. This step is crucial to correct for selection biases caused
by using a {\em count-rate} (not flux) limited sample. We firstly
formulate the absorption distribution function of AGNs and discuss its
luminosity and redshift dependence. Then, we present our results of an
intrinsic (before absorption) HXLF. Finally, we discuss the
contribution of AGNs to the CXB by constructing a purely ``observation
based'' population synthesis model. We adopt the cosmological
parameters of \cosmo\ = (70$h_{70}$ km s$^{-1}$ Mpc$^{-1}$, 0.3, 0.7)
as a default, but in some cases we also show results with \cosmo\ =
(50$h_{50}$, 1.0, 0.0) for comparison with previous works. Throughout
the paper the ``Log'' symbol represents the base-10 logarithm, while
``ln'' the natural logarithm.

\section{Sample}

To cover a wider luminosity and redshift range for studying the AGN
HXLF, we construct a flux-limited sample by adding published data from
the \heao\ survey and \chandra\ Deep Field North (CDFN) survey to the
\asca\ sources. The whole sample contains 247 hard band selected AGNs
(excluding BL Lac objects). The number of the sources and flux limit
(2--10 keV) in each survey is listed in Table~\ref{table-survey}. The
following subsections describe the details of each survey and
selection criteria of the sample.

\subsection{The {\it HEAO1} sample}

A flux-limited, hard X-ray selected sample from the \heao\ surveys is
included in our analysis as the brightest flux end sample. We use the
same sample as used for follow-up spectroscopic studies with \asca\
and \xmm\ by \citet{miy03}.  This enables us to take into account the
best spectral information at present. It consists of 49 sources,
(Part~I) 28 AGNs from the \heao\ A2 complete sample by \citet{pic82}
and (Part~II) 21 AGNs from a region-limited subsample of the MC-LASS
(\heao\ A1/A3) catalog by \citet{gro92}. Since our purpose is to
constrain the global spectral distribution of AGNs, detailed modeling
of individual sources is not crucial for our studies. Basically we
refer to results of a single power law fit with a neutral absorber at
the source redshift over the Galactic absorption, which are in most
cases acceptable. One source (Kaz 102 / V1803+676) shows a very flat
spectrum, which may be caused by an extreme warm absorber
\citep{miy03a}. We treat the spectrum of this object as an unabsorbed
power law with a photon index, $\Gamma$, of 0.96. In a few sources
there is evidence for a strong warm absorber, dual absorbers and/or a
soft excess \citep{miy03}, where significant residuals are seen from
an absorbed power-law model. For simplicity and consistency with
analysis of the other samples, we only assume a single, neutral
absorption even if the fit is not acceptable. Hence, the column
density we use is an ``effective'' one that may partially contain the
effects of warm or dual absorbers. A soft excess component, if any, is
neglected in calculating the luminosity.

In the Part~I sample we selected 28 emission-line AGNs from
\citet{pic82} brighter than 1.25 $R15$ c s$^{-1}$ in the first scan of
the \heao\ all sky survey at Galactic latitude of $|b| > 20^{\circ}$.
For convenience two Piccinotti AGNs (NGC~4151 and NGC~5548) located in
the overlapping region of the Part~II sample are included in the
latter sample. We refer to the new identification of the \heao\ source
H~0917-074 as the Seyfert~2 galaxy MCG--1--24--12 ($z=0.0198$)
revealed by \citet{mal02}. The $R15$ count rate is defined as a sum of
the two detectors, HED and MED, from the \heao\ A2 experiment
\citep{mar80}. The 1.25 $R15$ c s$^{-1}$ count rate corresponds to a
flux limit of $2.7\times10^{-11}$ \ergs\ (2--10 keV) for a power law
with $\Gamma=1.7$. The sample is 100\% complete. Except for Mrk~590
and H~0917-074, which are observed by \xmm\ \citep{miy03} and {\it
BeppoSAX} \citep{mal02} respectively, the AGNs have been observed with
\asca .

The Part~II sample comes from a complete flux-limited AGN sample
constructed from the MC-LASS catalog. The area covers a 55 degree
region from the North Ecliptic Pole. The count rate limit is 0.0036
LASS c s$^{-1}$, corresponding to $1.9\times10^{-11}$ \ergs\ (2--10
keV) for $\Gamma=1.7$. We exclude 3C~351 and H~1443+421 from the
original \citet{gro92} catalog because of possible source confusion
and/or misidentification problem\footnote{The \asca\ observation
revealed another X-ray source (BL Lac object) having a comparable flux
to 3C~351 in the same field of view. Although H~1443+421 was
identified as a broad line AGN at $z=1.4$ in \citet{gro92}, no bright
counter part is confirmed in the \rosat\ All Sky Survey
(RASS). Considering its especially large error in the original X-ray
source position, we suspect that it would be misidentification.}. The
completeness of the total Grossan sample is 86\% (or 85\% within our
limited region according to comparison between the \heao\ A1 and A3
catalogs). \citet{gro92} argues, however, that a majority of
unidentified sources turned out to be active coronae (Galactic stars)
or BL-Lac objects from a comparison with \einstein\ and \rosat\
data. Thus, we make the first working hypothesis that the
identification of the \heao\ Part~II sample is complete for AGNs other
than BL Lac objects, and do not apply any completeness
correction. Note that our results for the HXLF presented below are not
affected over the statistical error by whether the completeness
correction (85\%) is applied to this sample or not. We have spectral
data of \asca\ or \xmm\ for all the sources except H~1537+339, for
which we assume a power law with $\Gamma=1.7$ and no absorption. For
NGC~4151, whose X-ray spectrum is known to be highly variable and
complex, we adopt $\Gamma=1.5$ and \nh = $2\times10^{23}$ cm$^{-2}$ as
typical values from the results of \citet{wea94}. To convert the LASS
count rate into physical flux unit in our analysis, we create an
approximated energy response by scanning the quantum efficiency curve
of the A1 detector given by \citet{woo84}.

\subsection{The {\it ASCA} sample}

\subsubsection{The AMSS}

The AMSSn sample consists of 87 serendipitous sources in the AMSS
X-ray catalog \citep{ued01} in the northern sky (DEC$>20^{\circ}$),
detected with a detection significance larger than 5.5 $\sigma$ at a
flux limit of $3\times10^{-13}$ in the 2--10 keV band. The criteria
for the detection significance and flux limit were set relatively
conservative to lessen systematic errors due to the Eddington bias and
source confusion. Detailed description of the optical identification
is given in A03. They are identified with 78 AGNs, (including 3 BL Lac
objects), 7 clusters of galaxies, and 1 galactic star. One source
(1AXG~J133937+2730) is still unidentified\footnote{By a recent \xmm\
observation we detected another hard source located close ($<1'$) to a
$z$=0.908 quasar that was previously thought to be the optical
counterpart of 1AXG~J133937+2730 (see {\it note added in proof} of
A03). The new source, about 3 times brighter than the $z$=0.908 quasar
in the 2--10 keV band, is most likely a Seyfert~2 galaxy at
$z\approx0.1$ inferred from the optical extended morphology and the
hard X-ray spectrum, although we have not obtained its optical
spectrum yet.}. The AMSS catalog was constructed only from ``faint''
fields with a total GIS count rate smaller than 0.8 c s$^{-1}$. As
mentioned in \citet{ued99b}, this could cause a bias against bright
sources with fluxes larger than $\sim 10^{-11}$ \ergs\ (2--10
keV). Hence, to ensure the completeness we also apply an upper limit
of $5\times10^{-12}$ \ergs\ (2--10 keV) to define a good statistical
sample for the present study. Finally, we have 74 identified AGNs from
A03 after excluding the BL Lac objects and the brightest source
1AXG~J122135+7518. The survey area is calculated as a function of
limiting count rate by the same technique as described in
\citet{ued99b}. The total area covered for the AMSSn sample is 69
deg$^{2}$ at bright fluxes. Except for 1AXG~J233200+1945 and
1AXG~J234725+0053, for which \xmm\ follow-up observations have been
performed, we only have information of the hardness ratio $HR1$,
defined as $(C1-C2)/(C1+C2)$, where $C1$ and $C2$ represent the 2--10
keV and 0.7--2 keV vignetting-corrected count rate, respectively.

Furthermore, we are conducting an extension of the identification
program of the AMSS sources to the southern sky (we here tentatively
call it the ``AMSSs sample''; Akiyama \etal , in preparation),
including X-ray sources in a new Gas Imaging Spectrometer (=GIS;
\citealt{oha96}) source catalog constructed from \asca\ archival
data taken after 1997 (Ueda \etal , in preparation). The extended
sample currently available provides 20 additional identified AGNs to
the AMSSn sample with 2 unidentified objects.

\subsubsection{The ALSS}

The \asca\ Large Sky Survey (ALSS) is a wide-area unbiased source
survey covering continuous area of 5.5 \de\ near the North Elliptic
Pole \citep{ued98,ued99a}. From 34 sources detected in the 2--7 keV
band with the Solid-state Imaging Spectrometer (=SIS; \citealt{bur91})
above 3.5$\sigma$, \citet{aki00} identified 30 AGNs, 2 clusters of
galaxies, and 1 star. We use this sample with recent updates based on
follow-up X-ray observations\footnote{A \chandra\ observation of
AX~J131832+3259 (the only unidentified source in \citealt{aki00})
revealed no corresponding X-ray source within the \asca\ error circle,
suggesting that it was possibly a fake source. We ignore this source,
although strong time variability cannot be ruled out. Based on the
\xmm\ follow-up we change identification of AXJ~131021+3019 from the
original paper \citep{aki00} to a broad line AGN at $z$=1.152. The
X-ray spectrum can be fit with an unabsorbed power law with
$\Gamma=2.0\pm0.1$. We also performed an \xmm\ observation of
AXJ~131816+3240. This confirms the original identification (a
$z=1.649$ AGN) but reveals no significant absorption from a power law
with $\Gamma=2.1\pm0.2$.}. The completeness of the ALSS is 100\%. For
6 sources we refer to the result of follow-up X-ray observations by
\asca\ or \xmm . For the rest we use the result of a simultaneous fit
to the GIS and SIS spectra from the original \asca\ survey data.

\subsubsection{The \asca\ deep surveys}

We also use AGN samples from the \asca\ deep surveys in the Lockman
Hole field \citep{ish01} and in the Lynx field \citep{oht03}. In both
fields only the SIS data were analyzed, where higher sensitivity than
the GIS data was achieved thanks to its superior positional
resolution. The results of \rosat\ surveys in the same fields were
utilized to reduce source confusion in source detection. To estimate
the X-ray spectrum of each source, we also utilize recent data
obtained by \xmm\ or \chandra .

From the \asca\ Lockman Hole 2--7 keV survey we have 12 AGNs (6 type-1
AGNs and 6 type-II AGNs) at a flux limit of $2.7\times10^{-14}$ \ergs .
They belong to the ``hard-band selected sample'' defined by
\citet{ish01}, consisting of the 12 AGNs, 1 star, 1 cluster of
galaxies, and 1 unidentified source. The unidentified source
(designated ASCA-2 in \citealt{ish01}) is highly variable and was
detected only in one epoch out of the four \asca\ observations but
neither in the \rosat\ nor \xmm\ observations. Considering the unusual
nature of this source, we do not apply a completeness correction. The
survey area is calculated as a function of a limiting sensitivity by
the same technique applied to the ALSS \citep{ued99a} after excluding
a very low exposure region. The area at the brightest fluxes is 0.224
deg$^2$. Except for two sources (PSPC-504 and HRI-307) good quality
X-ray spectra in the 0.5--10 keV range are available from the \xmm\
observations \citep{mai02}. For the rest we use the \asca\ hardness
ratio between the SIS 1--2 keV and 2--7 keV count rates to obtain
spectral information.

We have 5 optically identified AGNs out of 6 hard-band selected
sources in the \asca\ Lynx 2--7 keV survey \citep{oht03}. Accordingly
we apply completeness correction using this ratio (5/6). The
sensitivity limit is $3.6\times10^{-14}$ \ergs\ (2--7 keV). Three
sources, including the type-II quasar candidate AX~J08494+4454
\citep*{aki02}, are located within the field of view of the \chandra\
ACIS observations performed in 1999. For these sources we perform
spectral analysis using the \chandra\ archive data, while we use the
hardness ratio between the SIS 0.7--2 keV and 2--7 keV count rates for
the others.

\subsection{The CDFN sample}
\label{sec-cdfn}

At the faintest end we define a flux limited sample in the 2--8 keV
band above $3.0\times10^{-15}$ \ergs\ (corresponding to
$3.8\times10^{-15}$ \ergs\ in the 2--10 keV band for $\Gamma=1.4$)
from the X-ray catalog of the \chandra\ Deep Survey North (CDFN) of a
1 Msec exposure \citep{bra01}. We refer to the published results of
optical identification by \citet{bar02} (we also utilize updated
information given in \citealt{bar03}), including both spectroscopic
and photometric redshifts. Following their definition, we divide the
X-ray sample into two, the bright sample and the deep sample. The
bright sample has a flux limit of $5\times10^{-15}$ \ergs\ (2--8 keV)
within a 10 arcmin radius region around the field center covering an
area of 294 arcmin$^2$. The deep sample is defined within a 6.5 arcmin
radius, for which we limit the flux range brighter than
$3.0\times10^{-15}$ \ergs\ (2--8 keV) to ensure high
completeness. Finally redshifts of 57 sources (including 8 photometric
redshifts) are known out of 61 sources. The completeness is 93\%. From
flux values listed in the \citet{bra01} table we calculate flux
hardness-ratio between the 0.5--2 and 2--8 keV bands, from which we
estimate the absorption and luminosity, using an appropriate response
of the ACIS instrument. We verified that our results are consistent
with the absorption values derived by \citet{bar02} assuming
$\Gamma=1.8$ (plotted in their Fig.\ 18). As argued in \citet{bar02}
the X-ray sample is complete down to its flux limits over the defined
area. Note that unlike the other samples, which are defined by {\em
count rate} limits, this CDFN sample is defined by a true {\em flux}
limit calculated by assuming a single power law with varying photon
indices \citep[see][]{bra01}. We have taken this into account when
correcting for selection biases of the sample.

\subsection{Sample Summary}
\label{sec-samplesummary}

In summary, we have in total 247 hard-band selected AGNs covering the
wide flux range from $10^{-10}$ to $3.8\times 10^{-15}$ \ergs\ (2--10
keV), 49 AGNs from the \heao\ surveys, 141 from \asca , and 57 from
\chandra\ with completeness of 100\%, 97\%, and 93\%,
respectively. AGNs at the overall flux range constitute a major part
($\simgt$70\%) of the CXB in the 2--10 keV band.

Our basic policy is to consider the HXLF of the entire AGN population
we detected (except for a minor population of BL Lac objects) to avoid
errors caused by further classification, which is not important in our
discussion of the CXB origin as a whole. Nevertheless, for convenience
of discussion, we classify each object into either a type-I or type-II
AGN based on the X-ray or optical properties. We define \xtII\ AGNs as
those showing a best-fit absorption column density at the source
redshift larger than $10^{22}$ cm$^{-2}$, and \xtI\ AGNs as less
absorbed sources. The method to obtain the column density is described
in the next section.

From the optical data we basically define \otI\ AGNs as those showing
broad emission lines and the rest as \otII\ AGNs. For the \heao\
Part~I AGNs we refer to the table compiled by \citet{sch97} except
NGC~2992, NGC~5506, and ESO~103, which we all classify as \otII . All
the \heao\ Part~II sources are classified as \otI , including Seyfert
1.5, by referring mainly to the NED database. For the \asca\ sources
located at $z<0.6$, where the H$\beta$ region is covered in the
optical spectroscopic observations (see A03), \otII\ AGNs are defined
as those where broad H${\beta}$ lines are not significantly detected
above 3$\sigma$. For those at $z>0.6$, we only classify those with
broad emission lines of \mgii , \ciii , and/or Ly$\alpha$ as \otI\
AGNs. For sources in the Lockman hole we adopt the same classification
as in \citet{leh01}. As for the CDFN sources, we refer to the last
column in Table~1 of \citet{bar02} if there are broad emission
lines. Thus, these optical classification schemes are not uniform for
the whole sample and sometimes ambiguous, depending on the quality of
the available optical spectrum and the wavelength coverage. In
\S~\ref{sec-otIIfrac} we discuss correlation between \otII\ AGNs and
\xtII\ AGNs.

\begin{figure}
\epsscale{1.0}
\plotone{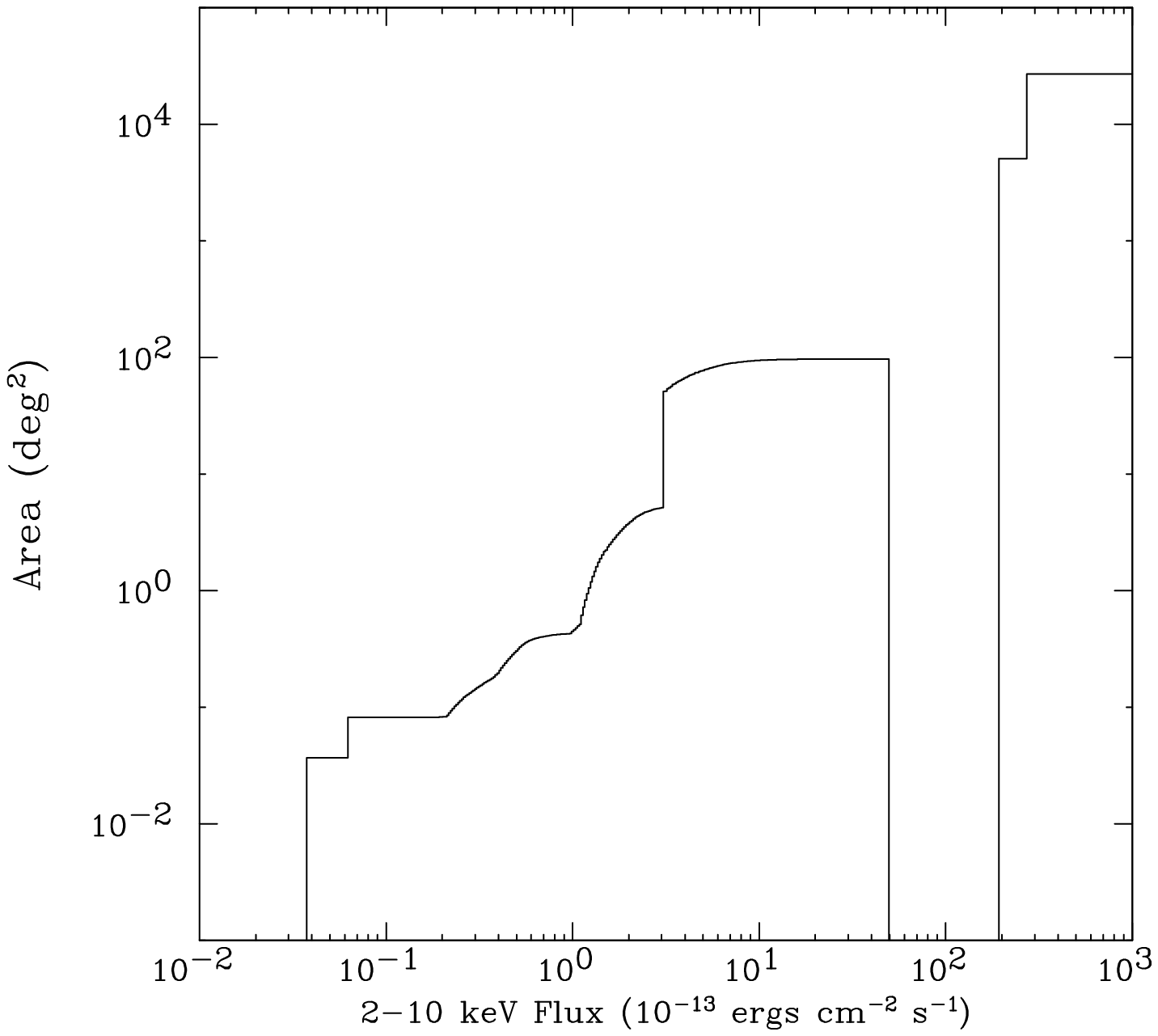}
\caption{
The total survey area as a function of a 2--10 keV flux for our whole sample.
\label{area}}
\end{figure}

Figure~\ref{area} shows the combined survey area as a function of
limiting fluxes in the 2--10 keV band for the whole sample. The gap
between $1.9\times10^{-11}$ and $5\times10^{-12}$ \ergs\ (2--10 keV)
is a result of setting the upper flux limit for the AMSS sample as
mentioned above. Because the area is originally given as a function of
count rate (except for the CDFN) in the survey band, we have assumed a
power law with $\Gamma=$1.7, 1.6, and 1.4 for \heao, \asca, and
\chandra, respectively, in plotting this figure. To make an effective
correction for completeness in the calculation of the HXLF, we modify
the survey area by multiplying these completeness percentages,
assuming that the luminosity/redshift distribution of unidentified
sources is the same as that of identified sources. This assumption may
not be always true because the incompleteness due to difficulty of
optical identification is likely to be caused by non-random
effects. Nevertheless, this does not cause any significant impacts on
our conclusions presented in this paper within statistical
uncertainties, thanks to the extremely high degree of the
completeness.

\section{Analysis}

\subsection{Goals}

In this paper, we derive absorption column-density distribution
function (``\nh\ function'') and the intrinsic hard X-ray luminosity
function from our sample. The ``intrinsic'' luminosity (hereafter \lx )
means that it is corrected for an absorption (the value before being
absorbed) in the rest frame 2--10 keV band. This enables us to discuss
more directly the cosmological evolution of AGNs. Note that, in
contrast, the SXLF by \citet{miy00} is given in the observers' frame
by assuming a single photon index of 2 for all the sources. As a
matter of fact, it has not been practical to correct for absorption of
a \rosat\ source because of the narrow bandpass limited below 2
keV. Consequently, however, the \rosat\ results could be subject to
biases against absorbed sources in that they miss them or
significantly underestimate the true luminosity, in particular at
lower redshifts where $k$-correction is not significant. The use of a
hard-band selected sample itself greatly improves these disadvantages.

\subsection{Absorption and Luminosity}
\label{sec-absandlum}

To estimate the intrinsic luminosity (\lx ) with the best accuracy, we
consider the spectrum of each object. After the photon index
($\Gamma$) and the absorption column density at the source redshift
(\nh) are determined, as described below, we convert the hard-band
count rate into the absorption-corrected flux in the rest frame
2--10 keV band, \fx , using the detector response. The Galactic
absorption estimated from the H~I observation \citep{dic90} is taken
into account field by field, even though its effect is mostly negligible. 
Then \lx\ is obtained as
\begin{equation*}
	L_{\rm X} = 4\pi d_{\rm L}^2 F_{\rm X},
\end{equation*}
where $d_{\rm L}$ is the luminosity distance. The statistical error in
\lx\ arises from the error in the hard-band count rate itself as well
as from those in the spectral parameters. The (propagated) errors in
\lx\ for the AMSS sources are $<$20\% and typically 5--15\%,
respectively. They correspond to an uncertainty of $<$0.1 in Log \lx,
which is negligible in our study.

Except for objects whose column density and photon index are
independently determined from follow-up X-ray observations, we use the
hardness ratio to estimate \nh\ or $\Gamma$. Here assumption must be
made on the spectral shape because only one observational quantity is
available. As an intrinsic spectrum before being absorbed, we assume a
power law with an exponential cutoff, in the form of $E^{-\Gamma} {\rm
exp} (\frac{E}{E_{\rm c}})$, with a Compton reflection component
composed of reprocessed X-rays through cold, optically thick matter
surrounding the emitter. We derive the corresponding \nh\ for each AGN
if the observed hardness ratio is larger than the value expected from
a ``template spectrum'', representing a typical intrinsic spectrum of
AGNs. Otherwise, we derive the corresponding $\Gamma$ value that
accounts for the observed hardness ratio assuming no absorption.

Considering the overall results from spectral analysis of nearby AGNs
\citep[e.g.,][]{nan94,tur97,geo98}, we adopt $\Gamma=1.9$ for the
template spectrum. This value is also consistent with the mean photon
index of the Lockman hole AGNs observed by \xmm\ \citep{mai02}. The
high energy cutoff ($E_{\rm c}$) of several hundred keV is found from
bright Seyfert galaxies \citep[e.g.,][]{zdz95}. In this paper we
assume $E_{\rm c}=$500 keV. Although the cutoff does not affect our
determination of \nh\ and \lx , it becomes important for predicting
the contribution of AGNs to the CXB above 10 keV
(\S~\ref{sec-comptonthick}).
The reflection component is commonly detected in the X-ray spectra of
nearby Seyfert galaxies \citep[e.g.,][]{nan94}, arising most likely
from the accretion disk. To calculate the spectrum, we use the
``pexrav'' model \citep{mag95} in the XSPEC package, assuming a solid
angle of 2$\pi$, an inclination angle of cos$(i)$=0.5, and the Solar
abundance for all elements. The ratio of the reflection component to
the direct one is about 10\% just below 7.1 keV (the K edge energy of
cold iron atoms) and rapidly decreases toward lower energies (0.1\% at
1 keV), while above 7.1 keV it has a maximum of about 70\% around 30
keV, producing a ``reflection hump''. This component makes the
apparent slope in the 0.5--10 keV slightly harder than the intrinsic
power law, depending on the redshift, but is almost negligible for
determination of \nh\ from an individual spectrum. There are arguments
that the relative strength of the reflection component (apparently)
decreases with the luminosity \citep[e.g.,][]{law97,ree00}, is smaller
in radio loud AGNs than radio quite ones \citep{zdz95}, and may be
even different between Seyfert 1 and Seyfert 2 galaxies \citep{zdz00}. 
For simplicity such possible dependence is neglected in this paper. We
also ignore iron K emission lines in the spectra, whose contribution
is less than few percent of the 2--10 keV continuum flux. These
effects neglected here are not important for our main conclusions of
the \nh\ function and the HXLF. Note that in the previous related
papers \citep{aki00,aki03} a single power law with $\Gamma=1.7$ was
adopted for the template spectrum. This photon index was obtained from
early X-ray missions such as \heao\ and \exosat\
\citep[e.g.,][]{tur89}, but is now considered to be in most cases an
apparent slope affected by warm absorbers and a reflection hump
\citep[e.g.,][]{nan94}. The choice of the $\Gamma$ value does not
essentially affect our results. For comparison we will also show
results of the \nh\ function obtained from the $\Gamma=1.7$
assumption.

\begin{figure}
\epsscale{1.0}
\plotone{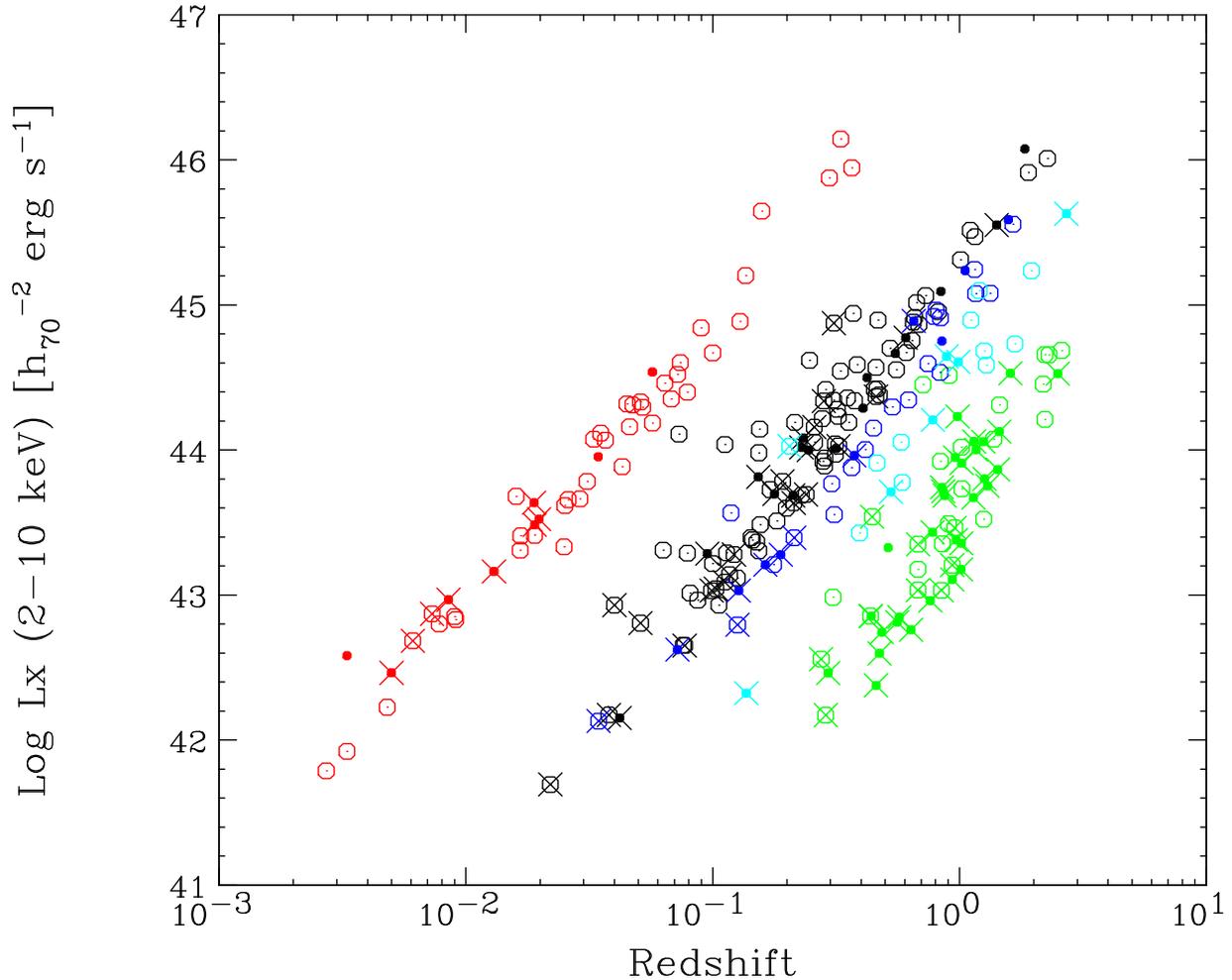}
\caption{ 
The redshift versus luminosity plot for our AGN sample (red: {\it
HEAO1}, black: AMSSn+AMSSs, blue: ALSS, cyan: {\it ASCA} deep surveys,
green: CDFN). The luminosity \lx\ is an ``intrinsic'' one in the
rest frame 2--10 keV band before being absorbed. Dots: {\it X-ray
type-II} AGNs (Log \nh\ $>$22). Crosses: {\it optical type-II} AGNs
(with no significant broad emission lines).
\label{z-l}}
\end{figure}

\begin{figure}
\epsscale{1.0}
\plotone{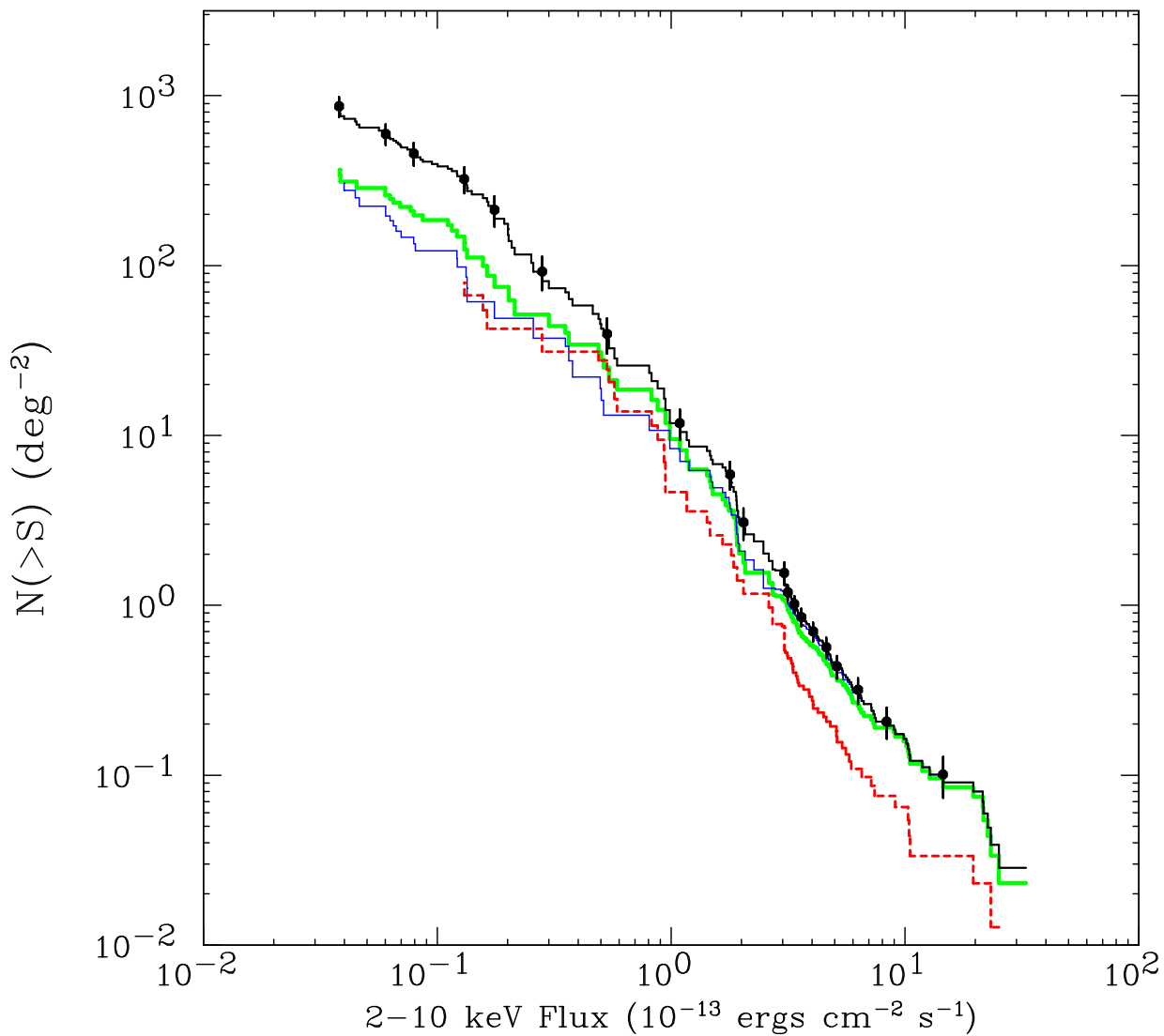}
\caption{
The observed \logn\ of the identified AGNs of our sample (uppermost
line, black). The attached errors bar indicate $1\sigma$ Poisson
errors in source counts. The thick sold line (green): \xtI\ AGNs. The
thin solid line (blue): AGNs at $z<0.8$. The dashed line (red): AGNs
with Log \lx\ $\geq$44.5.
\label{sample-logn}}
\end{figure}

The redshift versus luminosity distribution of the whole sample is
plotted in Figure~\ref{z-l}. \xtII\ AGNs are marked with dots and
\otII\ AGNs with crosses. Figure~\ref{sample-logn} shows the \logn s
for the identified AGNs with different compositions; the total, \xtI\
AGNs, AGNs at $z<0.8$, and those with $L_{\rm X}>10^{44.5}$
$h_{70}^{-2}$ \erg. In plotting this, we use an observed (after
absorption) flux in the 2--10 keV band for each source based on the
best-fit spectral parameters as determined above. For the CDFN sources
we simply use the original 2--8 keV fluxes given in \citet{bra01} by
multiplying a constant factor of 1.25 (i.e., assuming $\Gamma=1.4$) to
convert them into 2--10 keV fluxes.

\begin{figure}
\epsscale{1.0}
\plotone{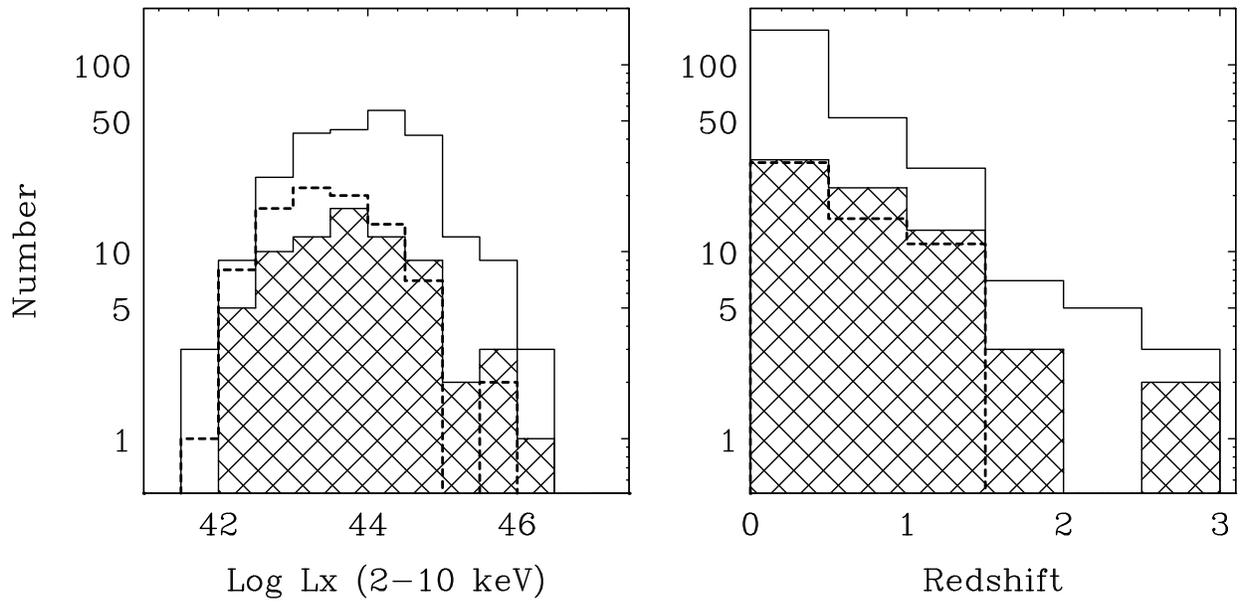}
\caption{
{\it left} (a): The luminosity distribution of the whole sample, compared with 
that of \xtII\ AGNs (shaded histogram) and that of \otII\ AGNs (dashed
histogram). {\it right} (b): The redshift distribution of the same samples 
as (a). 
\label{l-zhist}}
\end{figure}

Figure~\ref{l-zhist}(a) shows the observed \lx\ distribution of the
whole sample compared with that of \xtII\ AGNs (shaded histogram) and
\otII\ AGNs (dashed). Figure~\ref{l-zhist}(b) shows their redshift
distribution. Figure~\ref{l-zhist}(a) indicates that the fraction of
\otII\ AGNs is not constant but larger in the lower luminosity
range. The same tendency, though less evident, is also implied for
\xtII\ AGNs. It is important to note, however, that the observed ratio
of absorbed AGNs does not give a correct estimate of the true
fraction. In particular, there is a selection bias against detecting
hard sources in a {\em count-rate} limited sample: at a given \lx\ and
$z$, sources with larger \nh\ are more difficult to detect as they
give smaller count rates. Also, because of statistical errors in the
hardness ratio, the ``observed'' \nh\ distribution does not give the
true distribution. To correct for these biases we perform quantitative
analysis in \S~\ref{sec-nhf}, by means of the maximum likelihood fit
where the detector response is fully taken into account.

\subsection{Principle of the Statistical Analysis}

Below we explain the principle of how we determine the the \nh\
function and the HXLF by statistical analysis from our sample (for
simplicity we do not consider the statistical error in the \nh\
determination at this moment). The luminosity function, representing
the number density per unit comoving volume per Log \lx\ as a function
of \lx\ and $z$, is expressed as
\begin{equation*}
	\frac{{\rm d} \Phi (L_{\rm X}, z)}{{\rm d Log} L_{\rm X}}.
\end{equation*}
To describe the distribution of spectral parameters of AGNs at a given
luminosity and redshift, we introduce the \nh\ function, $ f (L_{\rm
X}, z; N_{\rm H}), $ a probability-distribution function for the
absorption column-density. The \nh\ function has a
unit of (Log \nh)$^{-1}$ and is normalized to unity over a defined
\nh\ region, that is
\begin{equation}
\int_{N_{\rm H,min}}^{N_{\rm H,max}} f (L_{\rm X}, z; N_{\rm H}) {\rm d Log} N_{\rm H} = 1.
\end{equation}
Generally, the form of the \nh\ function is dependent on the
luminosity and the redshift. Similarly, we can also define the
``photon index function'' $g (L_{\rm X}, z; \Gamma)$ per unit $\Gamma$
space. Throughout our paper we assume that there is no dependence of
$f$ on the photon index $\Gamma$.

If these functions are modeled by analytical expressions, we can use
the maximum likelihood (hereafter ML) method to search for the
best-fit parameters and their statistical errors \citep[see
e.g.,][]{miy00}. The basic idea is to make the probability of finding
the set of our observational results (i.e., distribution of the
redshift, the count rate, and the spectrum) highest from the given
survey conditions. Here we define the likelihood estimator $L$, to 
be minimized through the fitting, as
\begin{equation}
L = -2 \sum_{i} {\rm ln}
\frac{N(N_{{\rm H}i}, \Gamma_i,  L_{{\rm X}i}, z_i)}
{\int \int \int \int N( N_{\rm H}, \Gamma, L_{{\rm X}}, z) {\rm d Log} N_{\rm H} {\rm d} \Gamma {\rm d Log} L_{\rm X} {\rm d} z}
\label{eq-likelihood}
\end{equation}
where $i$ represents each object in our sample and $N$ is the expected
number of
detected sources per logarithm column density, per unit photon index,
per logarithm luminosity, and per unit redshift. Here
the expected number $N$ is calculated as
\begin{equation}
N(N_{\rm H}, \Gamma, L_{{\rm X}}, z) = 
f(L_{\rm X}, z; N_{\rm H}) g(L_{\rm X}, z; \Gamma) 
\frac{{\rm d} \Phi (L_{\rm X}, z)}{{\rm d Log} L_{\rm X}}
d_{\rm A}(z)^2 (1+z)^3 c \frac{{\rm d}\tau}{{\rm d}z}(z) \sum_{j} A_{\rm j} (N_{\rm H},\Gamma,L_{\rm X},z)
\label{eq-likelihood2}
\end{equation}
where $d_{\rm A}^2$ and $\frac{{\rm d}\tau}{{\rm d}z}$ is the angular
distance and the look back time per unit redshift, respectively, both
are functions of $z$. The symbol $j$ represents each survey and
$A_{\rm j}$ the survey area for a count rate expected from a source
with \nh , $\Gamma$, \lx , and $z$, which can be calculated through
the detector response and the luminosity distance. Eqs.\
(\ref{eq-likelihood}) \& (\ref{eq-likelihood2}) are a generalization
of the formula (4) \& (5) of \citet{miy00} respectively towards the
case where the spectrum of each source is taken into account in terms
of $\Gamma$ and \nh .

In the ML fit the 1$\sigma$ statistical error of each free parameter
is obtained as a deviation from the best fit value when the $L$ value
increases by 1.0 from its minimum. Unlike the $\chi^2$ fit, the
minimum value itself does not have a meaning in statistics and
therefore we cannot evaluate the absolute goodness of the fit. For the
HXLF we use the two dimensional Kolgomorov-Smirnov test (2DKS test;
\citealt*{fas87}) applied for the luminosity-redshift distribution
between the data and model: if the 2DKS probability is found to be
larger than 0.2 then the model is considered acceptable within the
statistics. Since the fit itself does not constrain the absolute
normalization of the HXLF, we estimate it from the total number of
observed sources.

As far as we use the above formula in the ML fit, the three (\nh ,
photon index, and luminosity) functions are coupled with one
another. In other words, ideally, all the parameters of the three
functions must be constrained simultaneously. This is not practical,
however, requiring huge computation time for the fit to
converge. Hence, in this paper, we take an approximated, step by step
approach as follows. (1) First, we determine the parameters of the
\nh\ function using the observed values of the luminosity, redshift,
and photon index of each source, without modeling the HXLF and the
photon-index function by an analytical form (the delta-function
approximation; see \S~\ref{sec-nhfana} for details). (2) Next, we
determine the model (and its best-fit parameters) of the HXLF through
the ML fit by fixing the \nh\ function determined in the first
step. For simplicity, we do not formalize the photon-index function
but use $\Gamma=1.9$ for the calculation of $L$, which is found to be
a sufficiently good approximation. The second step is repeated for
several different parameters of the \nh\ function chosen within the
statistical errors. We finally adopt the case when a combination of
the HXLF and the \nh\ function meet observational constraints best, as
described in \S~\ref{sec-HXLF}. The details of the first and second
steps are described in the subsequent two sections.

\section{The \nh\ function}
\label{sec-nhf}

\subsection{Analysis Method}
\label{sec-nhfana}

Below we derive the \nh\ function by representing the photon-index
function and the HXLF with a superposition of the delta functions
having a discrete peak at their observed values in the three
dimensional space of ($\Gamma$, \lx , $z$).  In this approximation
the formula of the likelihood estimator that constrains only the \nh\
function can be reduced to
\begin{equation}
L = -2 \sum_{i} {\rm ln}
\frac{f(L_{{\rm X}i}, z_i; N_{{\rm H}i}) \sum_{j} A_{\rm j} (N_{{\rm H}i},\Gamma_i,L_{{\rm X}i},z_i)}
{\int f(L_{{\rm X}i}, z_i; N_{{\rm H}}) \sum_{j} A_{\rm j} (N_{{\rm H}},\Gamma_i,L_{{\rm X}i},z_i) {\rm d Log} N_{\rm H}}.
\end{equation}
Here, the relative normalization of the delta function of each object
is adjusted to give the minimum $L$ value in
formula~(\ref{eq-likelihood}).

To make the best estimate of the \nh\ function, we also correct for
bias arising from statistical errors in the hardness ratio in the ML
fit. We derive \nh\ of each object at the source redshift from the
best-fit hardness ratio value but its propagated error can be very
large. As a result, the observed \nh\ function could be distorted from
the true one: for example, because the hardness-ratio range
corresponding to small absorptions (e.g., 20.5$<$ Log \nh\ = $<$21)
becomes extremely narrow at high redshifts, the probability of finding
objects in this \nh\ region is reduced when the hardness ratio is
subject to statistical errors. Since it is difficult to make direct
correction of the observed \nh\ function, we take the ``forward
analysis method'' to the observed data to constrain the parameters of
the (true) \nh\ function.  Similarly to do a spectral fit with limited
energy resolution, we introduce the ``\nh\ response matrix function''
$M_i(N^0_{{\rm H}}, N_{\rm H})$, representing the probability of
finding an {\em observed} value of $N_{\rm H}$ from the $i$-th object
if it had a {\em true} absorption of $N^0_{\rm H}$. The matrix is
normalized to unity between $20 \leq {\rm Log} N_{\rm H} < 24$. Then
the \nh\ function term $f(L_{{\rm X}i}, z_i; N_{{\rm H}})$ in the
above formula is replaced by

\begin{equation}
\int M_i(N^0_{\rm H}, N_{{\rm H}}) f(L_{{\rm X}i}, z_i; N^0_{{\rm H}}) {\rm d Log} N^0_{\rm H}.
\end{equation}
The \nh\ response matrix function is calculated for each object
based on the observed count-rate errors in the two bands, using the
hardness-ratio curve given as a function of \nh\ at the redshift of
the object. For objects whose \nh\ and photon index are independently
measured, we use the diagonal matrix assuming no statistical error.

\subsection{Results}

\begin{figure}
\epsscale{0.8}
\plotone{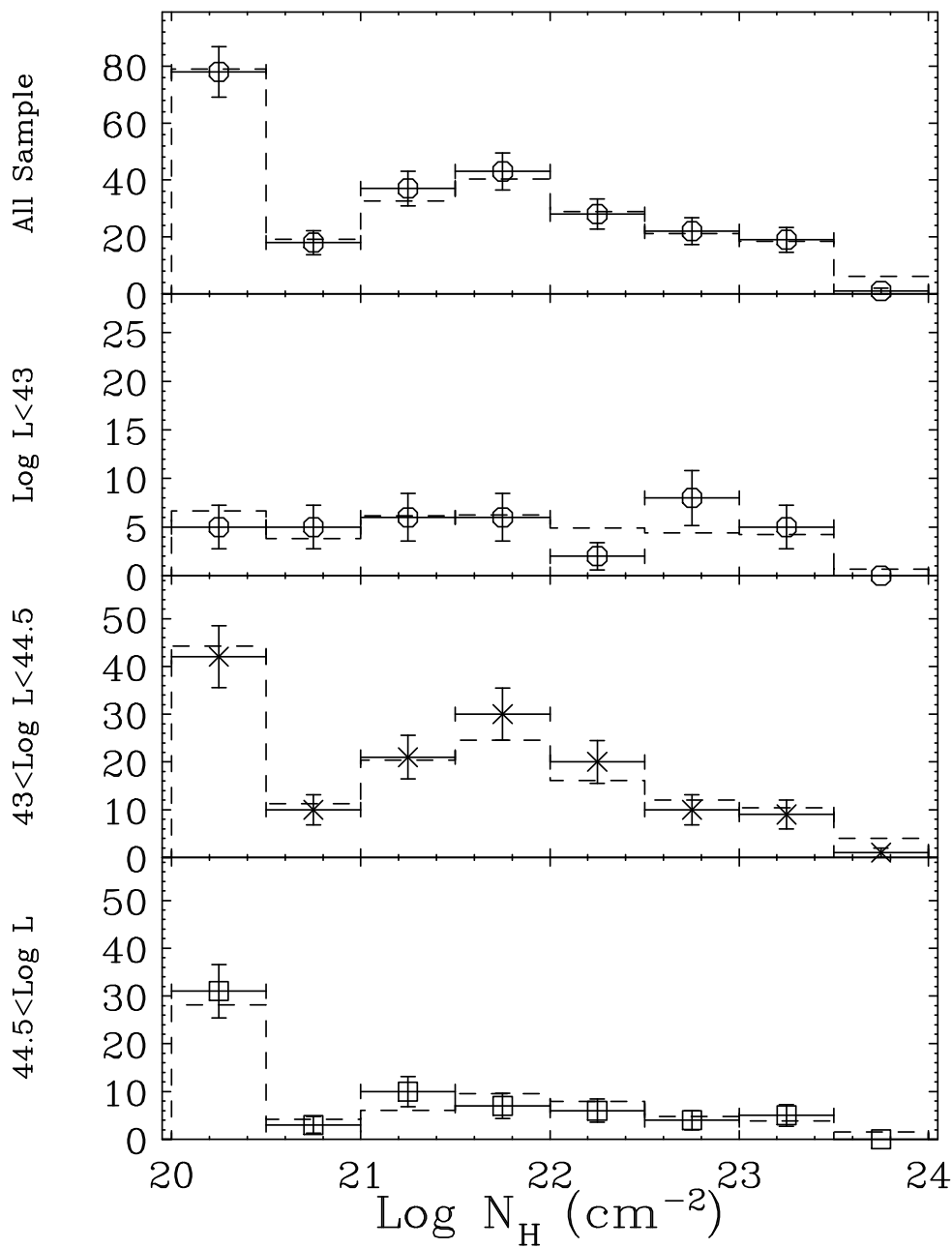}
\caption{
The observed \nh\ distributions (with 1$\sigma$ Poisson errors) are
compared with the prediction from the best-fit \nh\ function (dashed
histogram) in different luminosity ranges (from upper to lower panels:
total, Log \lx\ $<$ 43, 43 $\leq$ Log \lx\ $<$ 44.5, and Log \lx\ $\geq$
44.5).
\label{nhdist}}
\end{figure}

\begin{figure}
\epsscale{0.8}
\plotone{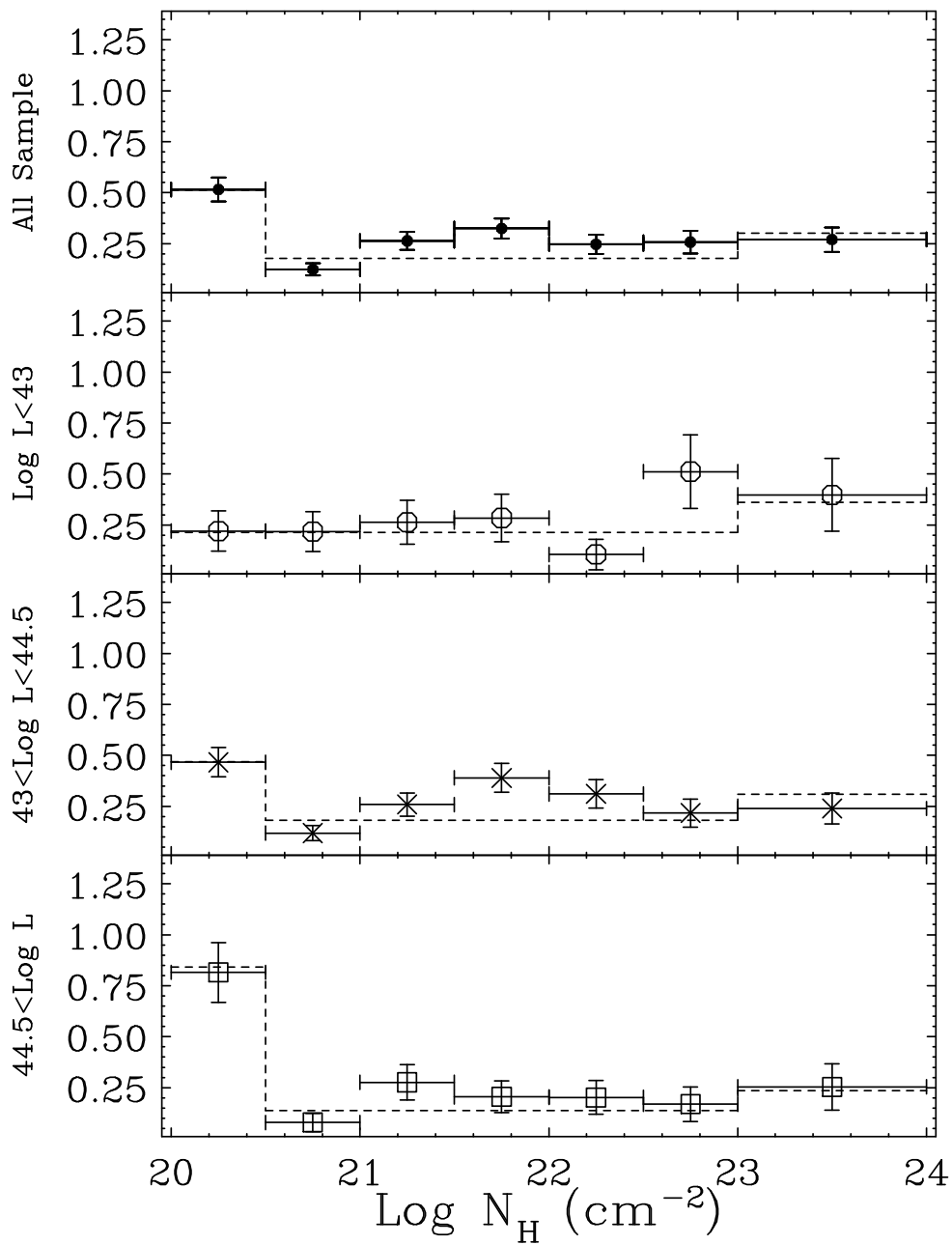}
\caption{
The observed \nh\ functions of our sample (with 1$\sigma$ statistical
errors) in different luminosity ranges (same as Figure~\ref{nhdist}.) 
The dashed lines represent the best-fit model of the \nh\ function,
obtained by correcting for biases due to the statistical error in \nh .
\label{nhfunc}}
\end{figure}

The histograms of Figure~\ref{nhdist} show the observed \nh\
distribution (in units of number per bin) in different luminosity
ranges (from the upper to lower panels, total, Log \lx $< 43$, $43 \leq$
Log \lx $<44.5$, and Log \lx $\geq44.5$), whereas those of
Figure~\ref{nhfunc} are the ``observed'' \nh\ function (the
probability distribution function in units of (Log \nh )$^{-1}$,
normalized to unity in Log \nh\ = 20 and 24) obtained only by
correcting for the dependence of survey area on \nh . As we have
mentioned, these plots are inevitably subject to statistical errors in
each \nh\ measurement, but are useful to make a first order
estimate. Column densities smaller than Log \nh\ $< 20$ are set to be
Log \nh\ = 20 as an effective zero value. Considering large errors
in the best-fit \nh\ value when Log \nh\ exceeds $\simeq$23, we here
merge the two bins of Log \nh\ = 23--24.

In this paper we simply model the \nh\ function by a combination of
three flat functions that have different values in the ranges of
$N_{\rm H} < 20.5$, $20.5 \leq {\rm Log} N_{\rm H} < 23$, and ${\rm
Log} N_{\rm H} \geq 23$. We find that this form gives a sufficiently
good explanation of the observed \nh\ distribution. Considering the
limited statistics, we fix the ratio of the \nh\ function between Log
\nh\ = 23--24 and Log \nh\ = 20.5--23, $\epsilon$, at 1.7, based on
the \nh\ distribution of nearby, optically selected Seyfert~2 galaxies
\citep{ris99}. We assume that $\epsilon$ is independent of the
luminosity or redshift. Introducing the $\psi$ parameter, the fraction
of absorbed AGNs (Log \nh\ = 22--24) to total AGNs (Log \nh\
$\leq$24), which is generally a function of both \lx\ and $z$, we can
write the form of the \nh\ function as
\begin{equation}
f(L_{{\rm X}}, z; N_{{\rm H}}) =
\begin{cases}
 2-\frac{5+2\epsilon}{1+\epsilon}\psi (L_{{\rm X}}, z) & (20.0 \leq {\rm Log} N_{\rm H} < 20.5)\\
 \frac{1}{1+\epsilon} \psi (L_{{\rm X}}, z) & (20.5 \leq {\rm Log} N_{\rm H} < 23.0)\\
 \frac{\epsilon}{1+\epsilon} \psi (L_{{\rm X}}, z) & (23.0 \leq {\rm Log} N_{\rm H} < 24.0).
\end{cases}
\end{equation}

\begin{figure}
\epsscale{1.0}
\plotone{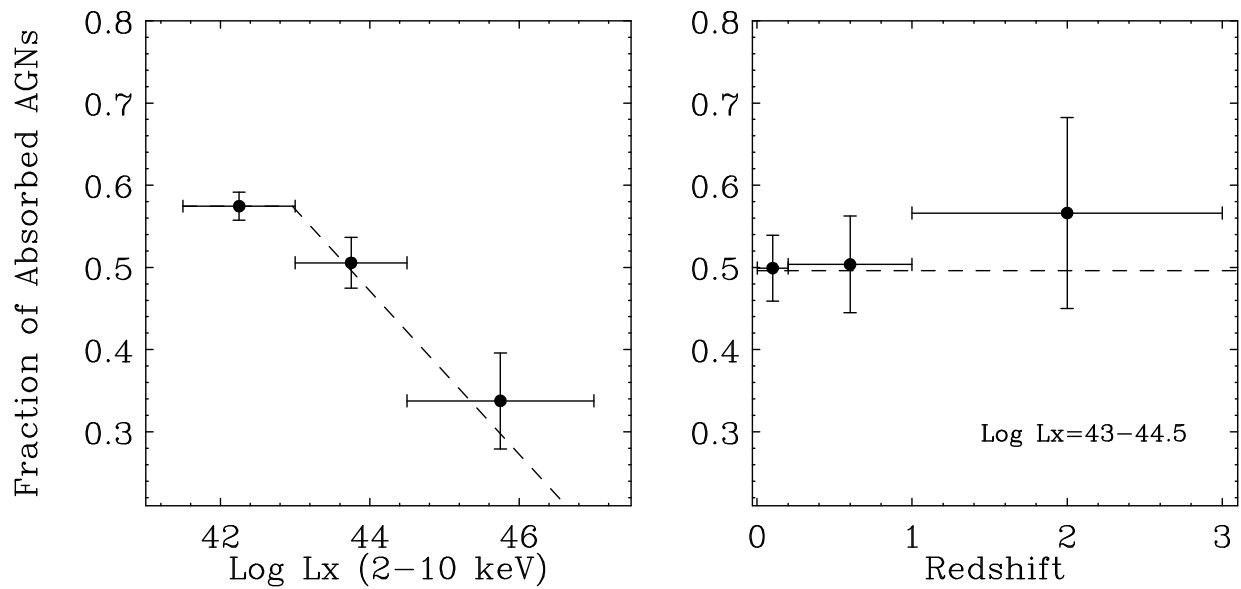}
\caption{
The fraction of absorbed AGNs with Log \nh\ $>22$ to all AGNs
with Log \nh\ $<24$, given as as a function of ({\it left} :a)
luminosity and ({\it right} :b) redshift.  The data points in (b) are
calculated from AGNs in the luminosity range of Log \lx\ =
43--44.5. Dashed lines represent the best fit model of the \nh\
function.
\label{absfrac1.9}}

\end{figure}

The comparison of the observed \nh\ functions in different luminosity
ranges indicates that the fraction of absorbed (non-absorbed) sources
is not constant, being smaller (larger) in higher luminosities. To
express the luminosity dependence, we formalize $\psi$ by a linear
function of Log \lx\ within a limited range:
\begin{equation}
\psi(L_{{\rm X}}, z) = {\rm min} [\psi_{\rm max}, {\rm max} [\psi_{44} - \beta ({\rm Log} L_{\rm X} - 44), 0] ]
\end{equation}
where
\begin{equation}
\psi_{max} = \frac{1+\epsilon}{3+\epsilon}.
\end{equation}
Here we set the upper limit for $\psi$ ($\psi_{\rm max}$=0.574 if
$\epsilon=1.7$) so that the value of the \nh\ function at Log \nh\
$<$20.5 does not become less than that of Log \nh\
$\geq$20.5. Introducing such ``cutoff'' is reasonable because a
significant population of unabsorbed AGNs is known at Log \lx\ $\simlt
41$ \citep[e.g.,][]{ter03}. We note that the main results will be
unchanged if the $\psi$ parameter is allowed to become larger than
$\psi_{\rm max}$ at low luminosities. The redshift dependence of the
absorption fraction is neglected. In fact, assuming that $\psi_{44}$
is proportional to $(1+z)^{\delta}$, we find that $\delta$ is
consistent with zero within the statistical error (see also
Figure~\ref{absfrac1.9}(b)). Because our sample covers the wide \lx -
$z$ range combined from surveys with different flux limits, it is
possible to constrain the luminosity dependence and the redshift
dependence independently.

\begin{figure}
\epsscale{1.0}
\plotone{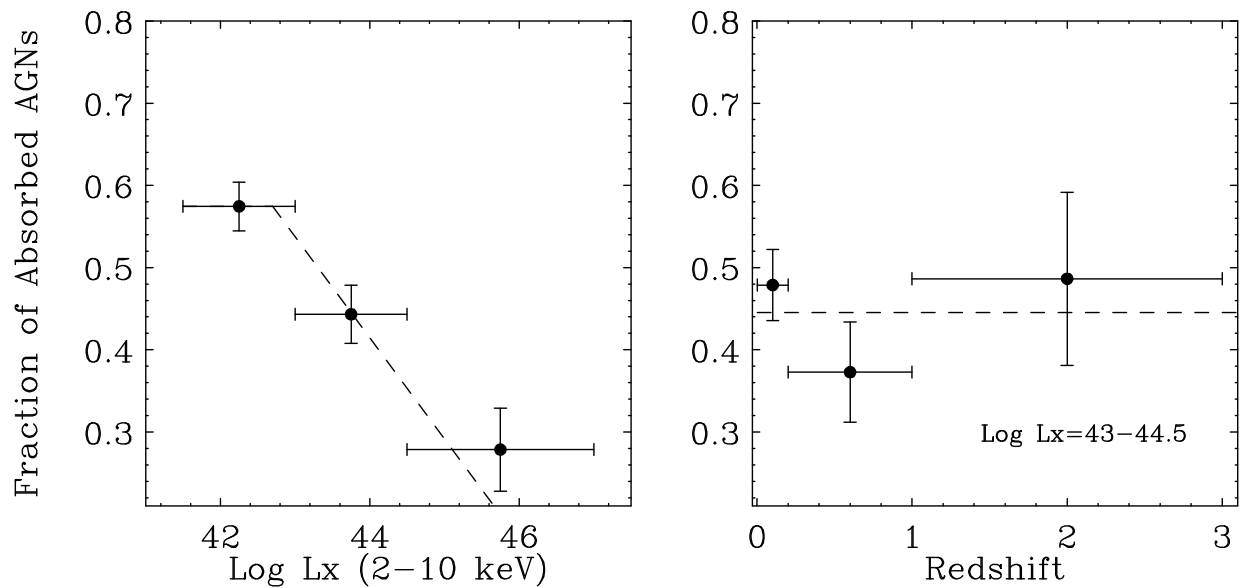}
\caption{
The same as Figure~\ref{absfrac1.9} obtained when a power law of
$\Gamma=1.7$ with no reflection component is assumed for the template
spectrum.
\label{absfrac1.7}}
\end{figure}

We perform the ML fit of the \nh\ function to our sample with the two
free parameters $\psi_{44}$ and $\beta$. The results are summarized in
Table~\ref{table-nhf}. The results obtained with \cosmo\ = (50, 1.0,
0.0) and those with the $\Gamma=1.7$ assumption for the template
spectrum are also given. To evaluate the goodness of the fit, the
observed \nh\ distributions are compared with the model prediction
(dashed line) in Figure~\ref{nhdist}. The 1-dimensional KS test yields
matching probabilities higher than 0.70. It is recognized that the
weak peak centered at Log \nh\ of $21.5-22$ is well reproduced through
the \nh\ response matrix function. We find that $\beta$ is
significantly larger than zero at $>3\sigma$ level, demonstrating the
luminosity dependence of the absorbed-AGN fraction: the fraction of
AGNs with Log \nh\ $>$22 to all GNs with Log \nh\ $<$24 decreases from
57$^{+0}_{-5}$\% at $L_{\rm X} = 10^{43}$ $h_{70}^{-2}$ \erg\ to
37$\pm$5\% at $L_{\rm X} = 10^{45}$ $h_{70}^{-2}$ \erg. In
Figure~\ref{nhfunc} we plot the best-fit model of the \nh\
function. Figure~\ref{absfrac1.9}(a) shows the averaged absorbed-AGN
fraction derived in three luminosity ranges at all redshifts together
with the best-fit model. Figure~\ref{absfrac1.9}(b) shows its redshift
dependence derived from the sample of $43<$ Log \lx $<44.5$ with the
best fit value calculated for its average luminosity. Although the
statistical error is large, no significant redshift dependence is
evident from our data. For comparison the same results obtained by
assuming $\Gamma=1.7$ (with no reflection component) are shown in
Figure~\ref{absfrac1.7}(a) and (b).

\subsection{Luminosity Dependence of the Fraction of X-ray Absorbed AGNs}

Our result indicates that simple extension of the ``unified
scheme'' to higher luminosity where the fraction of absorbed AGNs is
assumed to be constant needs to be modified. We call this a ``modified
unified scheme'', which should be taken into account in population
synthesis models of the CXB. This was originally suggested by
\citet{law82} and is consistent with the lack of type-II quasars seen
in the ALSS sample as well as in the \heao\ sample
\citep{aki00,miy03}.

To confirm this argument, we briefly discuss possible systematic
effects caused by relying on a single hardness ratio without making
detailed spectral analysis (which is practically impossible for most
of \asca\ sources). Firstly, we have ignored possible contribution of
soft components over the absorbed continuum, such as a scattered
component from the nucleus or a thermal emission by starburst
activities. Basically, it leads us to underestimate the fraction of
heavily absorbed sources. In a fixed energy band, the soft component
becomes more important at lower redshifts. Thus, this effect only
works to strengthen our argument because luminous objects are found at
higher redshifts in a flux-limited sample.
Secondly, we have assumed that the intrinsic power law index
($\Gamma$) does not depend on the luminosity. If there was correlation
where the photon index is larger at higher luminosities then our
result could be biased. We do not see such tendency, however, from our
sources whose $\Gamma$ and \nh\ are independently measured. Moreover,
as recognized from comparison between the $\Gamma=1.9$ and
$\Gamma=1.7$ results (Figure~\ref{absfrac1.9} and \ref{absfrac1.7}),
the luminosity dependence of the absorbed fraction is too large to be
explained by a systematic difference in $\Gamma$ unless it is much
larger than 0.2 between Log \lx\ = 43--44.5 and 44.5--46.5. These
considerations support our conclusion that the absorbed-AGN fraction
decreases with the luminosity.

We perform a more quantitative check of the overall effects of soft
components on the result of the \nh\ function. According to the \asca\
results of nearby Seyfert-II Galaxies \citep[e.g.,][]{tur97}, the
relative normalization of the soft component is, in most cases, less
than 5\% of the absorbed continuum. We find that the best-fit
parameters of the \nh\ functions do not change within errors even in
the extreme case that {\it all} the sources have a 5\% scattered
component.  Furthermore, even though the available data are limited,
the number fraction of absorbed AGNs with strong soft components
(i.e., those showing double peaks in the 0.5--10 keV band spectra) is
small in a flux-limited sample regardless of flux levels. The X-ray
spectroscopic study of the \heao\ sample \citep{miy03} and the \xmm\
result of the Lockman Hole sources (see Figure~3 of \citealt{mai02})
both indicate that it is only a few percent of the total sample (2 out
of 49 and 1 (the source designated \#50) out of $\approx$50,
respectively).

In the analysis we ignore ``Compton-thick'' AGNs with column densities
of Log \nh\ $>$24 assuming that such objects do not exist in our
sample detected below 10 keV. However, the sample may contain some
Compton-thick AGNs exhibiting reflection-dominated spectra below 10
keV. In this case we could underestimate not only \nh\ but also the
intrinsic luminosity by more than an order of magnitude. Although the
number density of Compton-thick AGNs detectable below 10 keV is poorly
known at present, we infer it unlikely that it constitutes a
significant fraction ($>$a few percent) in our sample, as indicated
by the result of \xmm\ and \chandra\ deep surveys
\citep[e.g.,][]{mai02,bar02}. Indeed, such a small percentage is
consistent with our estimate based on our population synthesis model
(\S~\ref{sec-pop}) when Compton-thick AGNs are included. Thus, as far
as we discuss only Compton-thin AGNs, we conclude that our results
should be reliable.

\subsection{Fraction of \otII\ AGNs}
\label{sec-otIIfrac}

\begin{figure}
\epsscale{1.0}
\plotone{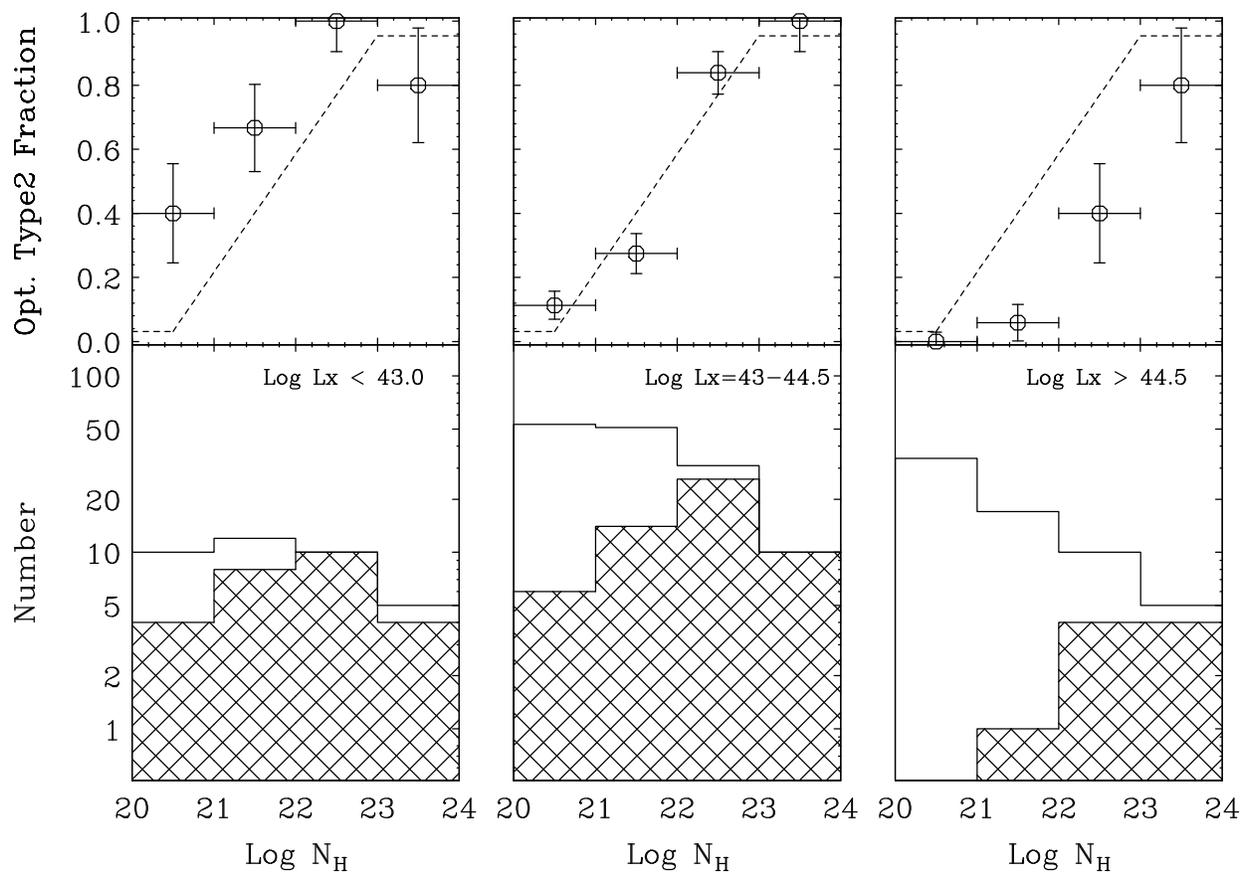}
\caption{
Upper panels: the fraction of \otII\ AGNs given as a function of \nh
. Lower panels: observed histogram of the total sample compared with 
that of only \otII\ AGNs (shaded histogram). The three figures
correspond to different luminosity ranges ({\it left}: Log \lx\ $<$43, 
{\it center}: 43$\leq$ Log \lx\ $<$44.5, {\it right}:  Log \lx\ $\geq$44.5).
The dashed line represents the best-fit analytical model determined in 
the whole luminosity range.
\label{agn2f}}
\end{figure}

It is possible to examine the fraction of \otII\ AGNs as a function of
\nh . One has to bear in mind, however, that the current definition of
\otII\ AGNs in our sample is heterogeneous and is even dependent on
the quality of the optical spectra because it is based on the
detection ``significance'' of broad emission lines
(\S~\ref{sec-samplesummary}). Hence, the results presented here should
be taken as upper limits for the \otII\ AGN fraction. Basically they
can be obtained by comparing the observed \nh\ distribution of \otII\
AGNs with that of the total AGNs because the survey area is common to
both types at the same \nh . Figure~\ref{agn2f} shows the fraction of
\otII\ AGNs, together with the observed histograms, as a function of
\nh\ given in three luminosity ranges. The dashed line represents the
best-fit analytical model determined in the whole luminosity range.

These figures indicate that there is good correlation between X-ray
and optical classification of AGNs, as expected: almost all AGNs with
Log \nh\ $>$23 are \otII\ AGNs, while there exists a small fraction of
\otII\ AGNs in \xtI\ AGNs. Apparently there seems to be luminosity
dependence in the fraction of \otII\ AGNs. At the low luminosity range
its fraction in AGNs in Log \nh\ $<$23 seems to be larger than the
average. However, these results may be highly subject to observational
biases: AGNs of lower luminosities are more likely to be contaminated
by galaxies which make it more difficult to detect broad emission
lines. Hence, we do not argue for the luminosity difference from these
data.

\section{The Hard X-ray Luminosity Function (HXLF)}
\label{sec-HXLF}

Using the \nh\ function obtained in the previous section, we
investigate the cosmological evolution of the HXLF of all AGNs
including both type-I and type-II AGNs. We note again that our HXLF is
defined for the absorption-corrected luminosity in the rest frame
2--10 keV band. In calculating the HXLF, unlike the \nh\ function, we
exclude objects at $z<0.015$ to avoid possible effects of the local
over-density and mis-estimate of their distances.

A practical goal here is to find an analytical expression that
describes the overall HXLF data well, most preferably, a continuous
function of \lx\ and $z$. Even though it does not have direct physical
meaning, having such formula makes it very convenient to calculate
various observational quantities and construct a population synthesis
model. We search for a model that not only fits the HXLF data but also
reproduces, when combined with the \nh\ function, other observational
constraints, such as source counts at fainter fluxes than the flux
limit in the 0.5--2 keV and 2--10 keV bands, the CXB intensity, and a
redshift distribution of AGNs from other survey data, by extrapolating
the form over the whole \lx\ and $z$ region. Here we refer to the
result of direct source counts and fluctuation analysis obtained from
the CDFN by \citet{miy02}. As for the absolute CXB intensity we adopt
the best-fit value by \citet{kus02}, $6.4(\pm0.7)\times10^{-8}$ \ergs\
Str$^{-1}$ in the 2--10 keV band (corresponding to a normalization of
9.7 keV cm$^{-2}$ s$^{-1}$ Str$^{-1}$ keV$^{-1}$ at 1 keV for a photon
index of 1.4), derived from a wide area of 50 deg$^2$ with the \asca\
GIS. This value is close to the median in the 90\% confidence error
region of a bayesian estimate by \citet{bar00} using the \asca\ and
{\it BeppoSAX} results (10.0$^{+0.6}_{-0.9}$ keV cm$^{2}$ s$^{-2}$
keV$^{-1}$ at 1 keV), but larger by $\simeq$20\% than the \heao\
measurement \citep{mar80} most probably because of cross-calibration error
\citep[see][]{bar00}. The use of the \asca\ value for self consistency
of our analysis is justified, because the contribution of \heao\
sources to the 2--10 keV CXB is less than 3\% and the absolute flux
calibration between \asca\ and \chandra\ is accurate within 10\%
\citep[see e.g.,][]{bar01}.

Once the analytical expression is chosen for the HXLF, the free
parameters are determined through the ML fit to our sample according
to formula (\ref{eq-likelihood}). After the ML fit, we check the
consistency of the model with the observational constraints listed
above and iterate these processes by changing the HXLF model or tuning
the fixed parameters (including those of the \nh\ function within the
errors). Since there are unlimited choices for an acceptable model
within statistics, we try to select as simple expression as possible.
Throughout the paper we adopt a smoothly-connected two power-law form
to describe the present-day HXLF,

\begin{equation}
\frac{{\rm d} \Phi (L_{\rm X}, z=0)}{{\rm d Log} L_{\rm X}} 
= A [(L_{\rm X}/L_{*})^{\gamma 1} + (L_{\rm X}/L_{*})^{\gamma 2}]^{-1}.
\end{equation}

\subsection{The PLE and the PDE}

To begin with, we tried the two simplest models, the pure luminosity
evolution (PLE) model and the pure density evolution (PDE) model.
By introducing the evolution factor
\begin{equation}
e(z) = 
\begin{cases}
   (1+z)^{p1}  &(z<z_{\rm c})\\
    e(z_{\rm c})[(1+z)/(1+z_{\rm c})]^{p2}  &(z \geq z_{\rm c}),
\end{cases}
\label{eq-evolv}
\end{equation}
the PLE model is expressed as
\begin{equation}
\frac{{\rm d} \Phi (L_{\rm X}, z)}{{\rm d Log} L_{\rm X}} 
= \frac{{\rm d} \Phi (L_{\rm X}/e(z), 0)}{{\rm d Log} L_{\rm X}},
\label{eq-PLE}
\end{equation}
while the PDE model is 
\begin{equation}
\frac{{\rm d} \Phi (L_{\rm X}, z)}{{\rm d Log} L_{\rm X}} 
= \frac{{\rm d} \Phi (L_{\rm X}, 0)}{{\rm d Log} L_{\rm X}} e(z).
\label{eq-PDE}
\end{equation}
The best fit parameters are summarized in Table~\ref{table-hxlf}
together with the adopted parameters of the \nh\ function. It is
interesting to note that we obtain the cutoff redshift above which the
evolution terminates of $z_{\rm c} \simeq 1.2$ in both models. This
value is smaller than the \rosat\ result, $z_{\rm c} \simeq 1.6$
\citep{miy00}. In terms of the 2DKS test performed over the whole
\lxz\ region, both models are found to be acceptable.

We find several difficulties, however, in adopting either the PDE or
PLE model as our basic description of the HXLF that meets all the
observational constraints. Firstly, the integrated intensity of AGNs
with Log \lx\ = 41.5--48 and $z<5.0$ calculated from the best-fit PLE
and PDE model overproduces the observed 2--10 keV CXB flux of
\citet{kus02} by a factor of 1.21 and 2.02, respectively. These values
significantly exceed the 90\% confidence region of the CXB intensity
by \citet{bar00}. More seriously, the predicted 0.5--2 keV source
counts in faintest flux levels largely overestimate the CDFN data by a
factor of $>$2.6 at $S \simeq 7\times10^{-17}$ \ergs. We find these
problems cannot be resolved by tuning $p2$ within a reasonable
range. These facts indicate that the number density of sources with
smaller luminosities and/or at higher redshifts than our sample have
to be reduced considerably from a simple extrapolation from the
best-fit PLE or PDE model.
Secondly, even though the significance is marginal, these models give
relatively poor description of the HXLF data at $z>0.8$ as non-random
residuals are left over a wide luminosity region. In the PLE, the
model systematically underestimates all the data of Log \lx\ $<45$ at
$z=0.8-1.6$ by a factor of $\simeq$2. Indeed, the 1-dimensional KS
test performed for the (local) luminosity distribution in $z$=0.8--1.6
from 54 objects yields a matching probability of only 0.03 between the
best-fit model and the data. The PDE model, on the other hand,
underestimates the HXLF data of Log \lx\ $\geq 44.5$ at $z=1.6-3.0$ by
a factor of $\simeq$4, resulting in a 1-dimensional KS probability of
0.02 (from 66 objects) for the redshift distribution in Log \lx\ $\geq
44.5$. These facts also suggest that the true evolution of the HXLF is
more complex than the PLE or PDE. Actually, there is no physical
reason why these simplest models should hold in the whole \lxz\ range.

\subsection{The LDDE model}

To find a more sophisticated description of the HXLF, we consider a
generalized luminosity-dependent density evolution (LDDE) model where,
by definition, the evolution term in the formula (\ref{eq-PDE}) is not
constant but a function of the luminosity. At first, to obtain an
overall idea about the LDDE behavior, we derive the cutoff redshift
($z_{\rm c}$) and the slope ($p1$) in the evolution factor from two
separate luminosity ranges, Log \lx\ $\geq44.5$ and $44.5>$ Log \lx\
$\geq 43$. Fixing the other parameters at the best-fit values of the
PDE in Table~\ref{table-hxlf} (except $p2=-1.5$ for consistency with
our final results derived below), we obtain ($z_{\rm c}$, $p1$) =
(1.90$^{+0.26}_{-0.31}$, 4.6$^{+0.6}_{-0.5}$) and
(1.01$^{+0.06}_{-0.10}$, 4.0$^{+0.5}_{-0.4}$), respectively (attached
is the 1$\sigma$ error for a single parameter). This indicates that
the cutoff redshift has strong luminosity dependence, being much
smaller in lower luminosities, while the slope can be regarded to be
constant within statistics. Note that this behavior is not the same as
the LDDE model adopted by \citet{miy00} for the SXLF, where $p1$ has
luminosity dependence but $z_{\rm c}$ is set to be constant.

\begin{figure}
\epsscale{1.0}
\plotone{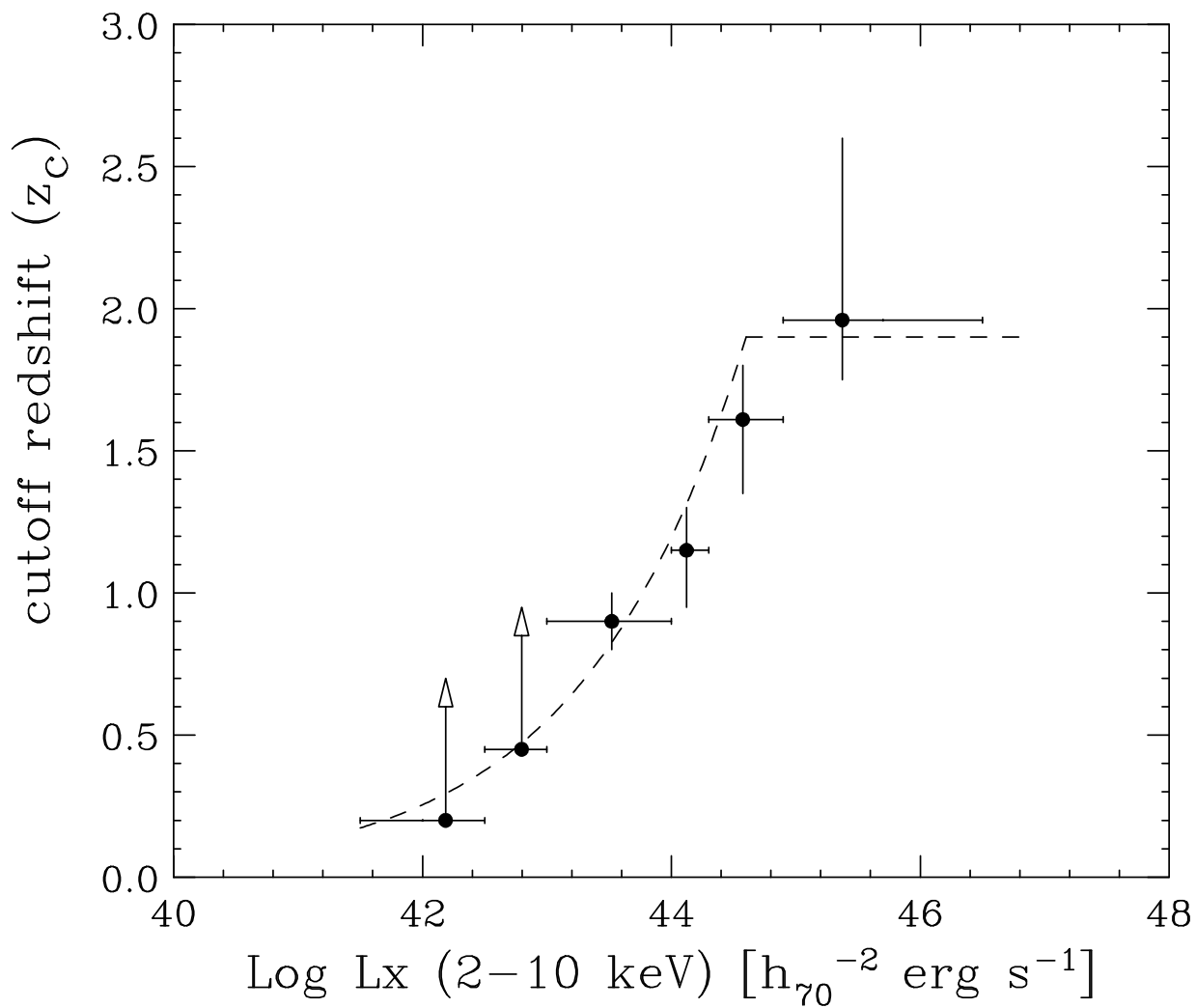}
\caption{
The cutoff redshift ($z_{\rm c}$) of the AGN density evolution 
determined as a function of the luminosity. Error bars correspond to
1$\sigma$ errors obtained through the ML fit. The arrows denote 90\%
lower limits. The position of the marker corresponds to the mean
luminosity in each region. The dashed line is the best-fit model
used in our HXLF formula.
\label{zc}}
\end{figure}

Based on this result, we determine $z_{\rm c}$ as a function of
luminosity in finer bins, by fixing $p1$ at 4.2. The results are shown
in Figure~\ref{zc}. In the luminosity regions smaller than Log \lx\ =
43 we obtain only lower limits, which are indicated by the arrows. The
figure clearly reveals that $z_{\rm c}$ rapidly drops from $\simeq1.9$
at Log \lx\ $\simeq 44.5$ toward lower luminosities. Because our
sample has been made highly complete at all flux levels, the
difference of $z_{\rm c}$ between luminosity ranges above and below
Log \lx\ $\approx 44.5$ is a robust conclusion even if we consider the
uncertainties on the redshift distributions of unidentified sources.

We finally find that the following LDDE model, where $z_{\rm c}$ is expressed
by a power law of \lx , well describes the current HXLF data as well
as other observational constraints:
\begin{equation}
\frac{{\rm d} \Phi (L_{\rm X}, z)}{{\rm d Log} L_{\rm X}} 
= \frac{{\rm d} \Phi (L_{\rm X}, 0)}{{\rm d Log} L_{\rm X}} e(z, L_{\rm X})
\end{equation}
where
\begin{equation}
e(z, L_{\rm X}) =
\begin{cases}
(1+z)^{p1} & (z<z_{\rm c}(L_{\rm X})) \\
e(z_{\rm c})[(1+z)/(1+z_{\rm c}(L_{\rm X}))]^{p2} & (z \geq z_{\rm c}(L_{\rm X}))
\end{cases}
\end{equation}
and
\begin{equation}
z_{\rm c}(L_{\rm X}) = 
\begin{cases}
 z_{\rm c}^* & (L_{\rm X} \geq L_a) \\
z_{\rm c}^* (L_{\rm X}/L_a)^\alpha & (L_{\rm X}<L_a).
\end{cases}
\end{equation}
The best-fit parameters are summarized in Table~\ref{table-hxlf} (we
fix $z_{\rm c}^*$ and Log $L_a$ at 1.9 and 44.6, respectively, and leave
$\alpha$ as a free parameter). In Figure~\ref{zc} the best-fit
function of $z_{\rm c} (L_{\rm X})$ is plotted by a dashed line. We
set $p2=-1.5$ independently of the luminosity, so that (1) the model
does not overproduce the source counts and that (2) it roughly
describes the decline in number density of luminous AGNs revealed by
the latest SXLF study \citep{has03}.

\begin{figure}
\epsscale{0.7}
\plotone{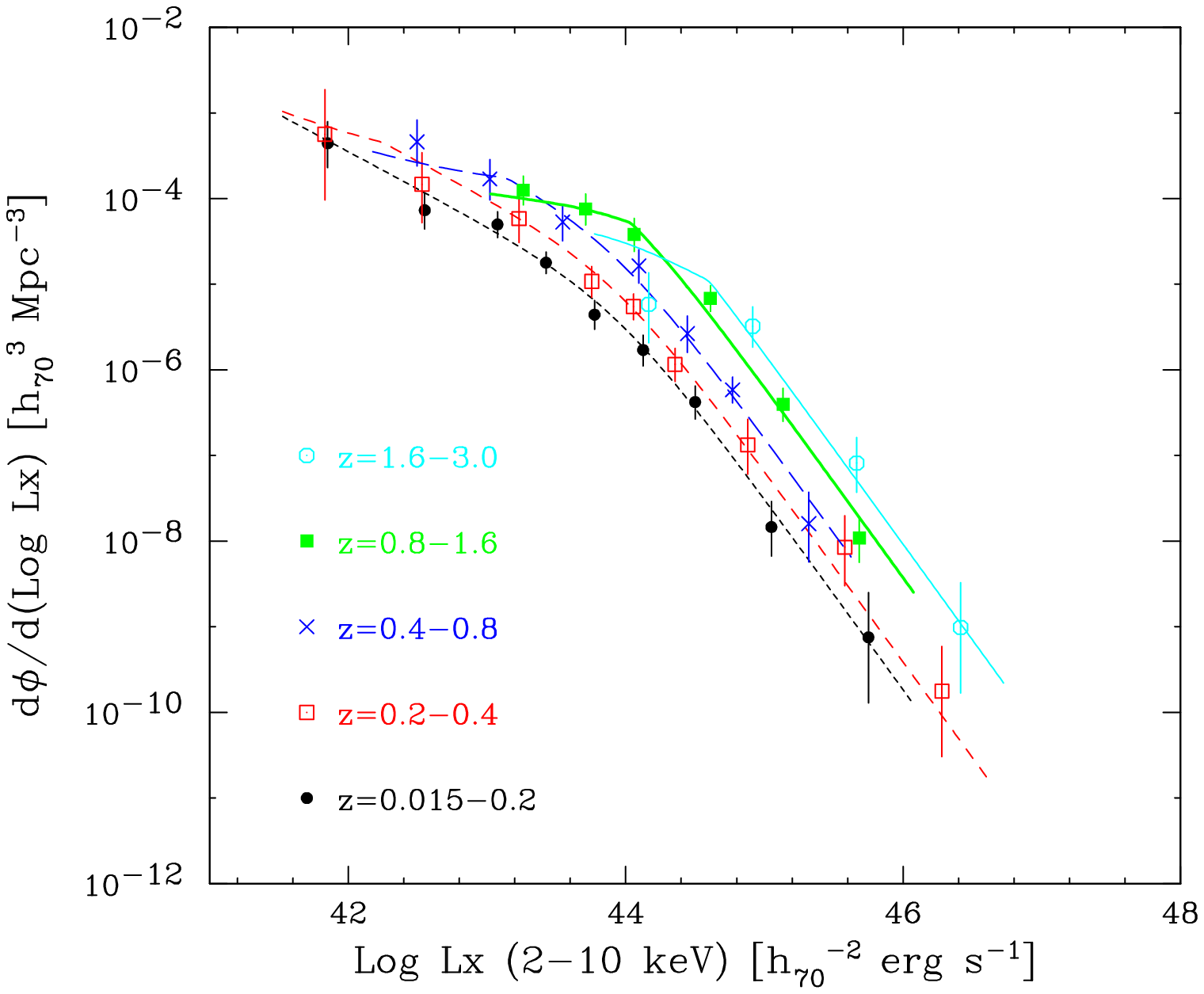}
\plotone{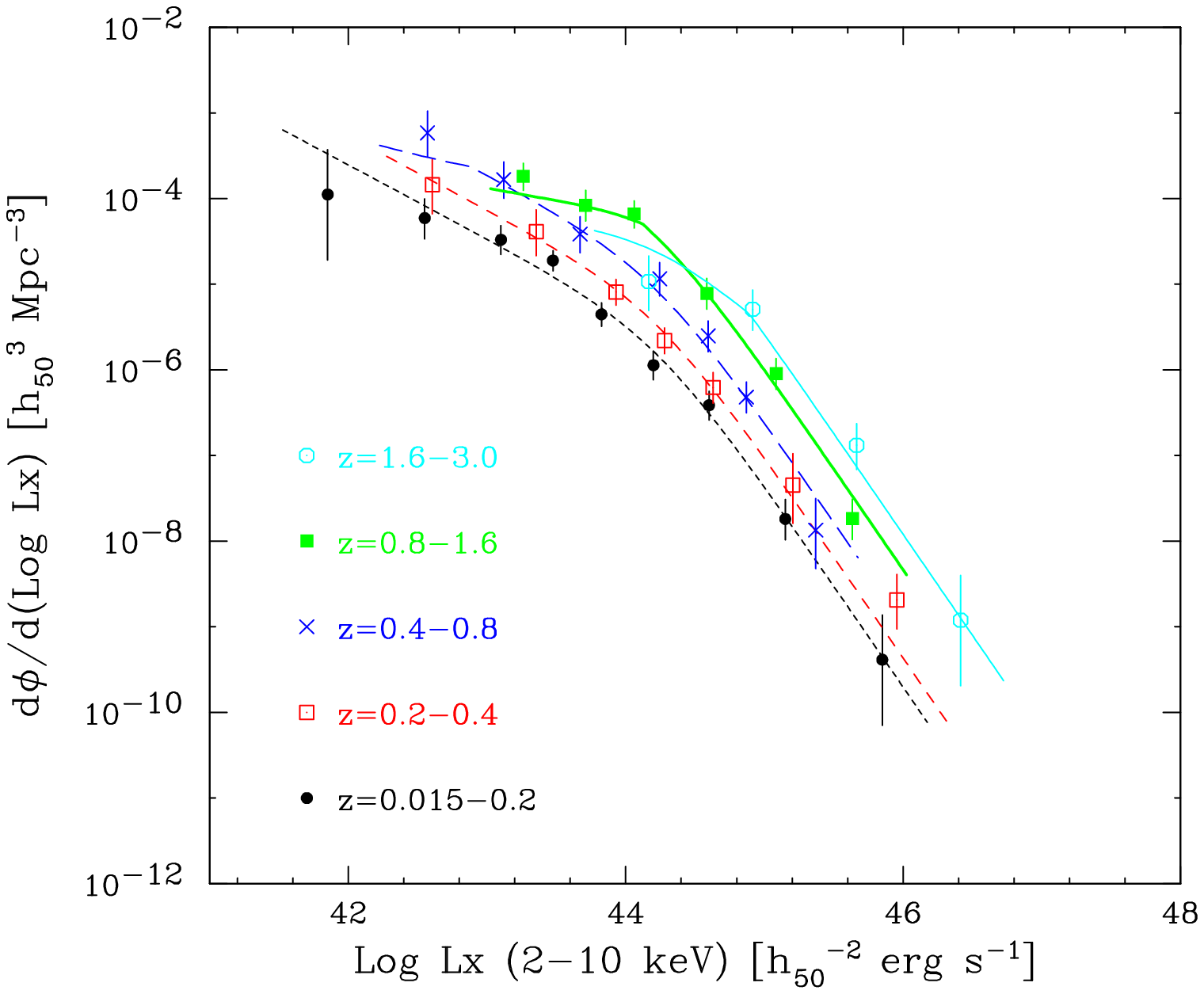}
\caption{
The intrinsic 2--10 keV hard X-ray luminosity function of AGNs. The
curves represent our best-fit model (the LDDE model). The data are
plotted according to the $N^{\rm obs}/N^{\rm mdl}$ method
\citep{miy01} with estimated 1$\sigma$ Poisson errors. The results are
given in the five redshift bins of $z$=0.015--0.2 (short-dashed,
black), 0.2--0.4 (medium-dashed, red), 0.4--0.8 (long-dashed, blue),
0.8--1.6(thick solid, green), and 1.6--3.0 (thin solid, cyan). Upper
(a): \cosmo\ = (70$h_{70}$, 0.3, 0.7). Lower
(b): \cosmo\ = (50$h_{50}$, 1.0, 0.0).
\label{lf}}
\end{figure}

Figure~\ref{lf} show the data of the HXLF in five redshift bins,
$z$=0.015--0.2, 0.2--0.4, 0.4--0.8, 0.8--1.6, and 1.6--3.0 with the
best-fit HXLF model calculated at the central redshift of each bin
($z$=0.1, 0.3, 0.6, 1.2, and 2.3, respectively). For plotting the HXLF
we adopt the ``$N^{\rm obs}/N^{\rm mdl}$ method'' \citep{miy01}, where
the best-fit model multiplied by the ratio between the number of
observed sources and that of the model prediction in each \lxz\ bin is
plotted. Although model dependent, this technique is the most free
from possible biases, compared with other methods such as the
conventional $1/V_a$ method. The attached errors are estimated from
Poissonian errors (1$\sigma$) in the observed number of sources
according to the formula of \citet{geh86}.

\begin{figure}
\epsscale{1.0}
\plotone{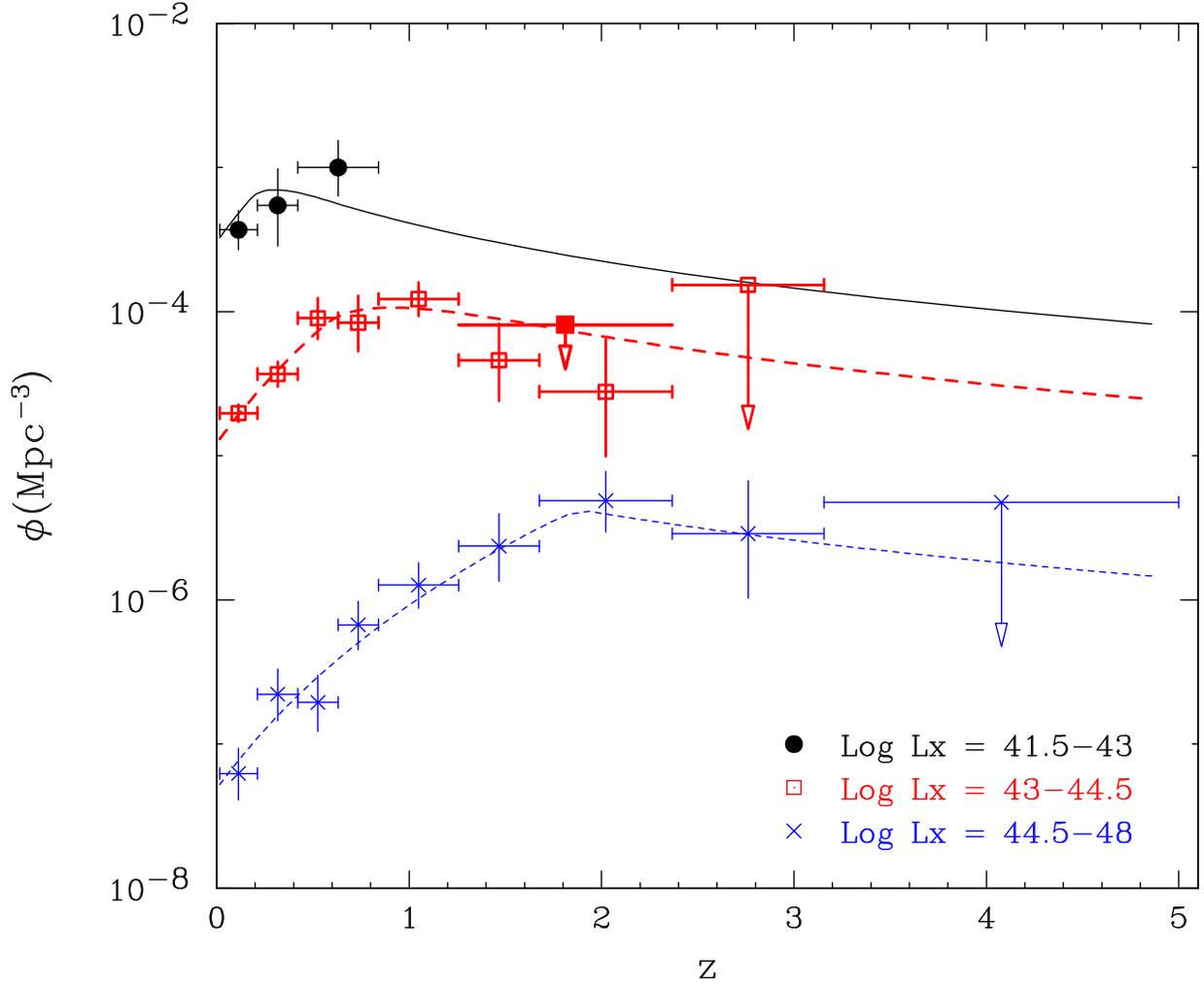}
\caption{
The comoving spatial density of AGNs as a function of redshift in
three luminosity ranges, Log \lx\ =41.5--43 (upper, black), 43--44.5
(middle, red), 44.5--48 (lower, blue).  The lines are calculated from
the best-fit model of the HXLF. The error are $1\sigma$, while the
long arrows denote the 90\% upper limits (corresponding to 2.3
objects).  The short arrow (marked with a filled square, red)
corresponds to the 90\% upper limit on the average spatial density 
of AGNs with Log \lx\ =43--44.5 at $z$=1.2--2.3 when all the unidentified
sources are assumed to be in this redshift bin.
\label{zf}}

\end{figure}

Figure~\ref{zf} shows the same HXLF result in terms of the (comoving)
spatial density as a function of redshift integrated in three
luminosity regions (Log \lx\ = 41.5--43, 43--44.5, 44.5--48). The
errors are 1$\sigma$ (same as in Figure~\ref{lf}), while the long
arrows denote the 90\% upper limits. It is clearly noticed that the
cutoff redshift increases with the luminosity, and as a result the
ratio of the peak spatial density to that of present day is much
smaller for AGNs with Log \lx\ $<44.5$ than for more luminous
AGNs. Note that these results are based on the ``effective''
correction for the sample incompleteness as described in
\S~\ref{sec-samplesummary}. To evaluate maximal errors due to the
incompleteness, we calculate the same plot assuming that all the
unidentified sources were located in a specific redshift as done by
\citet{cow03}. We find that the incompleteness could affect any data
points below $z<2.3$ only by a factor smaller than 2 except for those
at $z>1.2$ in the Log \lx\ = 43--44.5 range. The short arrow shows the
90\% upper limit on the average spatial density at $z=1.2-2.3$, which
still gives a tight constraint.

\subsection{Comparison with Observational Constraints}
\label{sec-HXLFcomp}

\begin{figure}
\epsscale{1.0}
\plotone{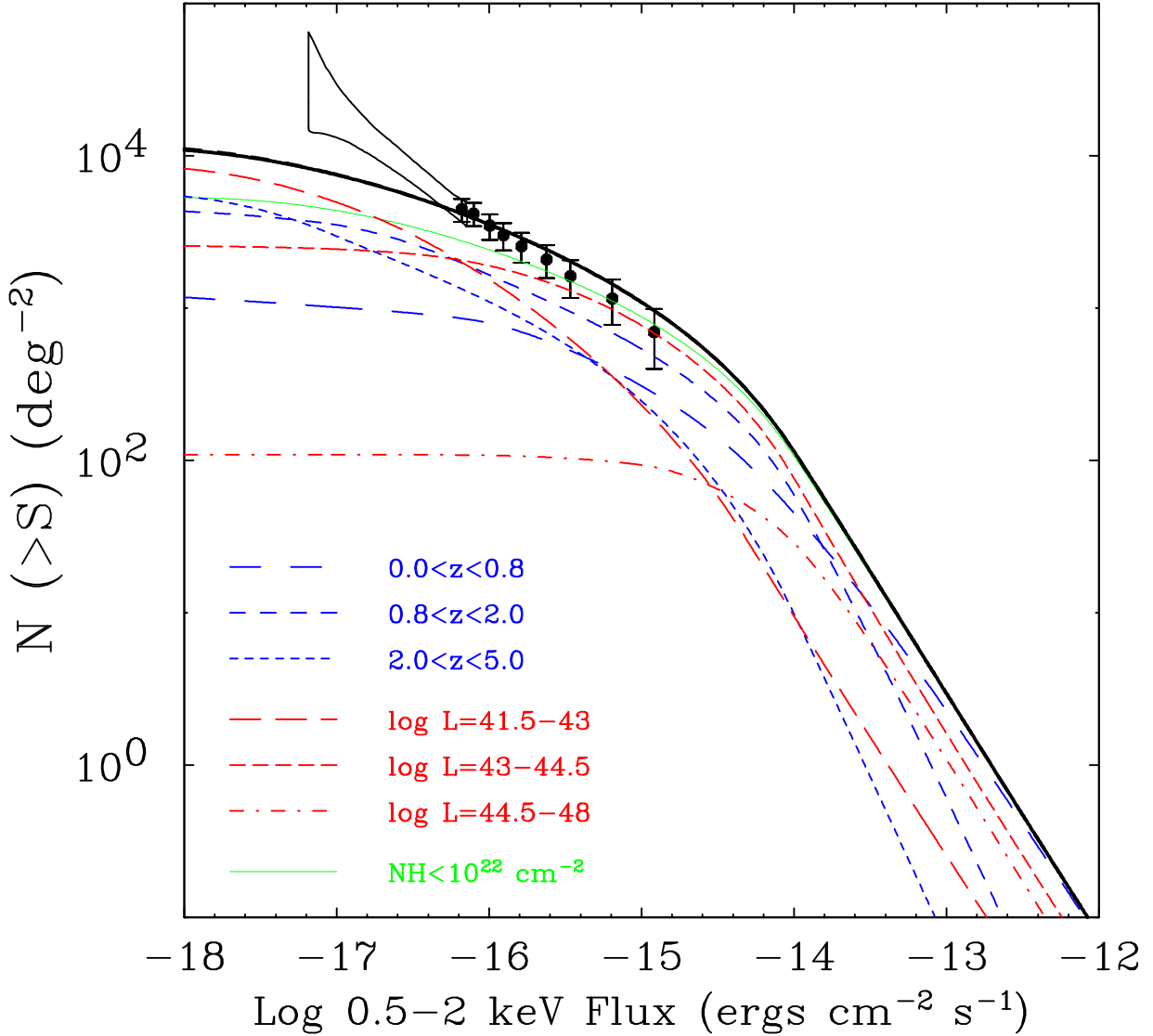}
\caption{
Predicted \logn s from our model of the HXLF and the \nh\
function. The four figures correspond to different survey bands, (a)
0.5--2 keV, (b) 2--10 keV, (c) 5--10 keV, and (d) 10--30 keV. Black
solid curve: the total contribution of only ``Compton-thin'' AGNs
integrated Log \lx\ = 41.5--48 and $z<$5.0. Black dashed curve: when
the same number of Compton-thick AGNs with Log \nh\ =24--25 as
those with Log \nh\ =23--24 is included. Green curve: that
of \xtI\ AGNs. Blue (red) curves: contribution from different
luminosity (redshift) ranges as indicated in the figure. The data
points and enclosed regions in (a) and (b) are the constraint from the
CDFN survey after \citet{miy02}. 
In the 5--10 keV band, the winding curve between Log $S$ = --14.6
and --13.4 and the data point at Log $S$ = --13.3 represents the
\xmm\ result of the Lockman hole and the {\it BeppoSAX} result,
respectively (after \citet{has01} and references therein).
\label{clogn}}
\end{figure}

Figure~\ref{clogn}(a)--(d) shows the prediction of source counts
(black solid curves) from the best-fit HXLF model and the \nh\
function in the four bands, 0.5--2 keV, 2--10 keV, 5--10 keV, and
10--30 keV. We also plot the source counts in the CDFN survey by
\cite{miy00} in the 0.5--2 keV and 2--10 keV bands, and that of \xmm\
in the Lockman hole field by \citet{has01} in the 5--10 keV band. It
is verified that our model reproduces the observed 2--10 keV source
counts above the flux limit, $3.8\times10^{-15}$ \ergs . The predicted
curve is, however, slightly below the 90\% error region at the
faintest fluxes $\simlt 10^{-15}$ \ergs. As we discuss later in
\S\ref{sec-comptonthick}, this discrepancy could be partially
explained by Compton-thick AGNs with Log \nh\ =24--25. In the 5--10
keV band, the observed source counts is well reproduced by our
model. In the 0.5--2 keV band, on the other hand, the predicted AGN
contribution significantly underestimates the result of fluctuation
analysis at fluxes of $8\times10^{-18}-8\times10^{-17}$ \ergs\ (0.5--2
keV). This situation does not change even if we add a soft component
with a relative normalization of 5\% to the continuum for every
AGN. Such discrepancy is also seen in the population synthesis model
by \citet*{gil01}. As discussed by \citet{miy02}, this fact indicates
the emergence of new populations, which could be attributable to
normal galaxies \citep*[e.g.,][]{ran03}. They may also contribute to
the faintest 2--10 keV source counts as well.

We confirm that our HXLF model can reproduce the redshift distribution
of another hard-band selected sample containing other \chandra\
sources than used in the present analysis. Figure~\ref{zdist} shows
comparison of the expected redshift distribution of AGNs (dashed
histogram) with the actual data (solid) at a flux limit of
$5\times10^{-15}$ \ergs\ in the 2--10 keV band taken from
\citet{gil03}. He compiled only spectroscopically identified sources
in the \chandra\ deep field south \citep{gia02}, the CDFN, the Lockman
Hole field, the Lynx field, and the Small Selected Area 13 field. The
identifications were highly complete at this flux limit. He also
excluded objects in certain redshift ranges where there are density
spikes due to the underlying large-scale structure. The 1-dimensional
KS test gives a matching probability of 0.75, which is well
acceptable.

Finally we examine the consistency of our HXLF and the \nh\ function
with the SXLF determined by \rosat\ data. We refer to the SXLF by
\citet{miy00}, which is calculated for all the soft X-ray selected
AGNs (except BL Lac objects) without optical classification, as is in
our case, but given for a luminosity in the observer 0.5--2 keV frame
with no absorption correction. To make direct comparison, we calculate
an expected SXLF by integrating contribution of AGNs with different
intrinsic luminosities and column densities from our HXLF and \nh\
function. In this step we firstly compute the \rosat\ PSPC count rate
in the 0.5--2 keV band and then convert it into an ``observed''
luminosity assuming $\Gamma=2$, using the detector response.

Figure~\ref{sxlf} shows the comparison between the observed SXLF data
(points with error bars) and the prediction from the HXLF model
(lines) in the redshift bins of $z$=0.015--0.2, 0.2--0.4, 0.4--0.8,
0.8--1.6, and 1.6--2.3. The data are taken from the numerical table of
\citet{miy01} for the same cosmological parameters, \cosmo\ = (70,
0.3, 0.7). In spite of the fact that they are determined from
completely independent surveys, we can see good agreement between the
two results: they match within a factor of 2 or less than 2 $\sigma$
level. Because of $k$-correction more absorbed sources can fall into
the soft X-ray band in higher redshifts, making the evolution appear
to continue until a higher redshift than that of the HXLF. The effect
is more important in the lower luminosity range because of the larger
fraction of absorbed sources. This is probably the reason why the
\rosat\ SXLF is well described by the LDDE model with a constant
cutoff redshift.

We note that the HXLF model tends to overestimates the SXLF at
$z$=0.4--1.6 in the high luminosity range above Log \lx\ $\approx$ 45.
A major reason is because the HXLF form we use is too simple, where a
common two power-law form is assumed over the whole redshift range. In
reality the slope $\gamma2$ seems to be larger (and/or $L_{*}$
smaller) at these redshifts, as indicated from the SXLF data (it is
also implied from our HXLF data; see Figure~\ref{lf}). This can be
connected to the fact that the predicted source counts in the 0.5--2
keV band slightly overestimates the \rosat\ result at fluxes below
$\approx10^{-13}$ \ergs\ (Fig.\ 8 of \citealt{miy00}). Similar
tendency is also noticed in the hard band in the flux range of
$10^{-12}-3\times10^{-13}$ \ergs\ (2--10 keV). Another possibility is
that the current \nh\ function model may underestimate the fraction of
absorbed sources, which could be dependent on redshift. Improvement of
modeling of the HXLF and the \nh\ function with more complex forms to
fully reflect these features is a future task, for which a combined
analysis of the HXLF and SXLF would be useful. Nevertheless, the
discrepancy at Log \lx\ $\simgt$ 45 is not a serious problem in
discussing the overall contribution to the CXB since the number
density drops rapidly with $\propto L_{\rm X}^{-2.2}$.

\section{Population Synthesis Model Update}
\label{sec-pop}

Our results of the HXLF and the \nh\ function are extremely useful to
establish a so-called population synthesis model of the CXB
\citep[e.g.,][]{mad94,com95,gil01}. The constructed model is most
directly determined from the observations in the hard band and should
update earlier works, which basically uses the result of the SXLF with
assumptions for the \nh\ function. Nevertheless, we may still have to
call it a ``model'' because below our flux limit it is based on
extrapolation. Detailed modeling with many additional parameters is
beyond the scope of this paper and we concentrate on basic
consequences that are directly obtained from the HXLF and the \nh\
function. The best-fit parameters of the LDDE and of the \nh\ function
in Table~\ref{table-hxlf} for \cosmo\ = (70, 0.3, 0.7) are used.  For
the intrinsic AGN spectra we assume the ``template spectrum'' adopted
in \S~\ref{sec-absandlum} (i.e., $\Gamma=1.9$, a solid angle of
2$\pi$, inclination of cos$(i)$=0.5, cutoff energy \ec\ of 500 keV,
and the Solar abundance). Several systematic effects caused by this
assumption are briefly discussed in \S~\ref{sec-reproCXB}.

In the calculation we integrate contribution of AGNs at $z<5$ within
the luminosity range of Log \lx\ $>41.5$. Setting the luminosity limit
is justified from comparison of our HXLF result with the upper limit
at $0.1<z<1.0$ by \citet{cow03}: the AGN spatial density does not
increase below the range of Log \lx\ = 41.5 and therefore its
contribution to the CXB should be negligible compared to that of the
total AGNs. We also neglect other populations such as clusters of
galaxies, normal galaxies, and BL Lac objects, all of which have
significantly softer spectra than the CXB spectrum. Results from
previous surveys indicate that the overall contribution of these
populations is very small affecting only the soft X-ray
background. Although we basically consider ``Compton-thin'' AGNs with
Log \nh\ $<24$ in this model, we finally discuss possible contribution
of Compton-thick AGNs in \S~\ref{sec-comptonthick}, which could have a
significant contribution to the CXB above 10 keV.

\subsection{The Composition of the CXB}

In Figure~\ref{clogn} we also plot the predicted source counts of
Compton-thin AGNs separately for different luminosity and redshift
ranges: (red) for Log \lx\ = 41.5--43, 43--44.5, and 44.5--48 in
$z<5.0$, and (blue) for $z$=0.0--0.8, 0.8--2.0, 2.0--5.0 in Log
\lx\ = 41.5--48. The summed contribution of all the Compton-thin AGNs
from Log \lx $=41.5-48$ and $z<5.0$ is shown by the thick solid
line (black), while that of only \xtI\ AGNs is shown by the thin solid
line (green). (The uppermost dashed line (black) corresponds to the
case when Compton-thick AGNs are included with an extrapolation of the
\nh\ function over Log \nh\ $>24$. See \S~\ref{sec-comptonthick}.) The
prediction in the 10--30 keV band can be examined with sensitive hard
X-ray surveys by future missions such as {\it Astro-E2}, {\it NeXT},
{\it Constellation-X}, and {\it XEUS}.

Figure~\ref{lzspec2-10}(a) and (b) shows the differential CXB
intensity in the 2--10 keV band per unit Log \lx\ and redshift, given
in different redshift and luminosity ranges, respectively. (The
uppermost curves represent the case when Compton-thick AGNs are
included.) It is seen that AGNs with Log \lx\ of $\simeq 43.8$ are the
largest contributors to the 2--10 keV CXB as a total, although the
peak luminosity decreases with the redshift. On the other hand, the
contribution of AGNs per unit redshift is peaked at
$z\approx0.6$. Similarly, the peak redshift increases with the
luminosity. These results can be fully understood as a consequence of
the LDDE behavior of the HXLF. Figure~\ref{lzspec10-30} shows the same
plots but in the 10--30 keV band. Since the effect of absorption is
almost negligible in this band, these plots enable us to understand
the CXB composition more directly. As recognized from the figure, the
main part of the $E > 10$ keV CXB is produced in the low redshift
universe, peaked at $z\approx 0.5$. It is also seen that the
contribution of Compton-thick AGNs becomes more important than in the
energy band below 10 keV.

\subsection{Reproduction of the Broad Band CXB Spectrum}
\label{sec-reproCXB}

A primary interest is whether our population synthesis model can
reproduce the CXB spectrum in the broader band including energies
above 10 keV. The uppermost dashed line (blue) in Figure~\ref{cxbspec}
shows the representative form of the CXB spectrum in $E I(E)$, where
$I(E)$ is the energy intensity per unit solid angle and $E$ the
energy. We plot a power law with $\Gamma=1.4$ in the 0.5--10 keV range
assuming the maximum normalization estimated by \citet{bar00}, 10.6
keV cm$^{2}$ s$^{-2}$ keV$^{-1}$ at 1 keV, and the empirical formula
determined by \citet{gru99} mainly from the \heao\ A2 and A4
experiments, in the 3--1000 keV range. Before making detailed
comparison, however, we have to recall possible systematic errors in
the measurement of the CXB spectrum. Figure~2 of \citet{gru99}
suggests that there are still uncertainties of $>$5\% in the
measurement of the relative CXB shape in the 20--40 keV and
$\simgt$100 keV ranges. Also the spectrum of the extragalactic portion
of the $E<$1 keV CXB is still controversial
\citep[e.g.,][]{par99}. Besides this, there is an uncertainty of at
least 16\% (as large as 30\% between different missions probably
because of cross-calibration errors) in the absolute normalization of
the CXB \citep[see][]{bar00}. In this figure we have increased the
original normalization of \citet{gru99} by 26\% to connect to the
0.5--10 keV plot.

The black thick curve in Figure~\ref{cxbspec} represents the
integrated spectrum of Compton-thin AGNs with Log \lx\ = 41.5--48 at
$z<5.0$ based on the template spectrum. Figure~\ref{cxbspec-nh} shows
contribution of AGNs with different column densities separately. The
model spectrum reproduces the relative shape of the CXB within an
accuracy of $\sim$20\%. To illustrate effects of changing parameters
of the template spectrum, we overplot the same results with the \ec\
value set to 400 keV (left red dashed curve) and 600 keV (right red
dashed cure), keeping the other parameters the same. When $\Gamma=1.9$
is assumed, it is recognized that \ec\ cannot exceed $\approx$600 keV
so as not to overproduce the CXB intensity above 100 keV. We also plot
the extreme case by the green dot-dashed curve when the reflection
component is absent. The comparison with the case of the template
spectrum clearly indicates the importance of the reflection component
in producing the hump structure of the CXB peaked at 30 keV, as was
suggested by several authors \citep[e.g.,][]{fab90,ter91}. The
spectrum without a reflection component significantly underestimate
the CXB intensity in the 5--100 keV range, producing a much broader
peak than the CXB itself. Finally, we note the necessity of
populations having non-thermal spectra (such as ``blazars'') with an
energy cutoff much higher than $\sim$ 1 MeV in reproducing the
Gamma-ray background above a few hundred keV, even though they are
minor populations in the X-ray band.

\subsection{Contribution of Compton-Thick AGNs}
\label{sec-comptonthick}

A detailed inspection of Figure~\ref{cxbspec} suggests that the model
based on the template spectrum slightly ($\approx$10--20\%)
underestimates the relative shape of the CXB spectrum around its peak
intensity. As a possibility to explain this discrepancy, we consider
contribution of Compton-thick AGNs. In fact, the result of
\citet{ris99} suggests that there are roughly twice (1.6$\pm$0.6) as
many AGNs with Log \nh\ $>24$ as those with Log \nh\ =
23--24. Accordingly, we simply extrapolate the \nh\ function above Log
\nh\ $>24$ keeping the same normalization up to Log \nh\ = 26,
assuming the ``modified unified scheme'' with no redshift
dependence. This extrapolation may overestimate the true number of
AGNs with Log \nh\ = 24--25, because in the \nh\ distribution of
\citet{ris99} only upper limits of \nh\ are obtained for most (80\%)
of objects in this bin.

When the absorbing matter becomes Compton-thick, we have to take into
account effects of Compton-scattering leading to significant decrease
of the emitted flux even in the hard X-ray bands \citep{wil99}.
Referring to their results of Monte Carlo calculation (with one Solar
abundance), we approximately take account of this effect by
multiplying energy dependent correction factors to the nominal
absorbed spectrum. We neglect any contribution of objects with Log
\nh\ $>$25 assuming that all X-rays are absorbed there before
escaping. Inversely speaking, we cannot really constrain the number
density of the {\em whole} Compton-thick populations by using the CXB
intensity as a boundary condition.

The thin solid curve (black) in Figure~\ref{cxbspec} represents the
integrated spectrum when Compton-thick AGNs are added according to the
above assumption. In Figure~\ref{cxbspec-nh} we plot the contribution
from Compton-thick AGNs with a red solid line (the upper red solid
line is the total of Compton-thin and thick AGNs). As expected, the
inclusion of Compton-thick AGNs reduces the gap between the model and
the CXB spectrum. The slight overestimate of the CXB around $\sim$100
keV is not an essential problem because it could be tuned by lowering
\ec. The spectral slope in the 2--10 keV band becomes slightly harder
than a $\Gamma=1.4$ power law, although we have to keep in mind that
we do not include any contribution of other ``softer'' populations
here. The uppermost dashed curves (black) in
Figure~\ref{clogn}(a)--(d) represent the predicted source counts when
Compton-thick AGNs are included. In the 2--10 keV band the total
source counts are increased by about 20\% at $S$ = $10^{-15}-10^{-16}$
\ergs , thus becoming more consistent with the data. In the 0.5--2 keV
the source count is not affected at all in our assumption (i.e., no
scattered component).

These results suggest that the presence of roughly equal number of
Compton-thick AGNs of Log \nh\ = 24--25 as those with Log \nh\ =
23--24 is still consistent with the observations. We infer that the
number density of AGNs with Log \nh\ = 24--25 assumed here roughly
corresponds to its upper limit in order not to overproduce the 10--30
keV CXB, although this constraint depends on the shape of the template
spectrum, in particular, on the relative strength of the reflection
component (parameterized by the solid angle of the reflector). Finally
we note that the reproductivity of the CXB spectrum is not yet {\it
perfect} in the sense that the model peaks at a somewhat lower energy
than that of the CXB. Here we do not pursue this discrepancy, however,
considering many systematic uncertainties arising from the overly
simple assumptions in our model (e.g., a single template spectrum, its
spectral model, etc) as well as the measurement error (5--10\%) in the
CXB spectral shape itself.

\section{Discussion}

\subsection{Summary of Our Work}

Using the highly complete hard-band selected AGN sample covering the
wide flux range of $10^{-10}- 3.8\times 10^{-15}$ (2--10 keV), we have
determined the \nh\ function and the intrinsic HXLF as a function of
redshift and luminosity in the range of Log \lx\ = 41.5--46.5 and $z$
= 0--3. This means that we have directly revealed the evolution of
most X-ray emitting, Compton-thin AGN populations that constitute a
major part of the hard X-ray background in the 2--10 keV band. In
other words, the CXB origin below 10 keV is now quantitatively solved
by superposition of AGNs with different luminosities, redshifts,
absorptions, and optical types.
Our HXLF and \nh\ function, with an extrapolation to fainter fluxes
than the flux limit, predicts various observational quantities that
are consistent with currently available data. Based on these results
we have constructed an observation based population synthesis
model. Even though we make many simple assumptions, it predicts
reasonably well the CXB spectrum in the 0.5--300 keV band. We find
that the presence of a significant amount of Compton-thick AGNs as
suggested in a local Seyfert 2 galaxy sample is consistent with the
observations.

\subsection{Implication from the \nh\ Function}

\subsubsection {Modified Unified Scheme}

We have revealed that the \nh\ function has significant luminosity
dependence in that the fraction of absorbed AGNs decreases with
luminosity, while its redshift dependence is not significant. We call
this picture ``modified unified scheme'' of AGNs, in contrast to the
pure unified scheme where all the AGNs have the same geometrical
structure regardless of its luminosity and redshift. A simple
interpretation is that the opening angle of the dust torus is larger
in more luminous AGNs. This may be attributable to the physical
process that the high radiation pressure from a nucleus of luminous
AGNs affects the physical structure of the torus.

\citet{miy00b} attempted to include this ``deficiency of type II
quasars'' in their population synthesis model. However, \citet*{gil99}
found that their model involving this effect underestimates the 5--10
keV source count at $S \simgt 5\times10^{-14}$ \ergs . \citet{gil01}
further explored models involving the cosmological evolution in the
ratio of absorbed and unabsorbed AGNs. One of them included this
``modified unified scheme'', but they disfavored this particular model
because it underpredicted 2--10 keV source counts. Mainly because we
now have a different form of the HXLF where the peak of the low
luminosity AGNs (with abundant \xtII\ AGNs) is at closer redshifts,
contributed by the higher normalization of the absolute CXB intensity
(see also e.g., \citealt*{pom00}), we have now obtained a solution
where this modified unified scheme is included, is consistent with
hard (5--10 keV and 2--10 keV) source counts, and the cosmological
evolution of the {\it X-ray type II} / {\it type I} ratio is not
necessarily required.

It is an important question for galaxy formation theories whether the
\nh\ function has any redshift dependence or not. For example,
\citet{fra02} and \citet{gan03} suggest the possibility that the
fraction of absorbed AGNs is significantly larger in $z \simlt 1$ than
in higher redshifts, assuming a link of obscured AGNs to starburst
galaxies. Our result derived from Compton-thin AGNs seems to rule out
strong redshift dependence of the absorption fraction as proposed by
these authors, confirming the argument by \citet{gil03}. However, at
least we do not have yet direct observational constraints on the
fraction of the Compton-thick AGNs as a function of redshift. To fully
establish our understanding of the whole AGN populations, further
studies using a larger sample as well as sensitive hard X-ray surveys
above 10 keV are necessary.

\subsubsection{Fraction of {\it Optical Type II} AGNs in the Local Universe}

Here we make a rough comparison on the fraction of \otII\ AGNs to
check consistency of our picture with the results of optical surveys
of local AGNs. The number ratio of Seyfert 1.8--2 galaxies to Seyfert
1--1.5 galaxies in the local universe is estimated to be 4.0$\pm$0.9
\citep{mai95}. The sample of \citet{ris99}, a sub-sample of the above
one, has an average luminosity of Log \lx\ $\approx 43$ as seen in
their Figure~1. At these luminosities, the fraction of AGNs with Log
\nh\ = 22--24 to the total Compton-thin AGNs is $\simeq0.57$ according
to the \nh\ function (Figure~\ref{absfrac1.9}). Estimates of the true
fraction of type-II AGNs can be made only after knowing of the number
density of the Compton-thick AGNs. Assuming that there exist 1.6 times
as many Compton-thick AGNs as those with Log \nh\ = 23--24
\citep{ris99}, and that the fraction of \otII\ AGNs is 10\%, 30\%, and
100\% at Log \nh\ $<$ 21, Log \nh\ = 21--22, and Log \nh\ $>$22,
respectively (Figure~\ref{agn2f}), we expect that the ratio of \otII\
to \otI\ AGNs is 3.6. This estimate is consistent with the optical
survey result.

\subsection{Implication from the HXLF}

\subsubsection{Comparison with Previous Results of the HXLF}

In \S~\ref{sec-HXLFcomp} we have already found that our HXLF result
with the \nh\ function is consistent with the SXLF obtained from the
largest sample of \rosat\ surveys \citep{miy00}. We here make
comparison of our result with earlier works of the HXLF (we refer to
the result with \cosmo\ = (50$h_{50}$, 1.0, 0.0) for consistency
between different papers). Even though we find that the HXLF is best
described by the LDDE model rather than other simpler forms, we here
use the results of the PLE fit as reference when the cosmological
evolution is discussed. \citet{ceb96} calculated a local HXLF mostly
from the \heao\ Grossan sample. \citet{boy98} derived a HXLF from a
small number of \asca\ sources and the Grossan sample, and more
recently \citet{laf02} obtain a HXLF, but of only \otI\ AGNs, from a
combination of the 5--10 keV {\it HELLAS} survey, the ALSS, and the
Grossan sample.

We confirm that the HXLF at $z<0.2$ (a sum of the \otI\ plus \otII\
AGNs) derived by \citet{ceb96} and by \citet{boy98} are consistent
with our data within the statistical errors. Good agreement of the
HXLF is confirmed in Log \lx\ $>44$ even though the error bars are
small, simply reflecting the fact that we have used essentially the
same \heao\ sample for determination of the local HXLF in this
luminosity range. \citet{boy98} obtained the evolution factor
$e(z)\propto (1+z)^{k}$ with $k=2.04^{+0.16}_{-0.22}$ assuming the
PLE. The reason why it is apparently smaller than ours
($p1=2.7\pm0.2$) is probably because the redshift cutoff above which
the evolution terminates is not included in their analysis. On the
other hand, since \cite{laf02} uses an \otI\ AGN sample, the
normalization they obtained is significantly smaller than our HXLF,
which contains both \otI\ and \otII\ AGNs. We see the tendency that
the discrepancy is larger in lower luminosities. This is expected
from the luminosity dependence of the fraction of \otI\ AGNs.  They
obtain the PLE parameter of $p1=2.52$ at $z<1.39$ (and zero above
then), which roughly agrees with our PLE-fit result ($p1=2.7\pm0.2$ at
$z<1.2$).

\subsubsection{Comparison with Optical Luminosity Function of Quasars}
\label{sec-olf}

In this subsection we compare our HXLF with an optical luminosity
function (OLF) of broad line quasars, a sub class of AGNs (i.e.,
luminous, \otI\ AGNs). Here we refer to the results of the 2dF quasar
survey by \citet{boy00}. They used more than 6000 quasars to construct
the OLF in the $B$-band absolute magnitude of $-26<$ \mb\ $<-23<$ at
$z$=0.35--2.3, which was found to be well described by the PLE but
with a different form for the evolution factor from formula
(\ref{eq-evolv}). The comparison is not so trivial as with SXLFs
because we must limit the HXLF to only \otI\ AGNs (or \xtI\ AGNs
approximately) to select the same population. Also, we need to assume
a relation between \mb\ and \lx .
The relation between an optical (ultra violet) luminosity 
and an X-ray luminosity is often parameterized by the
$\alpha_{\rm OX}$ parameter defined as 
$$
\alpha_{\rm OX} \equiv - \frac{{\rm Log} (l_{\rm O}/l_{\rm X})}{{\rm
Log} (\nu_{\rm O}/\nu_{\rm X})},
$$ where $l_{\rm O}$ and $l_{\rm X}$ is the monochromatic luminosity
(in units of \erg\ Hz$^{-1}$) at the rest-frame frequency $\nu$ of
2500$\AA$ and 2 keV, respectively. Many previous works indicate that
$\alpha_{\rm OX}$ is correlated with the optical luminosity (luminous
quasars being X-ray quiet), equivalent to the relation $l_{\rm X}
\propto l_{\rm O}^e$ with $e\simeq0.7-0.8$
\citep[e.g.,][]{kri85,avn86,wil94,gre95,yua98}, although there are
arguments against the presence of such an ``intrinsic'' correlation
\citep*{laf95,yua98b}. The comparison between an OLF and an X-ray
luminosity function provides an independent approach to constrain the
$l_{\rm O}-l_{\rm X}$ relation \citep[see ][]{kri85}. Accordingly, we
here search for an appropriate relation that realizes the best
matching between the HXLF and the OLF.

We find that, if $e=1$ (or constant $\alpha_{\rm OX}$) is assumed, the
apparent evolution of the HXLF of \xtI\ AGNs becomes significantly
slower than that of the OLF, whereas they become reasonably consistent
with each other if $e \simeq 0.7$. This value is somewhat smaller than
the previous result of the same approach by \citet{boy93} using their
SXLF ($e=0.88\pm0.08$), but is consistent with the results obtained by
\citet{wil94} and \citet{vig03}. Figure~\ref{olf} shows the comparison
of the quasar OLF (data points) by \citet{boy00} with the prediction
from the HXLF of \xtI\ AGNs (thick lines) in the \cosmo\ =
(50$h_{50}$, 1.0, 0.0) universe, where we have assumed the relation
$$
	\alpha_{\rm OX} = 0.1152 {\rm Log} l_{\rm O} - 2.0437,
$$
equivalent to $e=0.70$ and $\alpha_{\rm OX}=1.37$ at \mb\ = --22.64
(or at \lx\ = $10^{44} h_{50}^{-2}$ \erg )\footnote{We assume the
rest-frame spectrum of $F_{\nu} \propto \nu^{-0.44}$ in the optical
band \citep{van01} for the conversion of $l_{\rm O}$ to \mb , and
$\Gamma=1.9$ for $l_{\rm X}$ to \lx\ (2--10 keV).}. Corresponding \lx\
values are indicated in the upper label. For clarity we plot only the
data in four redshift bins, $z$=0.350--0.474, 0.613--0.763,
1.108--1.306, and 2.014--2.300. The dashed lines of the HXLF represent
extrapolated regions where no X-ray sample exists.

As noticed from the figure, they are consistent within a factor of 2
except for the smallest luminosity ranges (\mb\ $>-23$) where the OLF
is systematically smaller than the HXLF. The discrepancy may be
(partially) due to selection effects as contamination of galaxy lights
may affect the completeness of optical quasar surveys in low
luminosity ranges. Indeed, our results on the fraction of \otII\ AGNs
show the same tendency (Figure~\ref{agn2f}). To illustrate this
effect, we also plot the HXLF of \otI\ AGNs with thin lines in
Figure~\ref{olf}, assuming the approximated relation that the fraction
of \otI\ AGNs in \xtI\ AGNs decreases with the luminosity based on our
result (\S\ref{sec-otIIfrac}). As expected, this reduces, though not
perfectly, the discrepancy at \mb\ $>-23$. More detailed comparison
would require a larger sample for the HXLF as well as consideration of
possible selection biases, which we leave for future studies. Finally
we mention that, as recognized from Figure~\ref{olf}, the HXLF covers
wider luminosity ranges than the OLF, demonstrating the importance of
a combined analysis of X-ray surveys in various depths to understand
the overall evolution of AGNs.

\subsubsection{Accretion History of the Universe}

In this paper we have shown that the cosmological evolution of the
HXLF is best described by the LDDE model where the cutoff redshift
increases with the luminosity. This means that the luminous AGNs
(quasars) formed in earlier epochs than less luminous AGNs (such as
Seyfert galaxies), consistent with the claims by \citet{cow03} and
\citet{has03}. Our result directly constrains the formation history of
super massive blackholes (SMBHs) in galactic centers not only for
luminous, unobscured quasars that can be traced by optical surveys but
also for less luminous or obscured AGNs, main contributors to the bulk
of the CXB. Quantitative analysis of the growth curve of SMBHs and its
relation to the local mass function of SMBHs are topics of a separate
paper (Ueda \etal , in preparation).

The accretion history of the SMBHs may have a strong link to the star
formation history in their host galaxies. In this context, it is quite
interesting to compare the evolution of the AGN luminosity density
with that of the star forming rate density as a function of
redshift. \citet{kau00} showed that the optical luminosity function of
quasars, its cosmological evolution, and the cosmic star formation
history can be well reproduced simultaneously with some reasonable
assumptions in semi-analytic models. \citet{fra99} pointed out that
the history of luminosity density of luminous AGNs (\lx\
$>10^{44.25}h_{50}^{-2}$) derived from the \rosat\ SXLF more closely
resembles the formation history of early type galaxies rather than
that of the total star-formation rate. This seems reasonable if we
recall the tight correlation between the mass of a SMBH and the
luminosity of a ``spheroid component'' of its host galaxy in the local
universe \citep{mag98}. According to a more recent, semi-analytical
model by \citet{bal03}, the star formation of early type galaxies
peaks earlier (at $z \approx 3$) than that of late type galaxies (at
$z \approx 1.5$). This evolutionary difference is similar to our
result that the number density of luminous AGNs (quasars) decays
earlier than less luminous AGNs.

These facts may imply the following, first-order scenario. Luminous
AGNs have lived in galaxies that have large spheroid components at
present, such as ellipticals and S0s. Their SMBHs rapidly grew in
accordance with strong star bursts that happened in early epochs of
the universe (possibly as a result of mergers), followed by a rapid
decrease of their activities after $z\simlt 2$. On the other hand,
galaxies that now have only small spheroid (i.e., spirals) made star
bursts relatively later and/or slower than early type galaxies, and
accordingly the activity of smaller-mass (hence less luminous) AGNs
has continued until recently ($z<1$). If the evolutionary history by
\citet{bal03} is the case, the star forming activity peaks at somewhat
earlier epochs (at $z\approx 3$ for early type and $z\approx1.5$ for
late type galaxies) than the AGN activities of the same population in
the above scenario (at $z\approx2$ and $z\approx0.6$,
respectively). This would imply that the AGN activity of a galaxy
occur not in the same but a later phase than the major star burst. We
note, however, that an apparently weak AGN could be either a small
mass BH or a high mass BH with a small accretion rate, making the
actual story more complex. Obviously we need discussion combined with
the mass function of SMBHs, which shall be left for future work.

\section{Conclusion}

1. From a combination of hard X-ray surveys above 2 keV with various
depths and area performed with \heao , \asca , and \chandra , we have
constructed a highly complete AGN sample consisting of 247 sources in
the wide flux range covering $10^{-10} - 3.8\times 10^{-15}$ \ergs\
(2--10 keV). This provides us with an ideal opportunity to unambiguously
trace the evolution of both type-I and type-II AGNs in the range of
Log \lx\ of 41.5--46.5 and $z=0-3$.

2. For our purpose, we develop an extensive method of calculating the
{\em intrinsic} (before-absorption) hard X-ray luminosity function
(HXLF) and the absorption ($N_{\rm H}$) function. This utilizes the
maximum likelihood method fully correcting for observational biases
with consideration of the X-ray spectrum of each source.

3. We find that the fraction of absorbed AGNs decreases with
increasing luminosities, while the redshift dependence is not
significant within our data. This result requires modification of the
pure ``unified scheme'' of AGNs.

4. The HXLF shows different cosmological evolution between luminous
and less luminous AGN in terms of the cutoff redshift $z_{\rm c}$
above which density evolution terminates: quasars formed earlier than
lower luminosity AGNs. We find that the HXLF is well described with
the luminosity dependent density evolution model where $z_{\rm c}$
decreases from $\simeq 1.9$ at Log \lx\ $\simgt 45$ to $\approx 0.8$
at Log \lx\ $\simeq 43.5$.

5. The combination of the HXLF and the \nh\ function enables us to
construct a purely observation based population synthesis model of
the CXB. The model predicts observational constraints fairly well,
including source counts at faintest fluxes available in the ranges of
0.5--10 keV, the 0.5--2 keV luminosity function determined with \rosat
, and the broad band spectra of the CXB in the 0.5--300 keV band. The
presence of roughly the same number of Compton-thick AGNs with Log
\nh\ =24--25 as those with Log \nh\ =23--24 is consistent with the
observations. Our results are also consistent with the number ratio
between Seyfert~1 and Seyfert~2 galaxies in the local universe and
with the optical luminosity function of quasars at $z=$0.35--2.3 with
reasonable assumptions.

6. These results give a basis of observational constraints on the
accretion history of the universe, i.e., the formation history of
super massive black holes that reside in most of galaxies.

\acknowledgements

We thank G\"unther Hasinger, Yuichi Terashima, and Toru Yamada for
helpful discussion. We acknowledge the efforts by the CDFN team in
producing the catalogs included in the analysis. We are also grateful
to Adam Knudson for carefully reading the manuscript. This work has
been partially supported by NASA Grants NAG5-10043 and NAG5-10875 to
TM.

\clearpage

\clearpage

\clearpage

\clearpage

\clearpage

\clearpage

\clearpage

\clearpage

\clearpage

\clearpage

\clearpage

\clearpage

\begin{figure}
\epsscale{1.0}
\plotone{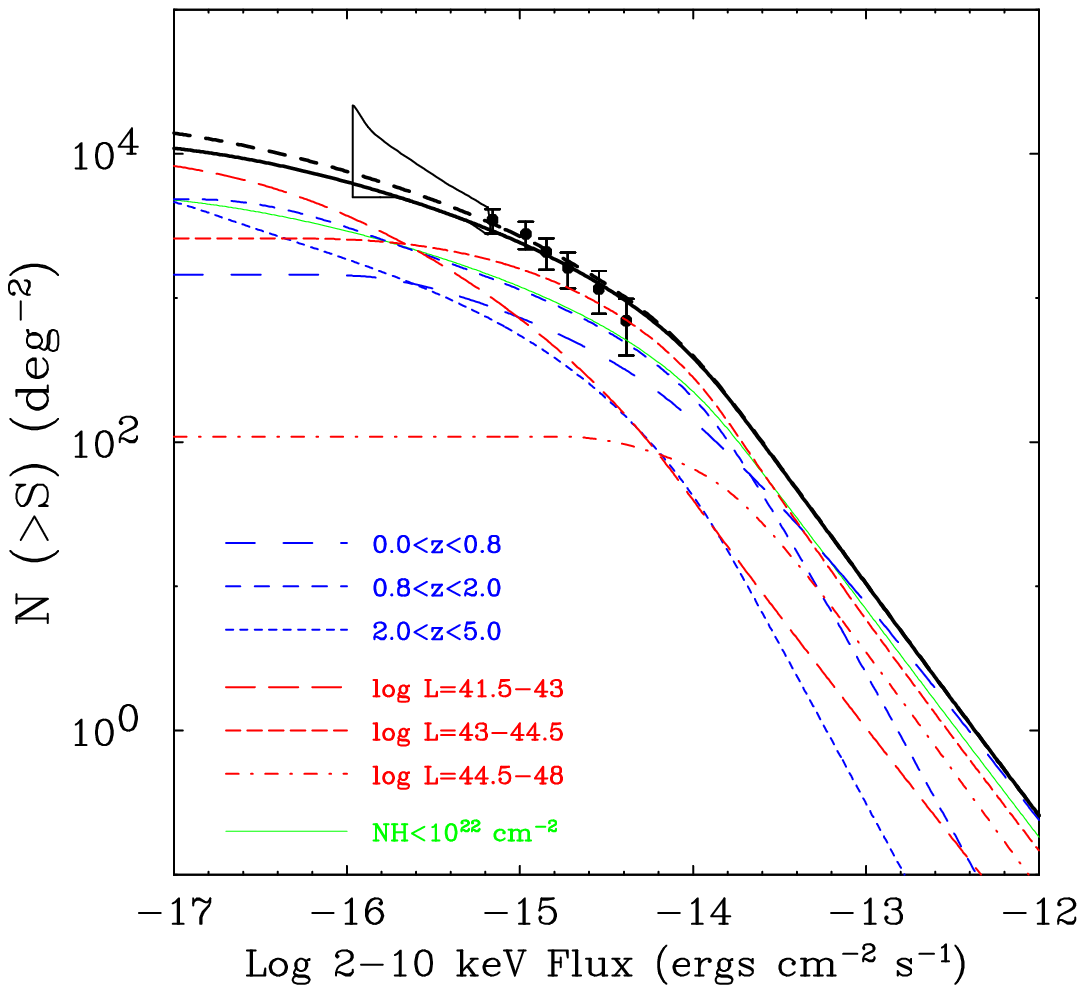}
\end{figure}

\begin{figure}
\epsscale{1.0}
\plotone{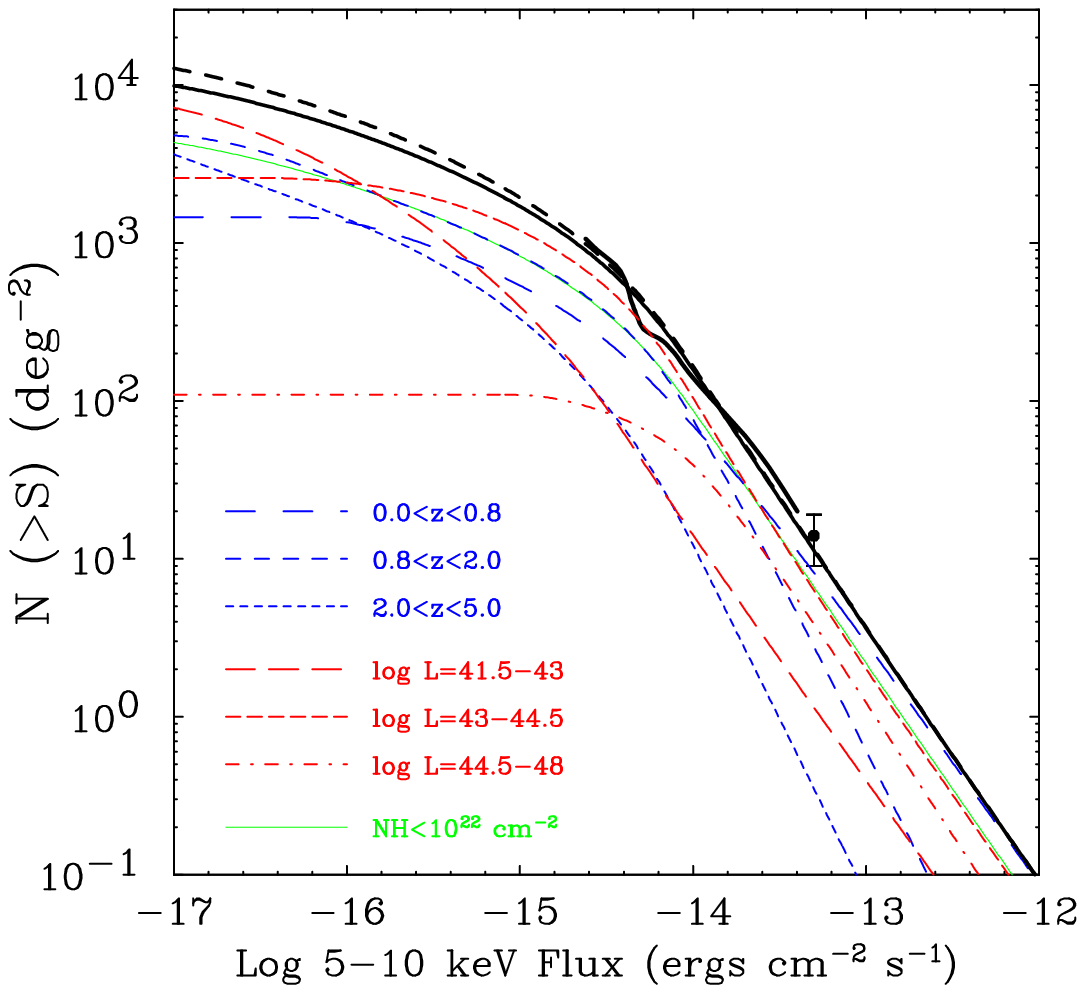}
\end{figure}

\begin{figure}
\epsscale{1.0}
\plotone{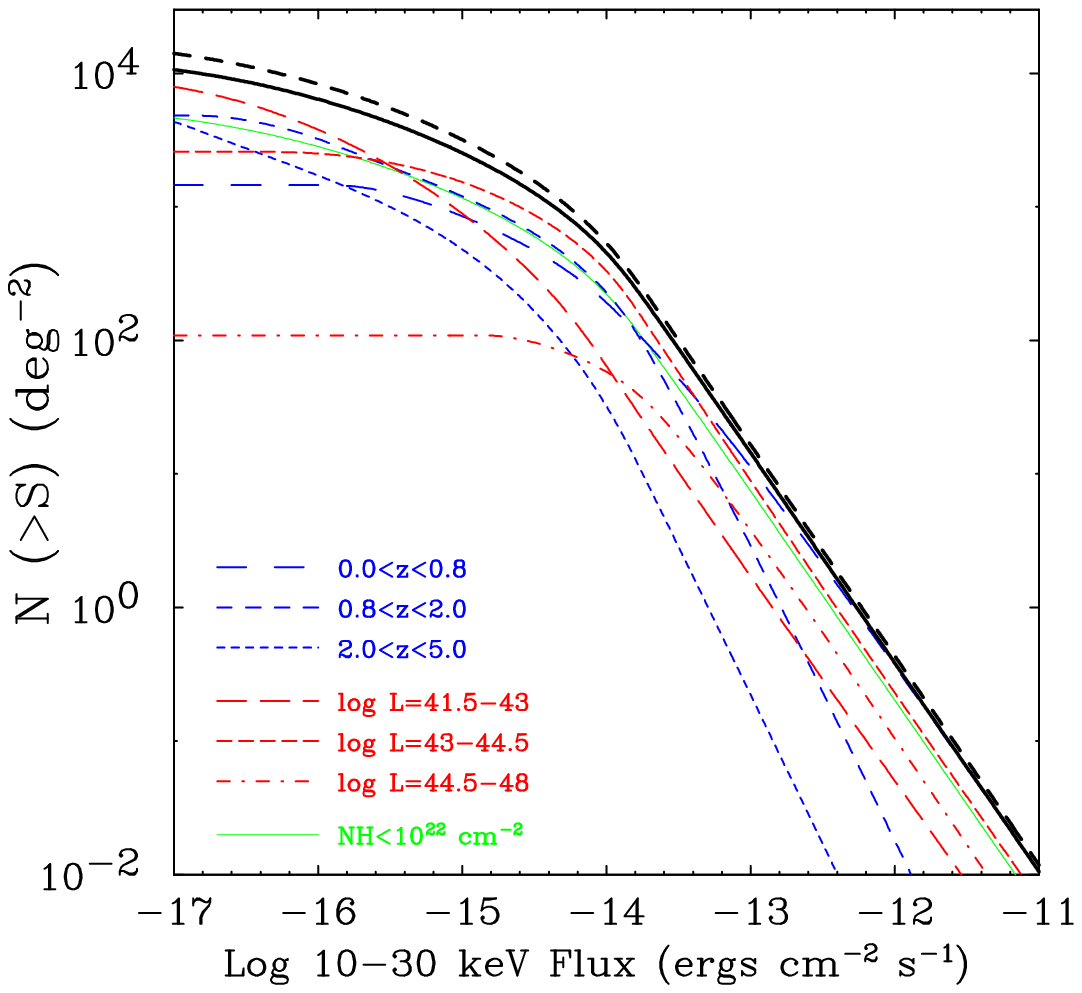}
\end{figure}

\clearpage
\begin{figure}
\epsscale{0.7}
\plotone{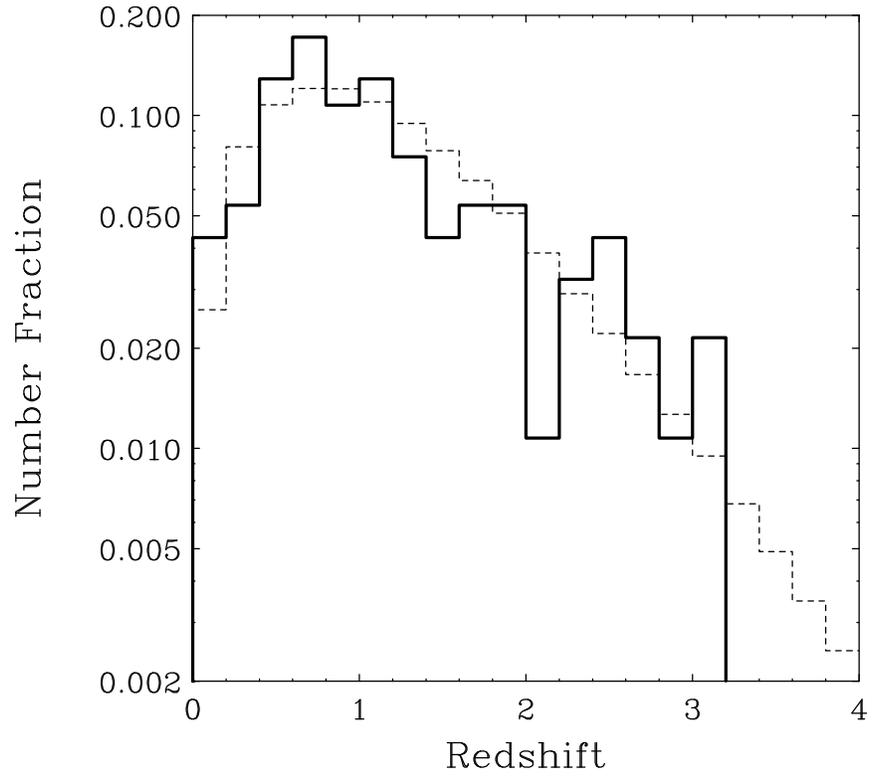}
\caption{
The redshift distribution of the hard-band selected sample compiled by
\citet{gil03} at a flux limit of $5\times10^{-15}$ in the 2--10 keV
band. The dashed histogram is the prediction from our model.
\label{zdist}}
\end{figure}

\clearpage
\begin{figure}
\epsscale{1.0}
\plotone{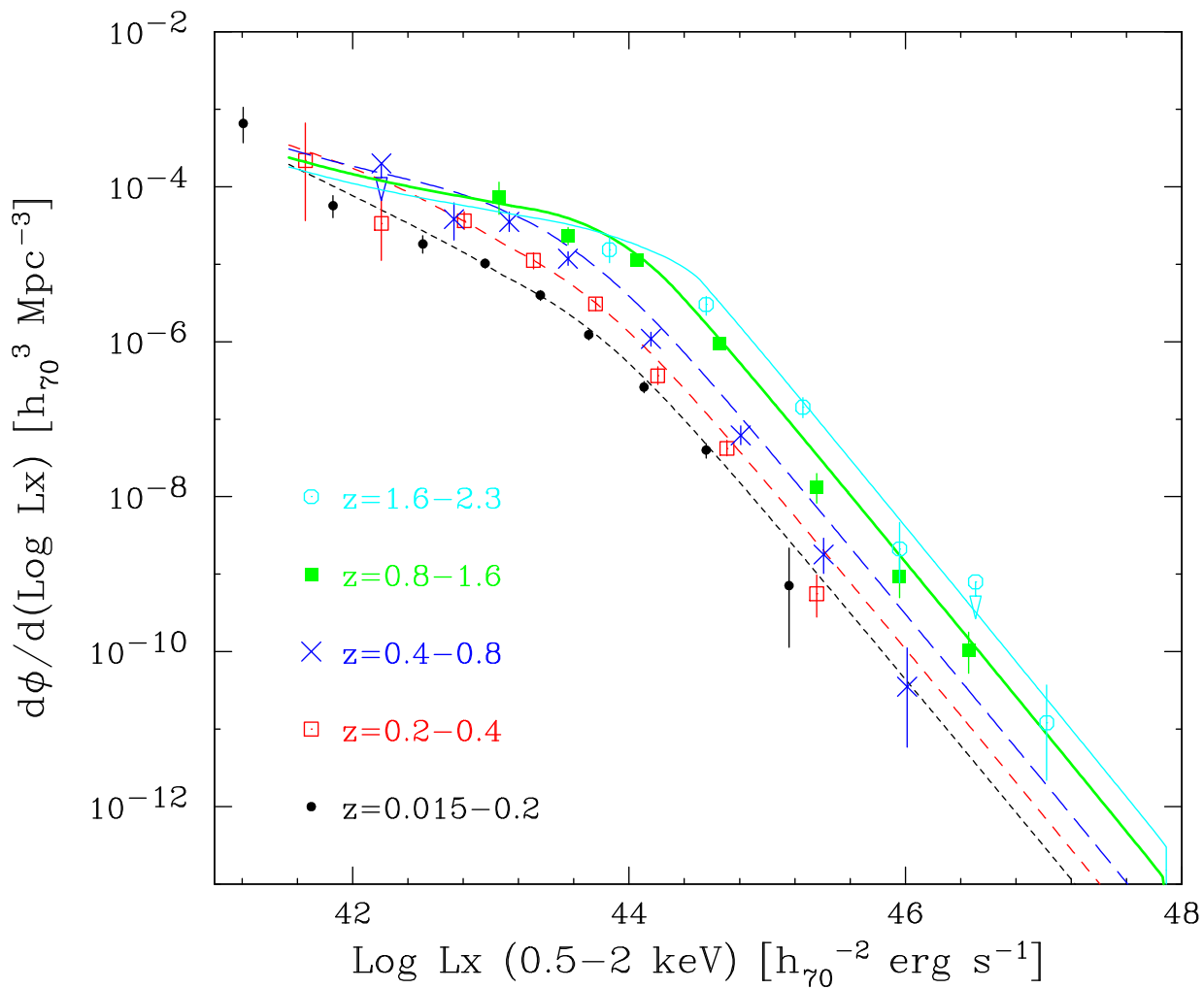}
\caption{Comparison of a predicted SXLF from our HXLF and the \nh\
function (lines) with the \rosat\ result (data points) by
\citet{miy01}, in the redshift bin of $z=$0.015--0.2 (short-dashed, 
black), 0.2--0.4 (medium-dashed, red), 0.4--0.8
(long-dashed, blue), 0.8--1.6 (thick solid, green), and 1.6--2.3 (thin solid, 
cyan).
\label{sxlf}}
\end{figure}

\clearpage
\begin{figure}
\epsscale{0.6}
\plotone{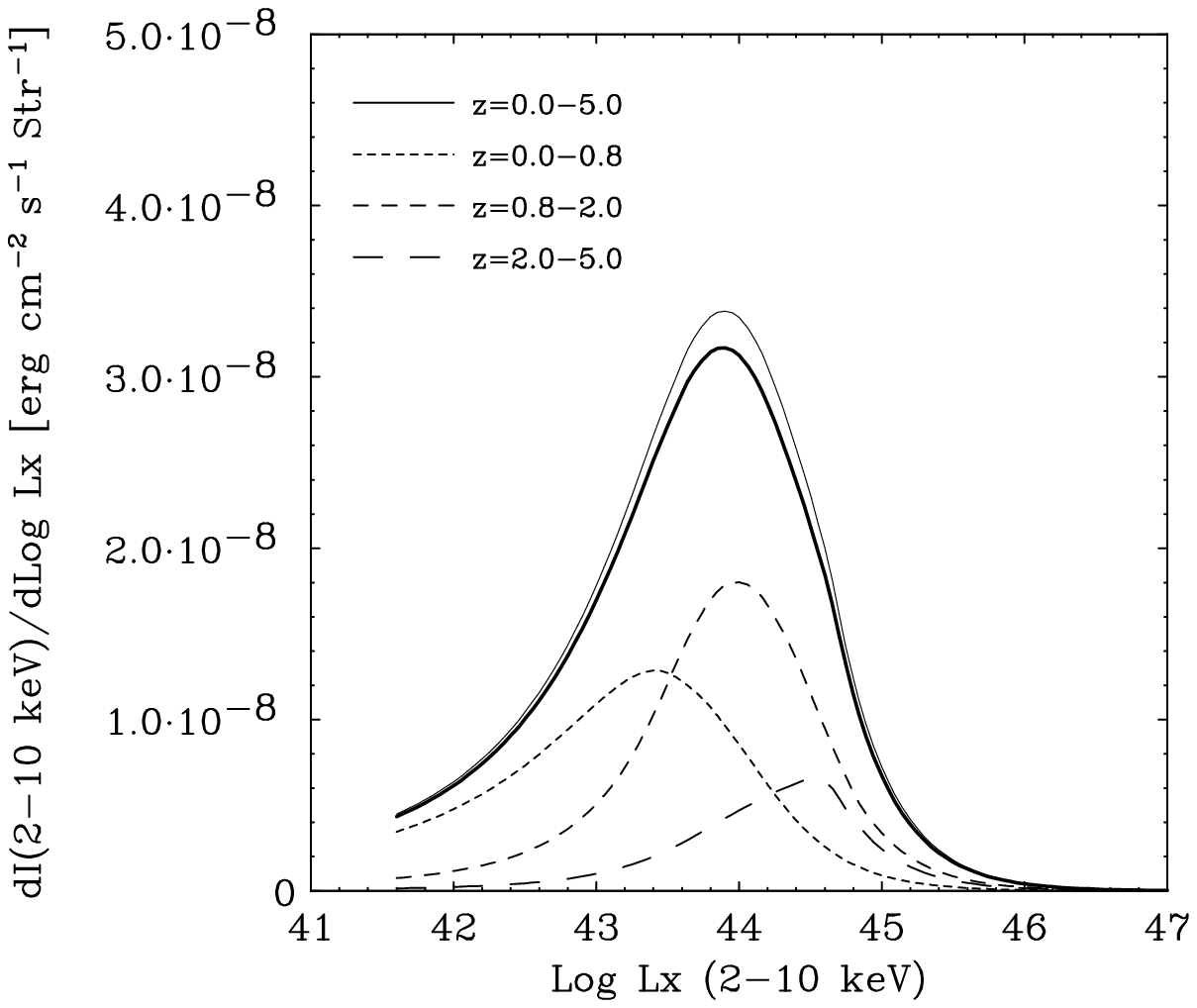}
\plotone{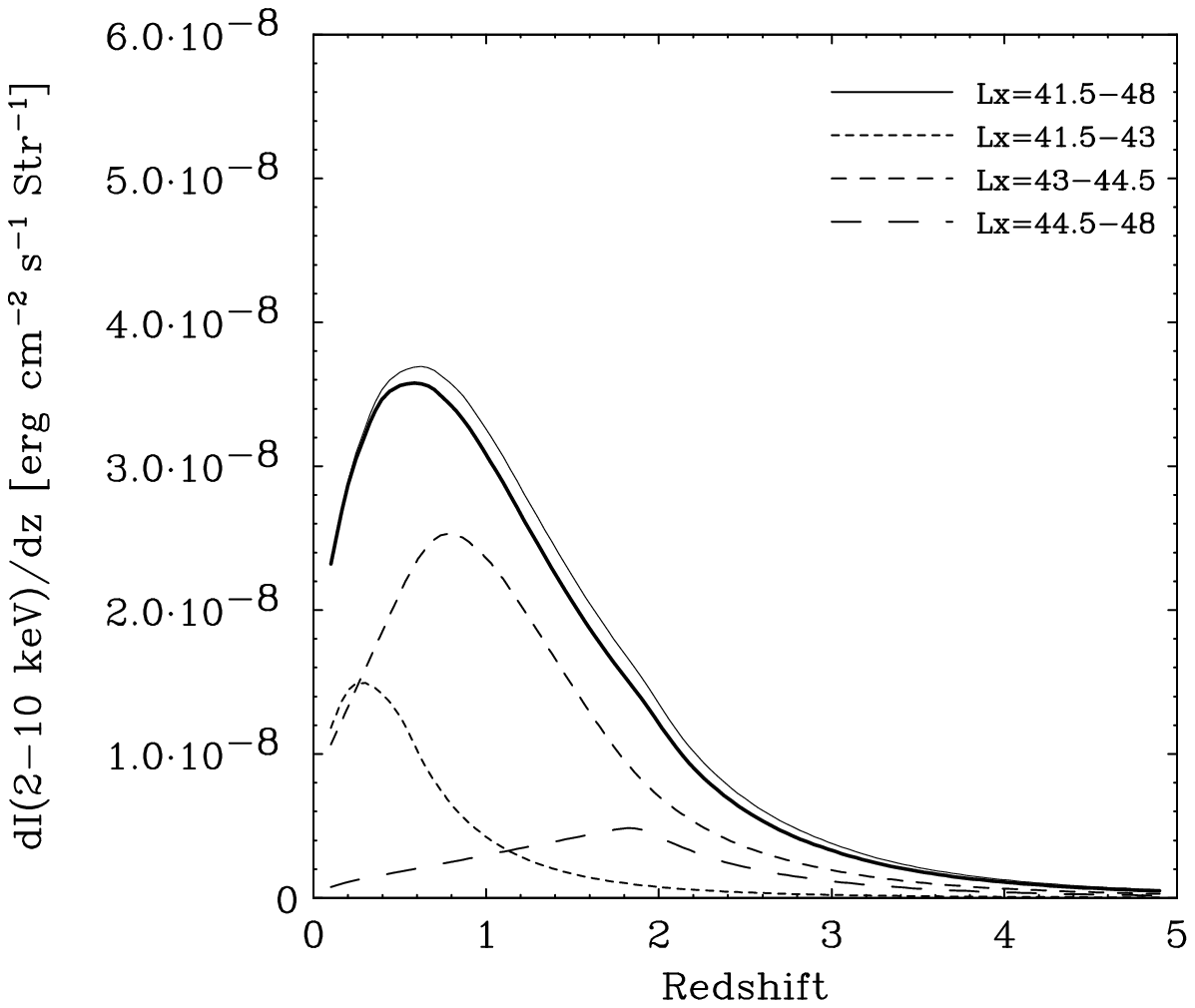}
\caption{
Differential contribution of AGNs to the 2--10 keV CXB intensity as a
function of ({\it upper}: a) luminosity and ({\it lower}: b) redshift,
based on the best-fit model of the HXLF and the \nh\ function. The
dashed lines show contribution from different redshift or luminosity ranges 
as indicated in the figures. The
uppermost curves correspond to the case when the same number of
Compton-thick AGNs with Log \nh\ =24--25 as those with Log \nh\
=23--24 are included.
\label{lzspec2-10}}
\end{figure}

\clearpage
\begin{figure}
\epsscale{0.6}
\plotone{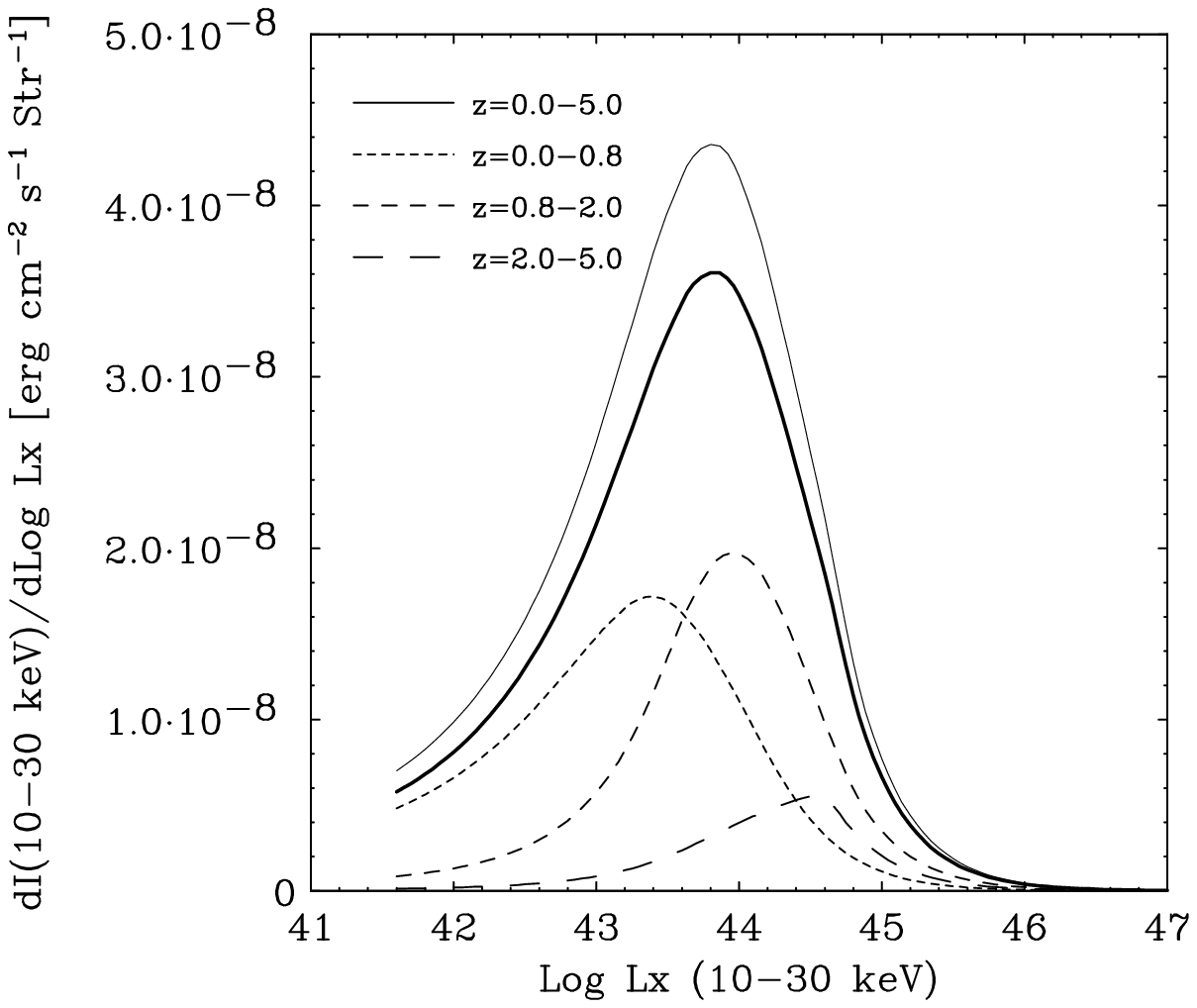}
\plotone{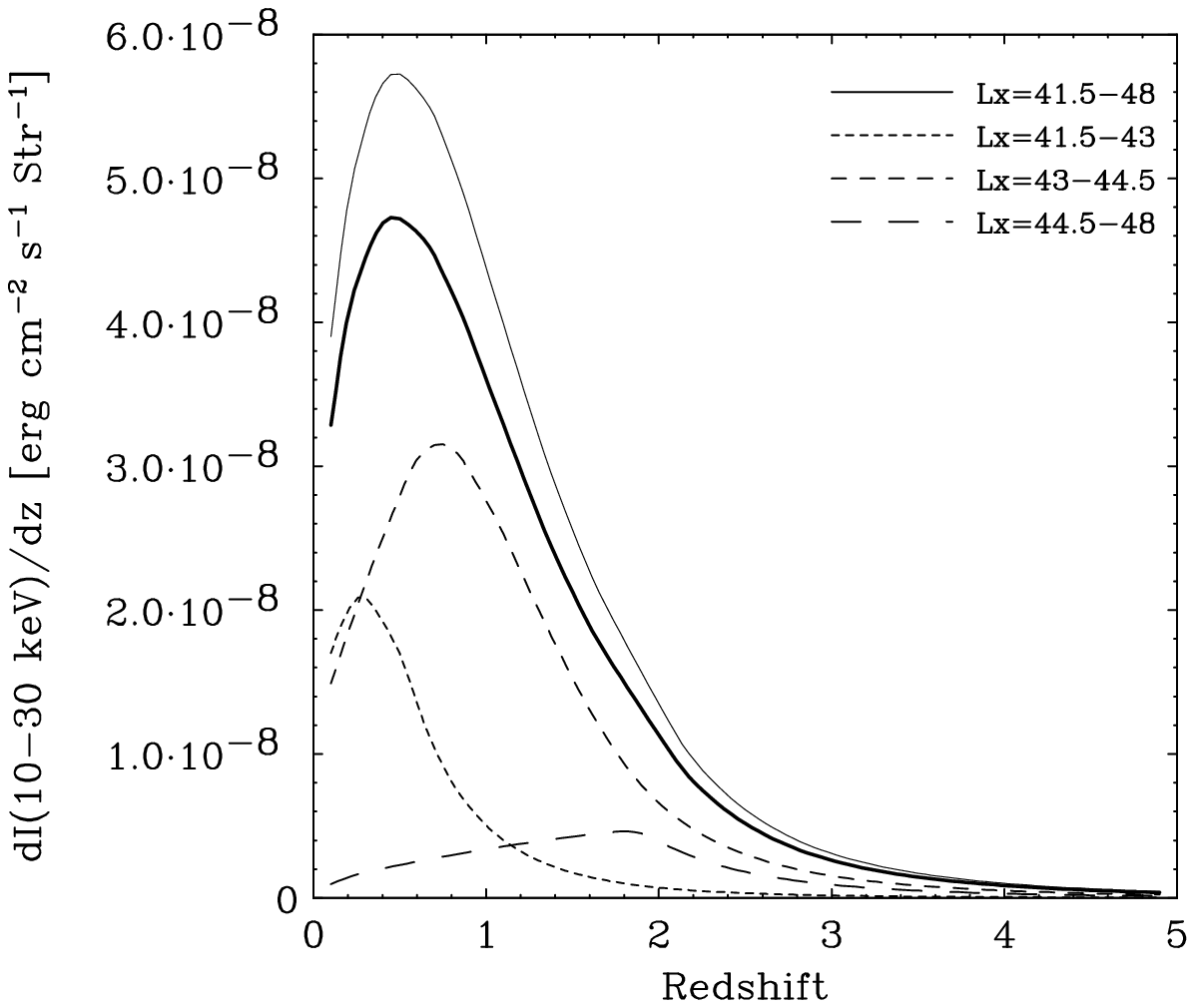}
\caption{
The same as Figure~\ref{lzspec2-10} but calculated for the 10--30 keV CXB.
\label{lzspec10-30}}
\end{figure}

\clearpage
\begin{figure}
\epsscale{1.0}
\plotone{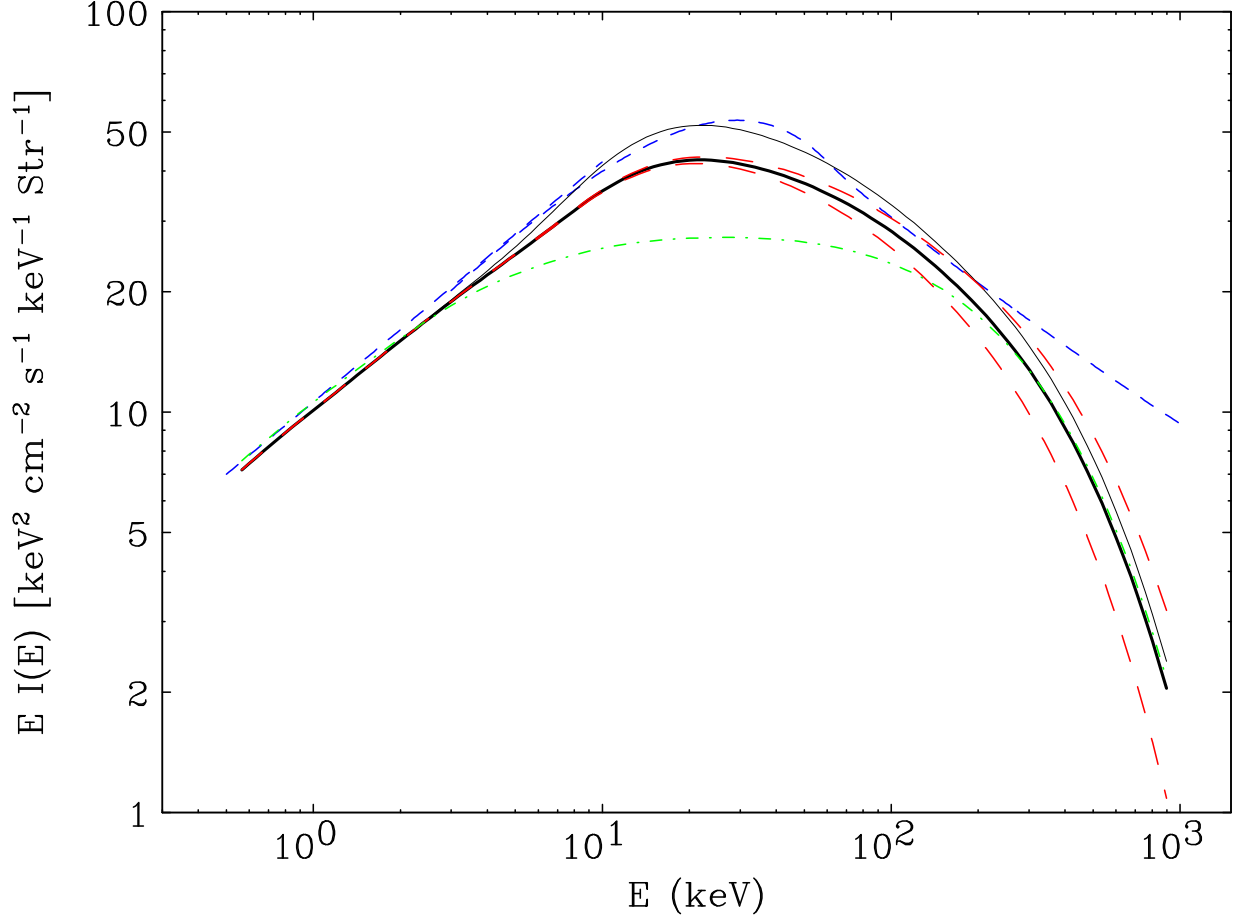}
\caption{
The integrated AGN spectra computed from our HXLF and the
\nh\ function where several different conditions are compared with the
observed CXB spectrum (uppermost dashed curves, blue). For the CXB
spectrum we plot a power law with $\Gamma=1.4$ in the 0.5--10 keV band
assuming the maximum normalization estimated by \citet{bar00}. In the
3--1000 keV band, the analytical formula by \citet{gru99} is plotted
with a normalization increased by 26\% from the original value. Thick
solid curve (black): the integrated spectrum of Compton-thin AGNs with
Log \lx\ = 41.5--48 at $z<$5.0.  Dot-dashed curve (green): that when
the reflection component is not included. Dashed curves (red): those
when the high energy cutoff is changed from $E_{\rm c}$= 500 keV to
400 keV (left) and 600 keV (right). Thin solid curve (black): that
when the same number of Compton-thick AGNs with Log \nh\ =24--25 as
those with Log \nh\ =23--24 are included.
\label{cxbspec}}
\end{figure}

\begin{figure}
\epsscale{1.0}
\plotone{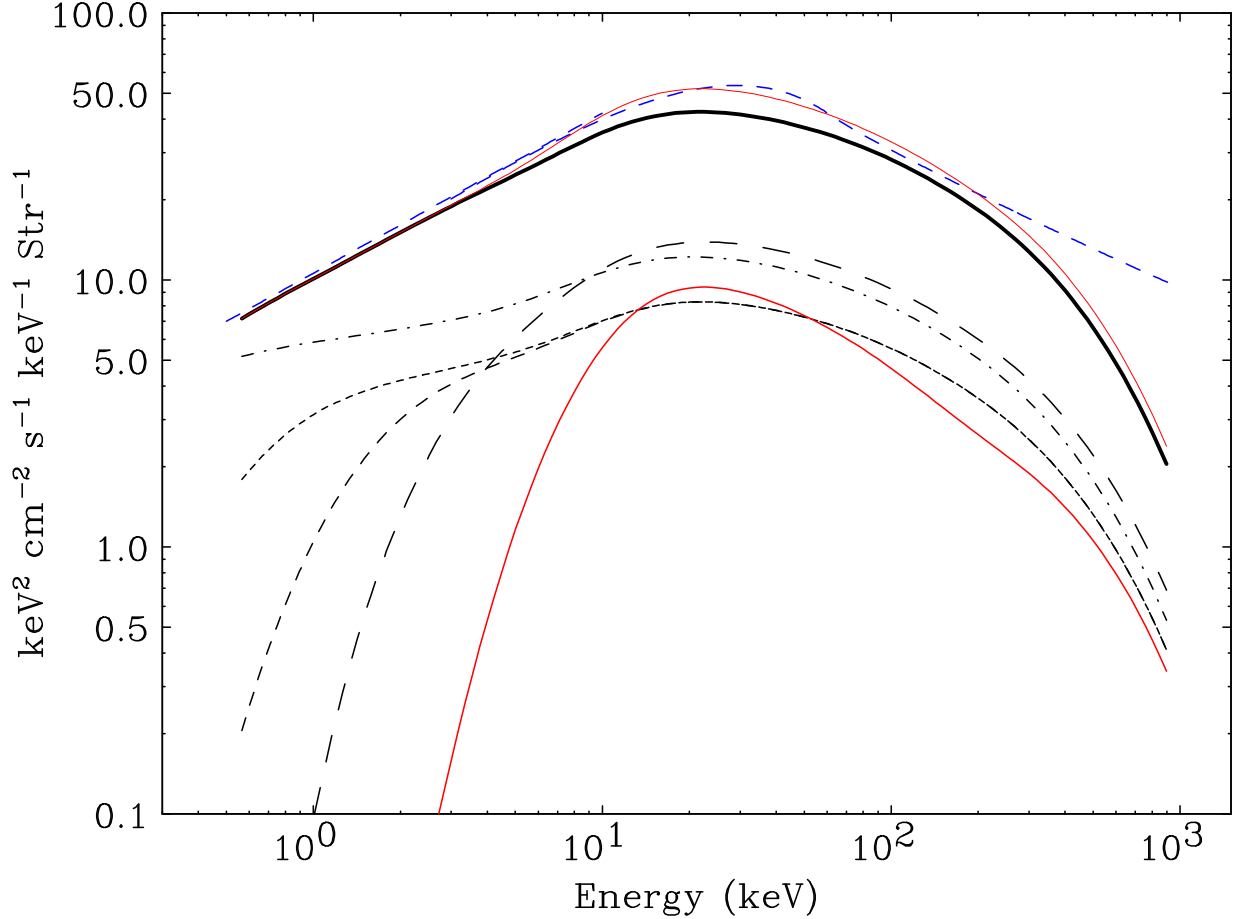}
\caption{Contribution to the CXB from AGNs with different \nh\ ranges. 
Uppermost dashed curves (blue): the CXB spectrum same as
Figure~\ref{cxbspec}. Thick solid curve (black): the integrated
spectrum of Compton-thin AGNs. Upper thin solid curve (red): the
integrated spectrum when the same number of Compton-thick AGNs with
Log \nh\ =24--25 as those with Log \nh\ =23--24 are included. 
Lower curves show separate 
contribution to the CXB from AGNs with Log \nh\ $<21$ (dot-dashed, 
black), Log \nh\ = 21--22 (short dashed), 22--23 
(medium-dashed), 23--24 (long dashed), and 24--25
(solid, red).
\label{cxbspec-nh}}
\end{figure}

\clearpage
\begin{figure}
\epsscale{1.0}
\plotone{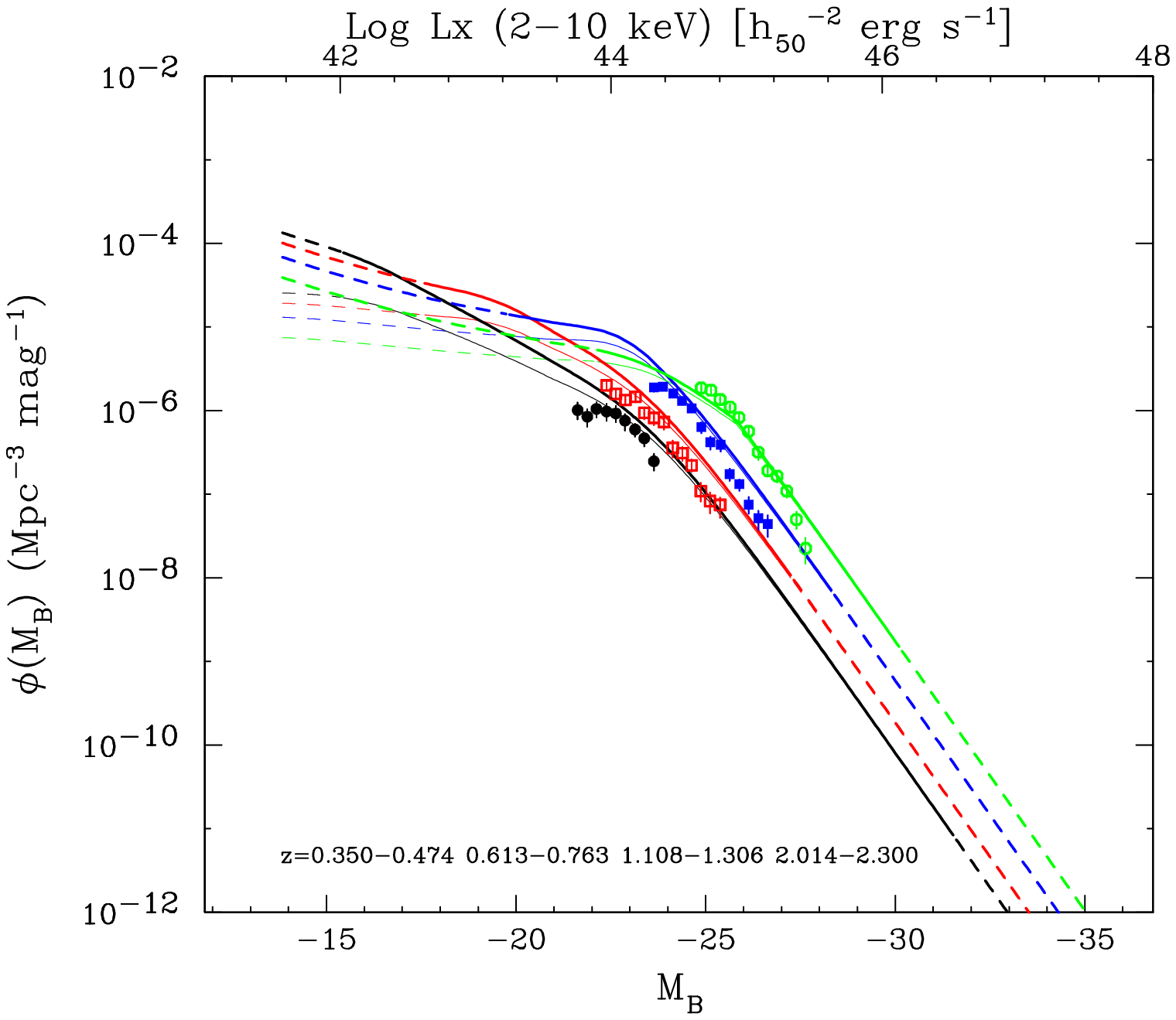}
\caption{Comparison of the optical quasar luminosity function (data
points) by \citet{boy00} with our HXLF of \xtI\ AGNs (thick lines) in
the \cosmo\ = (50$h_{50}$, 1.0, 0.0) universe. From left to right,
$z$=0.350--0.474 (black), 0.613--0.763 (red), 1.108--1.306 (blue), and
2.014--2.300 (green).  The relation $\alpha_{\rm OX} = 0.1152 {\rm Log}
l_{\rm O} - 2.0437$ is assumed between the 2 keV and $2500 \AA$
luminosities, corresponding to $l_{\rm X} \propto l_{\rm O}^{0.70}$
(see \S~\ref{sec-olf} for details). The thin lines represent an
estimated HXLF of only \otI\ AGNs. The dashed lines correspond to 
extrapolated regions where no X-ray sample exists.
\label{olf}}
\end{figure}

\clearpage

\begin{deluxetable}{lccl}
\tablenum{1}
\tablecaption{Surveys Used in the Analysis\label{table-survey}} 
\tablehead{
\colhead{Survey} & \colhead{No.\ of} &\colhead{Flux Limit (2--10 keV)\tablenotemark{a}} &\colhead{Reference}\\ 
\colhead{} & \colhead{Id.\ AGNs} &\colhead{[\ergs ]} &\colhead{}}
\startdata
{\it HEAO1} A2& 28 & $2.7\times10^{-11}$ & \citet{pic82}\nl
{\it HEAO1} MC-LASS& 21 & $1.9\times10^{-11}$ & \citet{gro92}\nl
AMSSn & 74 & $3.1\times10^{-13}$ & \citet{aki03}\nl
AMSSs & 20 & $3.1\times10^{-13}$ &  Akiyama et al., in prep.\ \nl
ALSS & 30 & $1.2\times10^{-13}$ & \citet{aki00}\nl
{\it ASCA} Lockman & 12 & $3.8\times10^{-14}$ & \citet{ish01}\nl
{\it ASCA} Lynx & 5 & $5.0\times10^{-14}$ & \citet{oht03}\nl
CDFN & 57 & $3.8\times10^{-15}$ & \citet{bar02}\nl
\tablenotetext{a}{
To convert the count-rate limit (or flux limit for the CDFN) in the survey
band to the 2--10 keV flux, $\Gamma$=1.7, 1.6, and 1.4 are assumed for
the \heao , \asca , and \chandra\ surveys, respectively.}
\enddata
\end{deluxetable}

\begin{deluxetable}{cccc}
\tablenum{2}
\tablecaption{The Best Fit Parameters of the \nh\ function\label{table-nhf}} 
\tablehead{
\colhead{Photon Index\tablenotemark{a}}&\colhead{1.9} &\colhead{1.9}& \colhead{1.7}\\
\colhead{Reflection\tablenotemark{a}} &\colhead{yes} &\colhead{yes} &\colhead{no}\\
\colhead{\cosmo } &\colhead{(70,0.3,0.7)} &\colhead{(50,1.0,0.0)} &\colhead{(70,0.3,0.7)}
}
\startdata
$\epsilon$ 	&1.7 (fixed) &1.7 (fixed) &1.7 (fixed) \nl
$\psi_{44}$	&0.47$\pm$0.03 &0.48$^{+0.04}_{-0.02}$ &0.41$\pm$0.03 \nl
$\beta$		&0.10$^{+0.04}_{-0.03}$ &0.09$^{+0.06}_{-0.03}$ &0.12$\pm$0.03 
\tablenotetext{a}{Parameters of the ``template spectrum'' assumed to 
derive \nh\ and \lx\ (see \S~\ref{sec-absandlum}). }
\tablecomments{Errors are 1$\sigma$ for a single parameter.}
\enddata
\end{deluxetable}

\begin{deluxetable}{ccccc}
\tablenum{3}
\tablecaption{The Best Fit Parameters of the HXLF models\label{table-hxlf}} 
\tablehead{
\colhead{Model} & \colhead{PLE} &\colhead{PDE} &\colhead{LDDE} &\colhead{LDDE}\\
\colhead{\cosmo } &\colhead{(70,0.3,0.7)} &\colhead{(70,0.3,0.7)} &\colhead{(70,0.3,0.7)} &\colhead{(50,1.0,0.0)}}
\startdata
$A$\tablenotemark{a} &$14.1\pm1.0$ &$2.64\pm0.18$ &$5.04\pm0.33$ &$1.92\pm0.13$ \nl
Log $L_{*}$\tablenotemark{b}& 43.66$\pm$0.17 & 44.11$\pm$0.23 &43.94$^{+0.21}_{-0.26}$ &44.23$\pm$0.19\nl
$\gamma_1$	&0.82$\pm$0.13 &0.93$\pm$0.13 &0.86$\pm$0.15 &0.86$\pm$0.13\nl
$\gamma_2$	&2.37$\pm$0.16 &2.23$\pm$0.15 &2.23$\pm$0.13&2.36$\pm$0.15\nl
$p1$ 		&2.70$^{+0.17}_{-0.25}$ &4.20$\pm$0.32 &4.23$\pm$0.39&4.43$^{+0.34}_{-0.27}$ \nl
$p2$		&0.0 (fixed) &0.0 (fixed) &$-1.5$ (fixed)&$-1.5$ (fixed)\nl
$z_{\rm c}$ (or $z_{\rm c}^*$) &1.15$^{+0.20}_{-0.07}$ &1.14$^{+0.13}_{-0.16}$ &1.9 (fixed)&1.9(fixed)\nl
Log $L_a$\tablenotemark{b} &\nodata&\nodata	&44.6 (fixed) &44.89 (fixed)\nl
$\alpha$	&\nodata&\nodata	&0.335$\pm$0.070 &0.243$\pm$0.040 \nl
\hline
$\epsilon$\tablenotemark{c} 	&1.7 &1.7 &1.7 &1.7 \nl
$\psi_{44}$\tablenotemark{c}	&0.5 &0.5 &0.5 &0.529 \nl
$\beta$\tablenotemark{c}	&0.1 &0.1 &0.1 &0.1  \nl
\hline
$P_{\rm 2DKS}$\tablenotemark{d}	&0.58	& $>$0.9 & $>$0.9 & $>$0.9\nl
$I_{\rm 2-10}$\tablenotemark{e} &1.21&2.02 &0.96 &0.97\nl
$N(>S_{0.5-2})$\tablenotemark{f} 
&2.6&6.1&0.94&0.98 \nl
\enddata
\tablenotetext{a}{In units of [$10^{-6}$ $h_{\rm 50}^3$ Mpc$^{-3}$] for the last column
and [$10^{-6}$ $h_{\rm 70}^3$ Mpc$^{-3}$] for the rest.}
\tablenotetext{b}{In units of [$h_{\rm 50}^{-2}$ \erg ] for the last column
and [$h_{\rm 70}^{-2}$ \erg ] for the rest.}
\tablenotetext{c}{The parameters of the \nh\ function adopted.}
\tablenotetext{d}{The 2-dimensional KS test probability for the \lxz\ distribution in the whole region.}
\tablenotetext{e}{The ratio of the predicted 2--10 keV flux density to 
the CXB intensity of $6.4\times10^{-8}$ \ergs\ Str$^{-1}$ \citep{kus02}.}
\tablenotetext{f}{The ratio of predicted source counts at
$S=7\times10^{-17}$ \ergs\ (0.5--2 keV) to the CDFN result \citep{bra01}.} 
\tablecomments{Errors are 1$\sigma$ for a single
parameter.}
\end{deluxetable}


\clearpage
\begin{thebibliography}{}

\bibitem[Akiyama et al.(2000)]{aki00}
        Akiyama, M., et al. 2000, \apj, 532, 700

\bibitem[Akiyama et al.(2002)Akiyama, Ueda, \& Ohta]{aki02}
        Akiyama, M., Ueda, Y., \& Ohta, K. 2002, \apj, 567, 42

\bibitem[Akiyama et al.(2003)]{aki03}
        Akiyama, M., et al. 2003, \apjs, in press (astro-ph/0307164) (A03)

\bibitem[Avni \& Tananbaum(1986)]{avn86}
	Avni, Y., \& Tananbaum, H. 1986, \apj, 305, 83

\bibitem[Balland et al.(2003)Balland, Devriendt, \& Silk]{bal03}
	Balland, C., Devriendt, J.E.G, \& Silk, J 2003, \mnras, 343, 107

\bibitem[Barcons et al.(2000)Barcons, Mateos \& Ceballos]{bar00}
	Barcons, X., Mateos, S., \& Ceballos, M.T. 2000, \mnras, 316, L13

\bibitem[Barger et al.(2001)]{bar01}
       Barger, A.J., Cowie, L.L., Mushotzky, R.F., \& Richards, E.A. \aj, 121, 662

\bibitem[Barger et al.(2002)]{bar02}
       Barger, A.J., et al. 2002, AJ, 124, 1839

\bibitem[Barger et al.(2003)]{bar03}
       Barger, A.J., et al. 2003, AJ, 126, 632

\bibitem[Boldt (1987)]{bol87}
	Boldt, E., 1987, Phys.\ Rep., 316, 215

\bibitem[Boyle et al.(1993)]{boy93}
	Boyle, B.J., et al. 1993, \mnras, 260, 49

\bibitem[Boyle et al.(1998)]{boy98}
	Boyle, B.J., et al. 1998, \mnras, 296, 1

\bibitem[Boyle et al.(2000)]{boy00}
	Boyle, B.J., et al. 2000, \mnras, 317, 1014, 2000

\bibitem[Brandt et al.(2001)]{bra01}
       Brandt, W.N., et al. 2001, \aj, 122, 2810

\bibitem[Burke et al.(1991)]{bur91}
       Burke, B.E., Mountain, R.W., Harrison, D.C., Bautz, M.W., Doty, J.P.,
       Ricker, G.R., \& Daniels, P.J. 1991, IEEE-ED, 38, 1069

\bibitem[Ceballos \& Barcons(1996)]{ceb96}
	Ceballos, M.T., \& Barcons, X. 1996, \mnras, 282, 493

\bibitem[Comastri et al.(1995)]{com95}
       Comastri, A., Setti, G., Zamorani, G., \&
       Hasinger, G.   1995, \aap, 296, 1

\bibitem[Cowie et al.(2003)]{cow03}
	Cowie, L.L., et al. 2003, \apj, 584, L57

\bibitem[Dickey \& Lockman(1990)]{dic90}
        Dickey, J.M., \& Lockman, F.J. 1990, \araa, 28, 215

\bibitem[Fabian et al.(1990)]{fab90}
	Fabian, A.C., et al. 1990, \mnras, 242, 14

\bibitem[Fabian \& Barcons(1992)]{fab92}
	Fabian, A.C., \& Barcons, X., 1992, \araa, 316, 429

\bibitem[Fasano \& Franceschini(1987)]{fas87}
	Fasano, G., \& Franceschini, A., 1987, \mnras, 225, 155

\bibitem[Franceschini et al.(1999)]{fra99}
	Franceschini, A. et al. 1999, \mnras, 310, L5

\bibitem[Franceschini et al.(2002)Franceschini, Braito \& Fadda]{fra02}
         Franceschini, A., Braito, V., \& Fadda, D. 2002, \mnras, 335, L51

\bibitem[Gandhi \& Fabian(2003)]{gan03}
	Gandhi, P., \& Fabian, A.C. 2003, \mnras, 339, 1095

\bibitem[Gehrels(1986)]{geh86}
        Gehrels, N. 1986, \apj, 303, 336

\bibitem[George et al.(1998)]{geo98}
      George, I.M., et al. 1998, \apjs, 114, 73

\bibitem[Giacconi et al.(2002)]{gia02}
      Giacconi, R., et al. 2002, \apjs, 139, 369

\bibitem[Gilli et al.(1999)Gilli, Risaliti, \& Salvati]{gil99}
      Gilli, R., Risaliti, G., Salvati, M.
      1999, A\&A, 347, 424

\bibitem[Gilli et al.(2001)Gilli, Salvati, \& Hasinger]{gil01}
      Gilli, R., Salvati, M., Hasinger, G.     
      2001, A\&A, 366, 407

\bibitem[Gilli(2003)]{gil03}
	Gilli, R. 2003, in New X-ray Results from Clusters of Galaxies and Black Holes, eds. C. Done, E.M. Puchnarewicz, M.J. Ward, Advances in Space Research, in press (astro-ph/0303115)

\bibitem[Green et al.(1995)]{gre95}
	Green, P.J., et al. 1995, \apj, 450, 51

\bibitem[Grossan(1992)]{gro92}
	Grossan, B., 1992, PhD thesis, MIT

\bibitem[Gruber et al.(1999)]{gru99}
	Gruber, D.E., et al. 1999, \apj, 520, 124

\bibitem[Hasinger et al.(2001)]{has01}
	Hasinger, G., et al. 2001, \aap, 36,5 L45

\bibitem[Hasinger(2003)]{has03}
	Hasinger, G. 2003, in the 13th Annual Astrophysics Conference in Maryland: The Emergence of Cosmic Structure, eds. Stephen S. Holt \& Chris Reynolds; AIP Conf.Proc. 666 (2003) 227-236 (astro-ph/0302574)

\bibitem[Ishisaki et al.(2001)]{ish01}
     Ishisaki, Y., et al. 2001, PASJ, 53, 445

\bibitem[Jones et al.(1997)]{jon97} 
Jones, L.R., et al. 1997, \mnras, 285, 547 

\bibitem[Kauffmann \& Haehnelt(2000)]{kau00}
	Kauffmann, G., \& Haehnelt, M. 2000, \mnras, 576, 311

\bibitem[Kriss \& Canizares(1985)]{kri85}
	Kriss, G., \& Canizares, C.R. 1985, \apj, 297, 177

\bibitem[Kushino et al.(2002)]{kus02}
	Kushino, A., et al. 2002, PASJ, 54, 327

\bibitem[La Franca et al.(1995)]{laf95}
	La Franca, F., et al. 1995, \aap, 299, 19

\bibitem[La Franca et al.(2002)]{laf02}
	La Franca, F., et al. 2002, \apj, 570, 100

 \bibitem[Lawrence \& Elvis(1982)]{law82} 
	Lawrence, A.~\&  Elvis, M.\ 1982, \apj, 256, 410 

\bibitem[Lawson \& Turner(1997)]{law97}
	Lawson, A.J., \& Turner M.J.L. 1997, \mnras, 288, 920

\bibitem[Lehmann et al.(2001)]{leh01}
	Lehmann, I., et al. 2001, \aap, 371, 833

\bibitem[Maccacaro et al.(1991)]{mac91}
      Maccacaro, T. et al. 1991, \apj, 374, 117

\bibitem[Madau et al.(1994)Madau, Ghisellin, \&Fabian]{mad94}
Madau, P., Ghisellini, G., Fabian, A.C. 1994, \mnras, 270, L17

\bibitem[Magdziarz \& Zdziarski(1995)]{mag95}
	Magdziarz, P., \& Zdziarski, A.A. 1995, \mnras, 273, 837

\bibitem[Magorrian et al.(1998)]{mag98}
	Magorrian, J., et al. 1998, \aj, 115, 2285

\bibitem[Mainieri et al.(2002)]{mai02}
      Mainieri, V., et al. 2002, \aap, 393, 425

\bibitem[Maiolino \& Rieke(1995)]{mai95}
    Maiolino, R., \& Rieke, G.H. 1995, \apj, 454, 95

\bibitem[Malizia et al.(2002)]{mal02}
Malizia, A, et al. 2002, \aap, 394, 801

\bibitem[Marshall et al.(1980)]{mar80}
	Marshall, F.E., et al. 1980, \apj, 1980, 235, 4

\bibitem[Miyaji et al.(2000a)Miyaji, Hasinger, \& Schmidt]{miy00}
     Miyaji, T., Hasinger, G., \& Schmidt, M. 2000a, \aap, 353, 25

\bibitem[Miyaji et al.(2000b)Miyaji, Hasinger, \& Schmidt]{miy00b} 
Miyaji, T., Hasinger, G., \& Schmidt, M.\ 2000b, Advances in Space Research, 25, 827 

\bibitem[Miyaji et al.(2001)Miyaji, Hasinger, \& Schmidt]{miy01}
     Miyaji, T., Hasinger, G., \& Schmidt, M. 2001, \aap, 369, 49

\bibitem[Miyaji \& Griffiths(2002)]{miy02}
	Miyaji, T., \& Griffiths, R.E. 2002, \apj, 564, L5

\bibitem[Miyaji et al.(2003a)]{miy03a}
	Miyaji \etal, 2003, \pasj, 55, L11

\bibitem[Miyaji et al.(2003b)]{miy03}
	Miyaji \etal, 2003b, in preparation

\bibitem[Nandra \& Pounds(1994)]{nan94}
	Nandra, K. \& Pounds, K.A. 1994, \mnras, 268, 405

\bibitem[Ohashi et al.(1996)]{oha96}
     Ohashi, T., et al. 1996, \pasj, 48, 157

\bibitem[Ohta et al.(2003)]{oht03}
     Ohta, K., et al. 2003, \apj, in press

\bibitem[Page et al.(1997)]{pag97} 
	Page, M.J., Mason, K.O., McHardy, I.M., Jones, L.R., \& Carrera, F.J.\
	1997, \mnras, 291, 324

\bibitem[Parmar et al.(1999)]{par99}
	Parmar, A.N. et al. 1999, \aap, 345, 611

\bibitem[Piccinotti et al.(1982)]{pic82}
	Piccinotti, G., et al. 1982, \apj, 253, 485

\bibitem[Pompilio et al.(2000)Pompilio, La Franca \& Matt]{pom00} 
	Pompilio F., La Franca, F., \& Matt, G. 2000, \aap, 353, 440

\bibitem[Ranalli et al.(2003)Ranalli, Comastri \& Setti]{ran03} 
	Ranalli, P., Comastri, A., \& Setti, G. 2003, \aap, 399, 39

\bibitem[Reeves \& Turner(2000)]{ree00}
	Reeves, J.N., \& Turner M.J.L. 1997, \mnras, 316, 234

\bibitem[Risaliti et al.(1999)]{ris99}
     Risaliti, G., Maiolino, R., \& Salvati, M. 1999, \apj, 522, 157

\bibitem[Schartel et al.(1997)]{sch97}
	Schartel, N., et al. 1997, \aap, 320, 696

\bibitem[Terasawa (1991)]{ter91}
	Terasawa, N. 1991, \apj, 378, L11

\bibitem[Terashima \& Wilson (2003)]{ter03}
	Terashima, Y., \& Wilson, A.S. 2003, \apj, 583, 145

\bibitem[Turner \& Pounds(1989)]{tur89}
	Turner, T.J., \& Pounds, K.A. 1989, \mnras, 240, 833

\bibitem[Turner et al.(1997)]{tur97}
	Turner, T.J., et al. 1997, \apjs, 113, 23

\bibitem[Ueda et al.(1998)]{ued98}
	Ueda, Y., et al. 1998, \nat, 391, 866

\bibitem[Ueda et al.(1999a)]{ued99a}
	Ueda, Y., et al. 1999a, \apj, 518, 656

\bibitem[Ueda et al.(1999b)]{ued99b}
    Ueda, Y., Takahashi, T., Ohashi, T., \& Makishima, K.
     1999b, \apj, 524, L11

\bibitem[Ueda et al.(2001)]{ued01}
     Ueda, Y., Ishisaki, Y., Takahashi, T., Makishima, K.,
     \& Ohashi, T. 2001, \apjs, 133, 1

\bibitem[Vignali et al.(2003)Vignali, Brandt \& Schneider]{vig03}
	Vignali, C., Brandt, W.N, \& Schneider, D.P. 2003, \aj, 125, 433

\bibitem[Vanden Berk et al.(2001)]{van01}
	Vanden Berk, D.E., et al. 2001, \aj, 122, 549

\bibitem[Weaver et al.(1994)]{wea94}
	Weaver, K.A., et al. 1994, \apj, 436, L27

\bibitem[Wilkes et al.(1994)]{wil94}
     Wilkes, B.J., et al. 1994, \apjs, 92, 53

\bibitem[Wilman \& Fabian(1999)]{wil99}
	Wilman, R.J., \& Fabian, A.C. 1999, \mnras, 309, 862

\bibitem[Wood et al.(1984)]{woo84}
	Wood, K., et al. 1984, \apjs, 56, 507

\bibitem[Yuan et al.(1998)]{yua98}
	Yuan, W., et al. 1998, \aap, 330, 108

\bibitem[Yuan et al.(1998b)Yuan, Siebert, \& Brinkmann]{yua98b}
	Yuan, W., Siebert, J., Brinkmann, W. 1998, \aap, 334, 498

\bibitem[Zdziarski et al.(1995)]{zdz95}
	Zdziarski, A.A., et al. 1995, \apj, 438, L63

\bibitem[Zdziarski et al.(2000)Zdziarski, Poutanen, \& Johnson]{zdz00}
	Zdziarski, A.A., Poutanen, J., \& Johnson W.N, 2000, \apj, 542, 703

\end{thebibliography}
\end{document}